\documentclass[fleqn,10pt]{wlscirep}
\usepackage[utf8]{inputenc}
\usepackage[T1]{fontenc}
\usepackage{subcaption} 
\usepackage{placeins} 
\usepackage{threeparttable} 
\usepackage{scalerel} 

\newcommand{\down}[1]{_{\scaleto{#1}{4pt}}} 
\newcommand{\up}[1]{^{\scaleto{#1}{4pt}}}

\title{Inferring and predicting Fried physical frailty phenotype deficits}

\author[1,*]{Glen Pridham}
\author[2]{Kenneth Rockwood}
\author[1,$\dagger$]{Andrew Rutenberg}

\affil[1]{{Department of Physics and Atmospheric Science}, {Dalhousie University}, {Halifax}, {B3H 4R2}, {Nova Scotia}, {Canada}}
\affil[2]{{Division of Geriatric Medicine}, {Dalhousie University}, {Halifax}, {B3H 2E1}, {Nova Scotia}, {Canada}}

\affil[*]{glen.pridham@dal.ca}
\affil[$\dagger$]{adr@dal.ca}


\begin{abstract}
We predict the Fried physical frailty phenotype health deficits (FPFP5: slow gait, weakness, weight loss, low activity, and exhaustion) using two measures of frailty: frailty index (FI) or frailty phenotype (FP). The FP theorizes that the FPFP5 are mutually dependent through shared etiology and positive feedbacks, so that the total number of FPFP5 deficits (NFPFP5) should be highly predictive of existing deficits. Alternatively, the FI theorizes that strong mutual dependencies exist between \emph{all} age-related health deficits, so that the FI would be more predictive. We investigated predictive models of FPFP5 using FI or NFPFP5 in the Health and Retirement Study (HRS), the English Longitudinal Study of Aging (ELSA), and the National Health and Nutrition Examination Survey (NHANES). We find that the FI, chronological age, and current deficit state are all important predictors of future FPFP5 deficits. Notably, the FI consistently out-performed NFPFP5, raising questions regarding FPFP5 causal connections and how best to measure the physical component of frailty. We discuss implications for both FPFP5 forecasting, and inference when data are missing or incomplete.
\end{abstract}

\begin{document}
\flushbottom
\maketitle
%
%
\thispagestyle{empty}


\section{Introduction}
Frailty is a health state characterized by reduced physiological function across multiple systems leading to an increased risk of adverse outcomes, such as disability, institutionalization, and death \cite{Fried2001-mr,Clegg2013-od,Hoogendijk2019-pk}. Frailty prevalence increases exponentially with age, affecting 30\%-50\% of individuals by age 82.\cite{Hoogendijk2019-pk} In addition to the primary effects on individuals, frailty carries a considerable cost to society \cite{Hoogendijk2019-pk}. Being able to predict decline trajectories is therefore of great practical importance from both a public health and clinical perspective. As multiple systems are involved, the nature of the decline can vary across individuals. Our goal here is to enhance our understanding of the physical component of decline trajectories and frailty through prediction of deficits in key attributes. 
Physical frailty can be characterized by the five key Fried physical frailty phenotype health deficits (FPFP5): weight loss, weakness, poor endurance/exhaustion, slow gait, and low physical activity \cite{Fried2001-mr}. Each deficit is concerning on its own, and potentially catastrophic in unison. For example, gait \cite{Peel2013-an} and grip strength \cite{Bohannon2019-tg} are key function indicators that predict numerous adverse outcomes including onset of disability --- ambulatory deficits are an essential category of disabilities \cite{Edemekong2021-gf}. Exercise can mitigate these effects \cite{Negm2019-rt}, but requires that individuals are able to exercise. Weight loss and reduced strength can prevent exercise, suggesting a positive feedback loop that could exacerbate all five of the FPFP5 \cite{Fried2001-mr}. It is therefore important for us to know how deficits in any one will lead to deficits in the others, and how best to predict future deficits in the FPFP5. 
The FPFP5 are believed to share a mutual underlying driver: homeostatic dysfunction of energy, skeletal muscle and stress response \cite{Fried2021-vh}.
This putative causal connection --- the frailty phenotype (FP) \cite{Fried2001-mr,Fried2021-vh} --- should make the FPFP5 deficits highly mutually-predictive \cite{Bandeen-Roche2020-za}. The severity of FP should also be highly predictive of constituent and future FPFP5 deficits, as quantified by the total number of FPFP5 deficits (NFPFP5)\cite{Kim2024-wx}. 

The FP is one of two leading theories describing frailty \cite{Hoogendijk2019-pk,Kim2024-wx}, the other being the frailty index (FI) \cite{Mitnitski2001-ch}. The contemporary FP picture contends that frailty emerges when age-related decline causes an individual's homeostatic dysfunction to reach a critical level, notably in the hypothalamic-pituitary-adrenal (HPA) stress-response, musculoskeletal and metabolic systems \cite{Fried2021-vh}. Maladaptive feedbacks then drive a vicious cycle of compounding health deficits, characterized by the FPFP5. In contrast, the frailty index (FI) picture posits that age-related health deficits promote further deficits \cite{Mitnitski2006-fq}, making the FI a useful summary health measure that captures this effect. Associated with each picture is an algorithm to estimate the degree of frailty \cite{Kim2024-wx}. Hence we have two prospective theories, each with a widely-available associated measure that should be a salient predictor of FPFP5 deficits. The practical difference between the two pictures is that the FP is estimated by a small set of specific, physical deficits \cite{Fried2001-mr}, whereas the FI is estimated by a large set of non-specific, age-related deficits (30+ from multiple domains) \cite{Searle2008-xi}. The widespread use of these pictures makes the data needed widely available across studies and relatively easy to measure, making both pictures potentially useful for addressing both public health and clinical perspectives.
 


Measures of function commonly have missing data, especially performance measures.
While the averaging process used in the FI has some intrinsic robustness against missingness \cite{Pridham2022-ko}, the FP is more susceptible to bias due to missing values. This is especially so in weight, gait and grip strength measurements in large-scale studies --- since they are frequently missing. When individuals are not measured they are at high risk of informative missingness \cite{Hardy2009-nl}, e.g.\ individuals too frail to measure, potentially leading to bias unless properly imputed \cite{Pridham2022-ko}. This may not be a serious limitation in a clinical setting, but can greatly limit applicability and reproducibility of population studies \cite{Theou2015-uz}. This complicates translational medicine, since the clinical FP may not coincide with the FP from such studies. An important auxiliary goal of our work is to determine how well we can infer missing values --- particularly using an FI based on questionnaire data with low missingness. If the FI is a good predictor, we can use it to impute unknown values. Multiple imputation can be used to estimate the uncertainty associated with the imputation.
Using the longitudinal Health and Retirement Study (HRS), the longitudinal English Longitudinal Study of Aging (ELSA), and the cross-sectional 2001-2002 National Health and Nutrition Examination Survey (NHANES) we analyzed future prediction and contemporaneous inference based on the FI or NFPFP5 (number of FPFP5 deficits). We use NFPFP5 as the ordinal measure of FP severity, since it has been shown to perform better than other FP gradings \cite{Kim2022-ou}. We directly compare the ability of the NFPFP5 to predict functional deficits to that of an FI generated from other variables. This is achieved using ELSA and HRS to compare future prediction of the FPFP5 at the followup time. The approach is also applied to cross-sectional data from NHANES, using a leave-one-out approach wherein 4 of the FPFP5 are used to predict dysfunction in the 5th. We additionally forecast survival for HRS and NHANES as an alternative followup outcome (ELSA in supplemental).

\section{Materials and Methods}
We used data from three national studies: HRS (longitudinal), ELSA (longitudinal), and NHANES (cross-sectional). Our goal is to understand relationships between variables. Definitions necessarily varied across the studies for both the FI and FP. We considered separate exclusions for predicting health deficits versus survival. When predicting health deficits we considered only individuals with all three measurements: weight, weakness and activity, and at least one of grip or gait (both grip and gait are proxies for sarcopenia \cite{Fried2001-mr} so imputation should pick up one from the other). We did not apply this cut when predicting survival, instead we considered all individuals irrespective of FPFP5 measurements. Our focus is on age- and frailty-related decline and hence we consider only individuals aged $60+$; this also avoids issues with gated variables \cite{Pridham2022-ko}. We also dropped any individuals with top-coded ages, since we are interested in age-dependence (i.e.\ ages capped at max values, age 90+ for ELSA and age 85+ for NHANES).

The criteria used for defining FPFP5 deficits are summarized in Table~\ref{tab:fpvar}. NHANES necessarily and notably differed from the other two studies in three major ways: (i) NHANES did not include weight loss as part of the weight deficit --- we tried to compensate for this by using a higher BMI cut, (ii) NHANES used self-reported weakness instead of grip strength measurement, and (iii) NHANES used an objective question of physical ability for exhaustion based on difficulty walking between rooms, rather than a subjective feeling of being exhausted or unable to ``get going''. 

For the FI we used self-reported limitations in activities of daily living (ADLs) and instrumental activities of daily living (IADLs), chronic disease diagnoses, healthcare utilization, signs, and symptoms. The specific variables used differed across the studies (lists in supplemental). We generated an FI entirely using self-reported questionnaire data (default), but also considered a modified FI that includes the FPFP5 deficits, as indicated. In all comparisons the NFPFP5 had access to as much or more of the FPFP5 specific information than the FI (i.e.\ the NFPFP5 was never disadvantaged whereas the FI was in some tests).
See supplemental for full details.

\begin{table}[!h]
    \centering
    \caption{FPFP5 deficit definitions.} \label{tab:fpvar}
\begin{threeparttable}
    \begin{tabular}{l|lll}
        Deficit & ELSA & HRS & NHANES  \\ \hline
        Weight & $\text{BMI}\leq 18.5~kg/m^2$ or lost 10\%  & $\text{BMI}\leq 18.5~kg/m^2$ or lost 10\%  & {$\text{BMI}\leq 22.5~kg/m^2$} \\ 
        Weakness & Bottom quintile\tnote{1,x} & Bottom quintile\tnote{1,x} & {Self-reported, objective} \\
        Gait & Bottom quintile\tnote{1,x} & Bottom quintile\tnote{1,x}  & Bottom quintile\tnote{x}  \\
        Exhaustion & Self-reported, subjective & Self-reported, subjective & {Self-reported, objective} \\
        Activity & Self-reported, objective & Self-reported, objective  & {Self-reported, subjective}  \\ 
        \hline
    \end{tabular}
\begin{tablenotes}
\item[1] cuts were learned from first wave (wave 2 for ELSA and wave 8 for HRS).
\item[x] sex-stratified quantiles.
\end{tablenotes}
\end{threeparttable}
\end{table}

We used waves 8-14 (2006-2018) from HRS via the RAND preprocessed files \cite{hrs} (only these waves had all needed variables). Waves are measured every 2 years, but gait and grip are only measured every 4 years. We thus analyzed two sets of preprocessed data: one for predicting FPFP5 deficits (feature selection and prediction of decline) and another for survival, with spacings of 4~years and 2~years, respectively. We excluded individuals who entered the study aged below 60 ($\sim17000$) for a total of 13848 for survival study. For FPFP5 deficit analysis we further rejected individuals missing both gait and grip measurements, leaving 5619 individuals. After preprocessing, the median gap between measurements was $4.1$~years (interquartile range: $0.3$~years)  

We used waves 4 and 6 from ELSA \cite{elsa} (only these waves had all needed variables). We excluded $1055$ individuals missing both gait and grip measurements, and an additional 13 with top-coded age. ELSA survival estimates were based on end-of-life interviews, which capture only a fraction of the deaths due to a variety of response rate and fieldwork issues \cite{NatCen_Social_Research2015-nw}. This means that we necessarily underestimate the mortality rate because we are forced to assume that any individual without an end-of-life interview was censored instead of dying. We therefore present ELSA survival data only in supplemental (results qualitatively agree with NHANES and HRS but with a much lower mortality rate). After preprocessing, the median gap between measurements was $4.0$~years (interquartile range: $0.0$~years). 

We used the 2001-02 NHANES with linked public mortality data \cite{nhanes} (2001-02 was the last year to report gait speed). We excluded 577 individuals with top-coded age ($236$) and/or missing both gait measurement and self-reported weakness ($474$).

\subsection{Data}
Demographics for the study data are summarized in Table~\ref{tab:demo}. Demographics and prevalences were similar across waves and datasets, with the exceptions of weight loss (low BMI) and exhaustion in NHANES --- which were slightly different measurements as described above. In all three studies FP prevalence was in or near the typical range (7-10\%) \cite{Fried2021-vh}.

\begin{table}[!h]
    \centering
\begin{threeparttable}
    \caption{Summary of datasets\tnote{1}. Prevalence or mean (standard deviation).} \label{tab:demo}
    \begin{tabular}{llll}
         & HRS & ELSA & NHANES \\ \hline
        Individuals & 5619 & 3053 & 1295 \\ 
        Entries & 8965 & 3053 & 1295 \\ 
        Age & 74 (5) & 69 (7) & 70 (7) \\ 
        Females & 56.1\% & 56.1\% & 50.3\% \\ 
        FI & 0.15 (0.11) & 0.12 (0.10) & 0.13 (0.12) \\ 
        NFPFP5 & 0.91 (1.02) & 0.71 (1.00) & 0.67 (0.98) \\ 
        FP frailty\tnote{2} & 8.2\% & 7.3\% & 6.5\% \\ 
        Weight loss & 5.7\% & 4.6\% & 12.6\% \\ 
        Weakness & 26.9\% & 15.0\% & 23.3\% \\ 
        Slow gait & 15.1\% & 16.2\% & 10.7\% \\ 
        Exhaustion & 27.0\% & 22.0\% & 6.9\% \\ 
        Low activity & 16.4\% & 13.4\% & 13.2\% \\ \hline
    \end{tabular}
\begin{tablenotes}
\item[1] Datasets are post-imputation and are the populations used to analyze functional decline. When analyzing survival we included more individuals as noted in the main text (HRS also included more timepoints).
\item[2] FP~frailty is defined by NFPFP5 $\geq 3$.
\end{tablenotes}
\end{threeparttable}
\end{table}

\subsection{Missing data}
We imputed using Multivariate Imputation by Chained Equations (MICE) \cite{mice} with the classification and regression tree (CART) model. We imputed 15 times for each dataset and pooled using Rubin's rules \cite{Murray2018-gp}. CART handles any type of data and non-linearities; it's also been shown to perform well with NHANES health deficit data \cite{Pridham2022-ko}. Missingness was clearly not completely at random: individuals with all 5 FPFP5 measurements had a 53\% and 81\% lower risk of death in NHANES and HRS, respectively (both $p<2\cdot 10^{-16}$). Failure to impute under these circumstances may lead to biased results \cite{Schafer2002-bj} (e.g.\ in the FI \cite{Pridham2022-ko}). (We permitted either grip or gait to be missing to try and minimize sampling bias, since they have high missingness and can be plausibly imputed since they are mutually associated.)

\subsection{Outcome modelling}
Analysis was done in \texttt{R} version 4.1.1. For parametric models, we used logistic regression \cite{R_Core_Team2021-uq} to predict FPFP5 deficits and Cox proportional hazard regression for survival \cite{surv}. We refer to model inputs as predictor variables. To avoid confusion, we refer to any model where the outcome is a current state as \textit{inference}, whereas future states are referred to as prediction. While we started with linear models, we observed from the calibration curves that the relationship was sub-linear and so we also considered first transforming the predictor by the square root (which, in general, gave a visually excellent fit). Logistic regression used the weighted exogenous sampling method  \cite{King2001-mq} with a balanced prior, this is necessary to avoid trivial models for the rare deficits (see supplemental of \cite{Pridham2023-yj} for details). Logistic models were tested using 10-fold, 10-repeat cross-validation \cite{caret}. We were less concerned about overfitting with the survival models (no weights, linear models) and thus the estimates are direct fit estimates with asymptotic errors. See Supplemental Section~S5 for model diagnostics.

In addition we considered non-parametric models, using the area under the receiver operating characteristic curve (AUC) for FPFP5 deficits \cite{Robin2011-ou} and the linearly-interpolated Kaplain-Meier for survival \cite{surv}. AUC errors were estimated by bootstrap (2000 resamples; default) \cite{Robin2011-ou}, always assuming higher is worse (age, FI or NFPFP5). (Note that for survival we always treat the first measurement as time 0, rather than stratifying by chronological age.)

\subsection{Notation}
Confidence intervals (95\%) are included for tabulated values where available. The lower bound is represented by a subscript and the upper bound by a superscript. Logistic regression coefficients are summarized by their odds ratios (the ratio of the odds that the event happened over didn't happen). Survival regression coefficients are summarized by the hazard ratio for binary variables and the hazard ratio per standard deviation for continuous variables.

\section{Results}
\subsection{Feature selection indicates that the FI is better than NFPFP5 for predicting and inferring FPFP5 deficits}
We start by looking at the relationship between the three major predictor variables: the FI, NFPFP5 and chronological age. Both the FI and NFPFP5 grow exponentially with age, with strong sex-effects (males always lower; Supplemental Figure~S1). The correlations between FI or NFPFP5 vs age were modest, with Spearman $\rho < 0.3$ (supplemental). In contrast, the FI and NFPFP5 were moderately to strongly correlated, with Pearson $\rho = 0.52$ (HRS), $0.58$ (ELSA), or $0.70$ (NHANES) (including the FPFP5 in the FI definition increased these to $0.65$, $0.71$, or $0.82$, respectively).

In feature selection, our question is which summary health measure best predicts FPFP5 deficits. We compare the FI, NFPFP5, and chronological age. Chronological age serves as a reference and a null test. If chronological age out-performs the frailty measures then we can infer that the frailty measures may only be useful due to their age-dependence. Our outcomes are the FPFP5 health deficits (gait, activity, exhaustion, weakness and weight loss) at the followup timepoint. Since the NFPFP5 has the advantage of prior knowledge of each deficit, we consider 3 tests versus the FI: (Test 1) NFPFP5 vs FI where the FI includes the FPFP5 in its estimation, (Test 2) leave-one-out NFPFP5 vs FI where we have excluded self-prediction from the NFPFP5 (e.g.\ slow gait is predicted by the number of deficits in the remaining 4), and (Test 3) NFPFP5 with all measures vs the default FI that excludes the FPFP5. We performed only Test 2 to cross-sectional data for in-wave inference, since the other tests would include knowledge of the outcome in the predictor variable. 

\begin{figure*}[!ht]
    \centering 
    \begin{subfigure}[t]{0.49\textwidth}
        \centering
        \includegraphics[width=\textwidth]{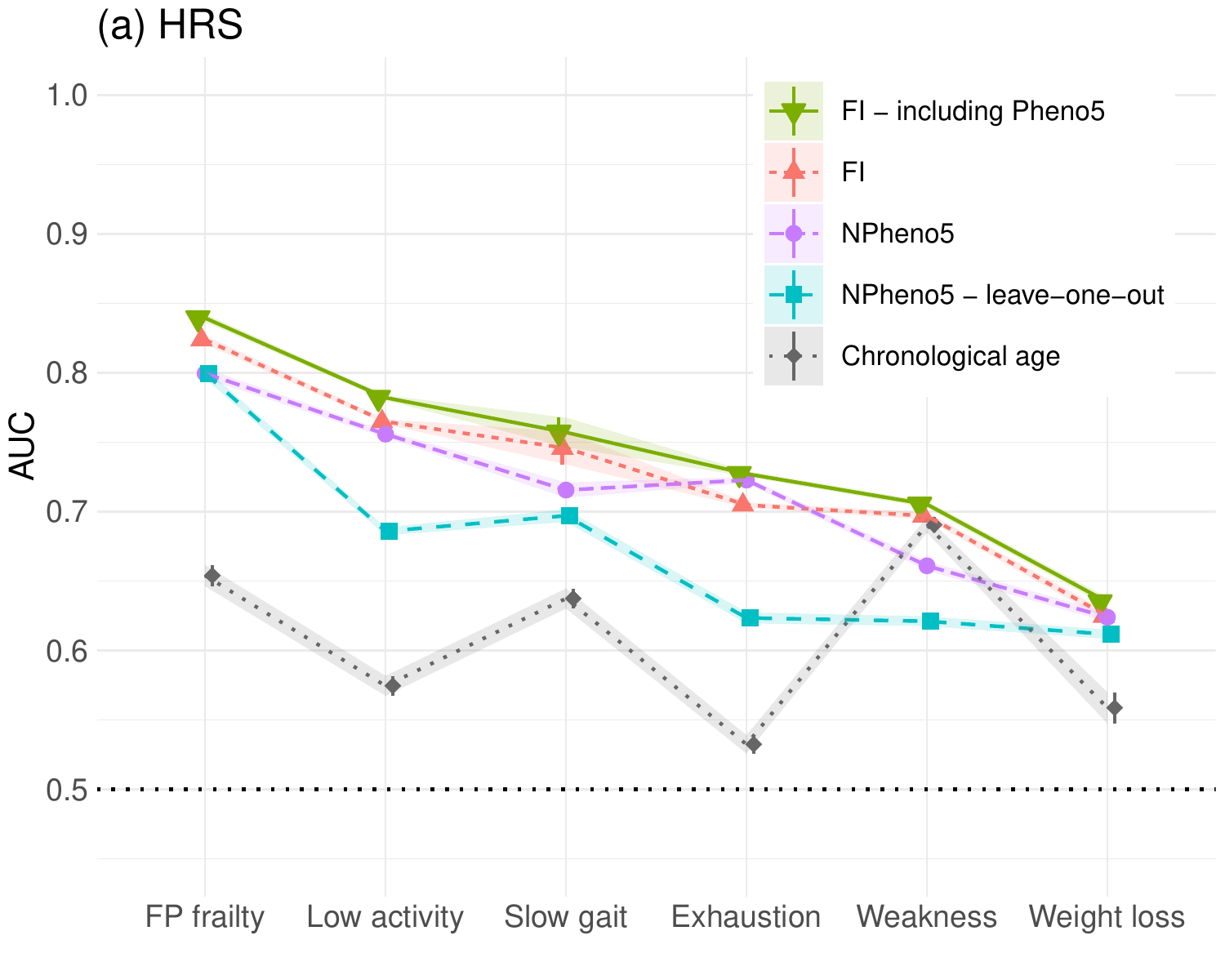} 
    \end{subfigure}
    ~
    \begin{subfigure}[t]{0.49\textwidth}
        \centering
        \includegraphics[width=\textwidth]{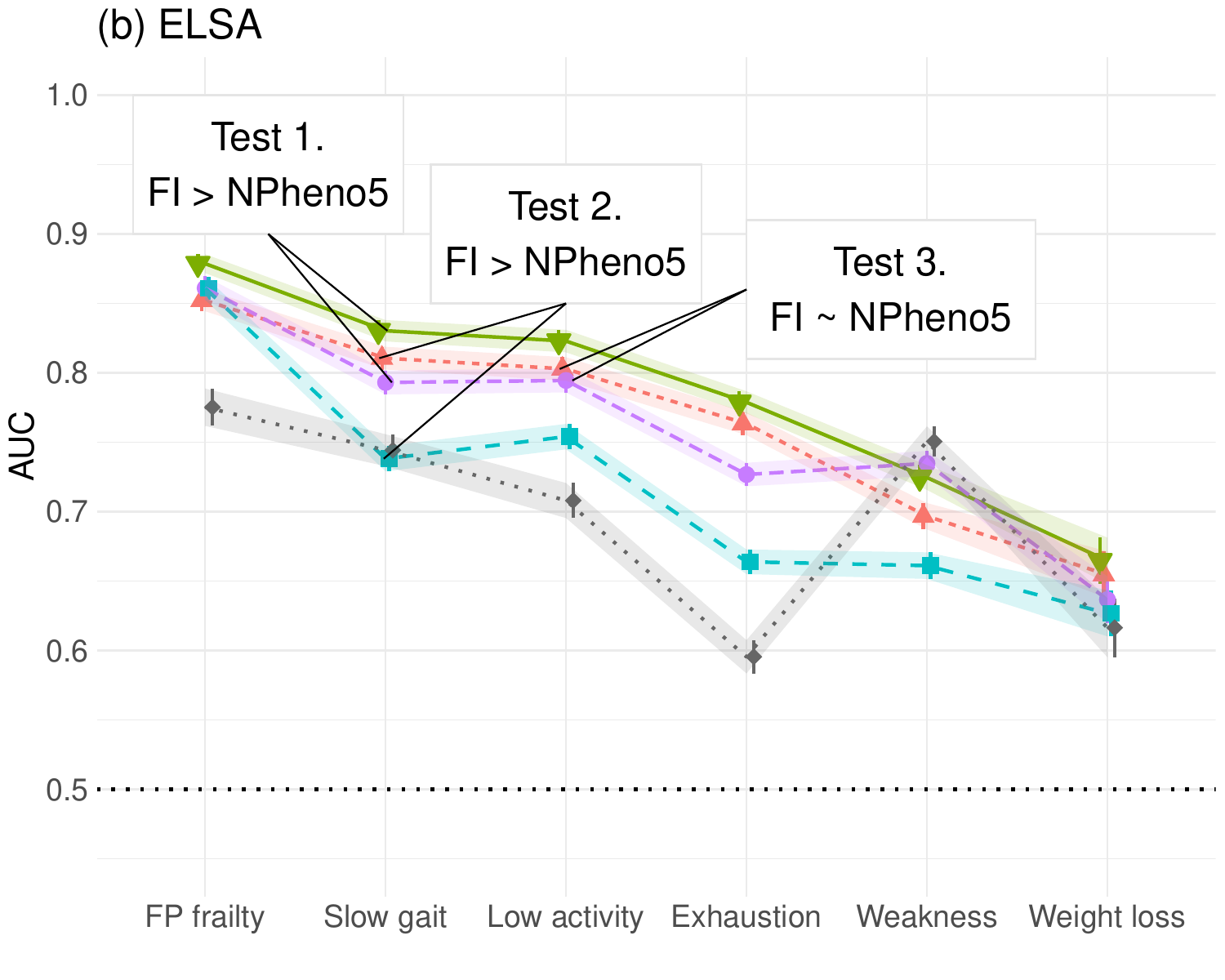} 
    \end{subfigure}%
    \caption{\textbf{The FI predicts future FPFP5 deficits better than NFPFP5.} The FI that includes the FPFP5 performed the best (green upside-down triangles). Similarly, when the default FI (red triangles) was compared to leave-one-out NFPFP5 (cyan squares), the FI was superior. Both the NFPFP5 and FI typically out-performed chronological age, except for predicting weakness (deficit grip strength). Only NFPFP5 including all deficits (purple circles) performed comparably to the default FI. The AUC is the probability that a metric will correctly rank positive individuals as higher than negative individuals \cite{Hanley1982-pm} (dotted line at 0.5 indicates a random guess). Leave-one-out excludes the outcome deficit from the predictor. Error bars are standard errors.} \label{fig:auc}
\end{figure*}

The FI out-performs NFPFP5 for followup deficit prediction, Figure~\ref{fig:auc}. The FI and NFPFP5 contain a great deal of overlapping information, as recapitulated by the Pearson correlation which was $>0.5$ in all studies. The default FI and full NFPFP5 perform similarly-well when predicting future health deficits in the FPFP5, despite the default FI lacking knowledge of current FPFP5 deficits (i.e. Test 3). For example exhaustion is better predicted by the FI and weakness is better predicted by NFPFP5 in ELSA, but the converse is true in HRS. Nevertheless, the default FI performed slightly better, exceeding the error interval in 4 predictions, whereas the NFPFP5 was better in just 2.  The FI with FPFP5 (green points) performs the best, while the NFPFP5 without the predictor variables (blue points) performs the worst. As a result for all three tests, the FI performed as well or better than NFPFP5. This indicates that the FI is the best health measure at predicting FPFP5 health trajectories.

For the cross-sectional NHANES data, only Test 2 is non-trivial: leave-one-out NFPFP5 vs the FI. Leave-one-out NFPFP5 again performed poorly (Supplemental Figure~S2). The FI is a very strong inference engine for FP~frailty, exhaustion and weakness ($\text{AUC}>0.9$), good at inferring gait and activity ($\text{AUC} >0.8$), but of little use at inferring low BMI ($\text{AUC} \sim0.5$). This demonstrates that when FPFP5 data are missing, the FI could be used to estimate the missing value or FP~frailty status (except for low BMI). 

In all three datasets, the FI performed as well or better than NFPFP5, including when the FI had no access to FPFP5 deficit information. 
Sex-stratified analyses (see supplemental) confirm both the prediction and inference patterns and the superior performance of the FI.
We therefore select the FI as our primary predictor. 

\begin{figure*}[!ht]
        \includegraphics[width=\textwidth]{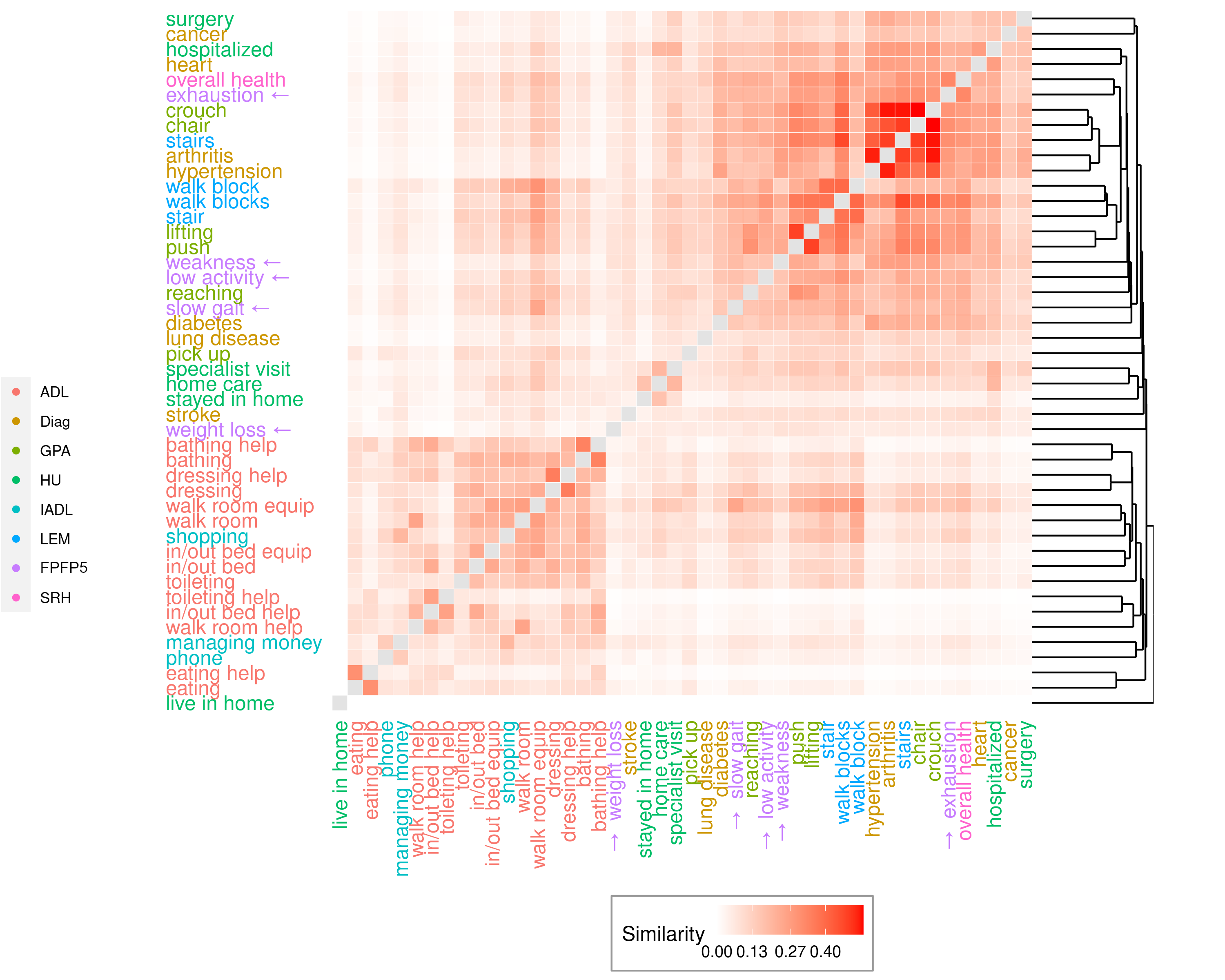} 
    \caption{\textbf{There was no distinct FPFP5 cluster among HRS health deficits (purple labels; arrows)}. Red tiles are similar deficits whereas white are different, as measured by the Jaccard similarity coefficient ($1-\text{Jaccard distance}$). Variables are  hierarchically clustered \cite{R_Core_Team2021-uq} (right). There is no FPFP5 cluster, instead there are a large number of similar deficits that cluster with the FPFP5 and which can be used to improve prediction and inference.  ADL: activity of daily living, Diag: medical diagnosis, GPA: general physical activity, HU: health care utilization, IADL: instrumental ADL, LEM: lower extremity mobility, SRH: self-reported health. See supplemental for ELSA and NHANES.} \label{fig:clustering}
\end{figure*}

The success of the FI appears to be due to the similarity of the underlying health deficits, Figure~\ref{fig:clustering}. The FPFP5 don't form a unique cluster but rather are interspersed among many other health deficits. This offers an explanation for the success of the FI. There are a large number of health deficits that are closely related to, but not included with, the FPFP5. By using all of them in prediction and inference we can better exploit the underlying health states governing the FPFP5.



\subsection{The FI and baseline deficit status are useful for predicting future FPFP5 deficits or death}
\begin{figure*}[!ht]
    \centering
    \begin{subfigure}[t]{0.5\textwidth}
        \centering
        \includegraphics[width=\textwidth]{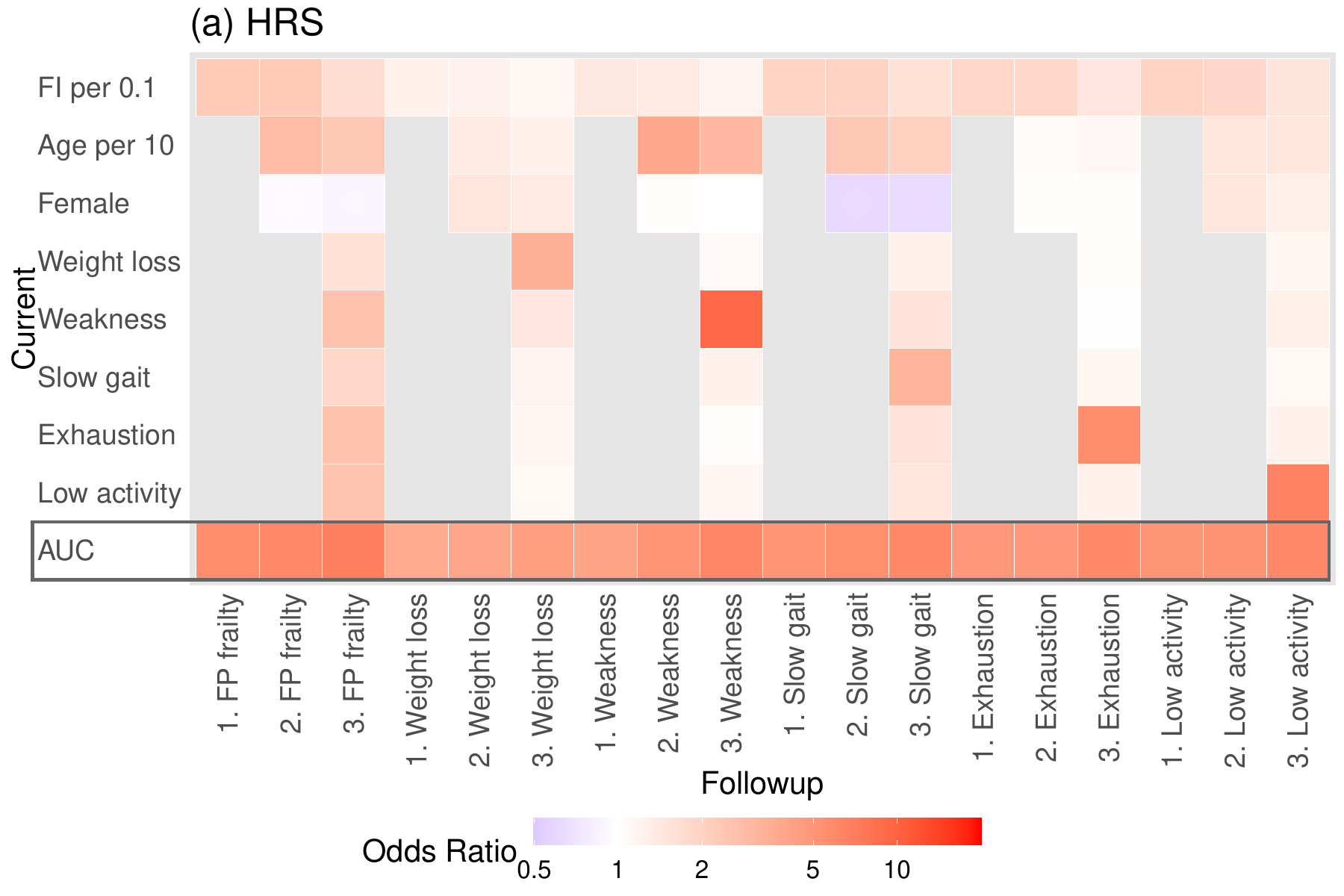} 
    \end{subfigure}
    ~
    \begin{subfigure}[t]{0.5\textwidth}
        \centering
        \includegraphics[width=\textwidth]{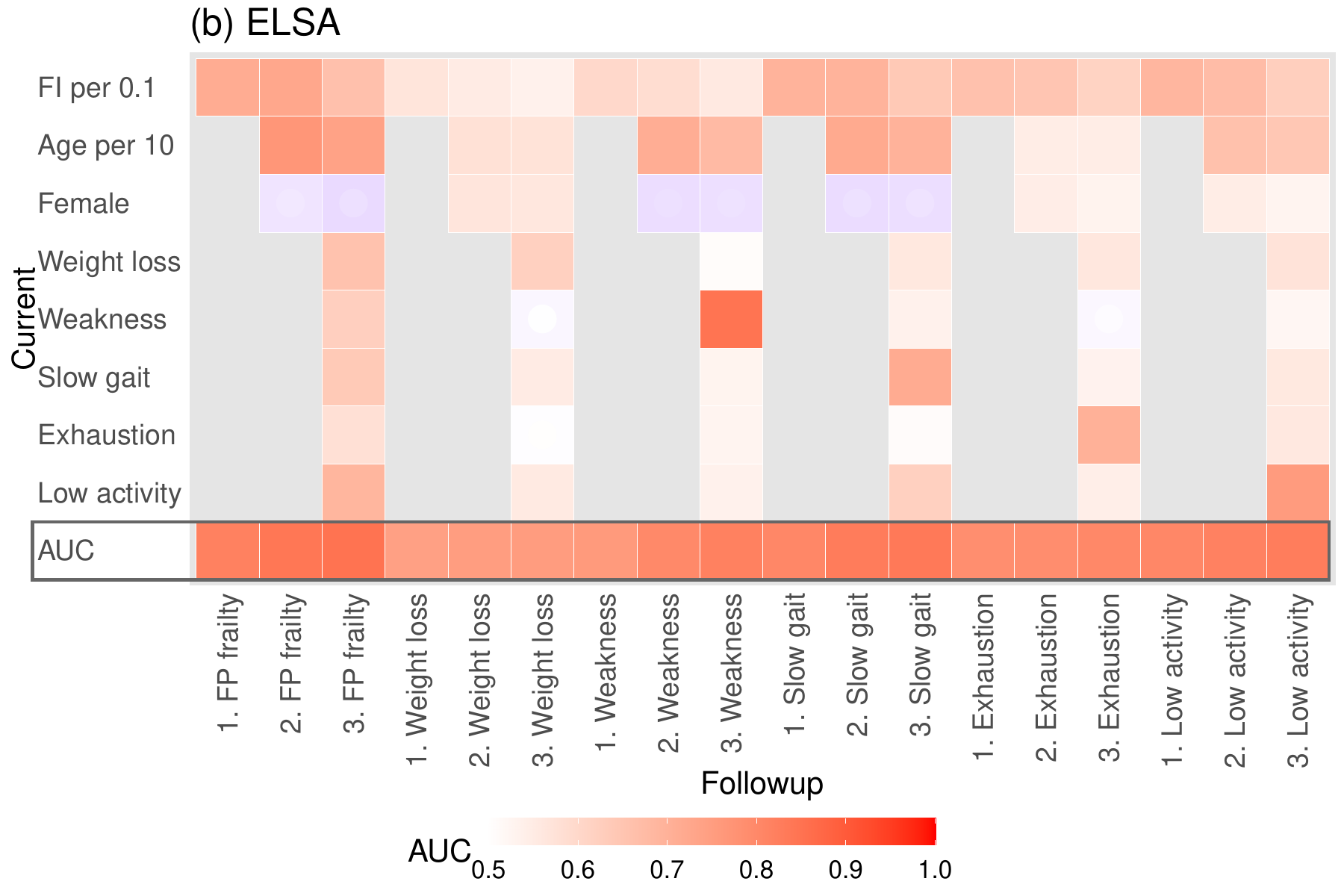} 
    \end{subfigure}%
    \caption{\textbf{Predicting future FPFP5 deficits using multivariate logistic regression.} Prediction performance is very similar between HRS (\textbf{a}) and ELSA (\textbf{b}). Each model treats the indicated followup value as the outcome and includes the current values as predictors, columns alternate between model 1, 2, and 3 as indicated. The bottom AUC row indicates overall model performance (0.5: guess to 1: oracle). The remaining tiles are odds ratios on log$_{10}$ scale. Larger odds ratio means higher probability of having the deficit at followup.  Nested circles are the lower limit of the 95\% CI closest to 0 (typically not visible: indicating high significance). Estimates are 10-fold, 10-repeat cross-validated result from linear multivariate logistic regression. See supplemental for specific values. Note: legends are shared by both figures. Greyed out tiles are not applicable. Predictors: Model 1: FI only; Model 2: FI , age and sex; Model 3: FI, age, sex, and previous FPFP5.} \label{fig:logreg}
\end{figure*}

Having determined that the FI is the better summary measure of deficit risk, we use it to predict future decline.  We start with the FI and build three nested linear-logistic regression models to predict future functional status at followup, approximately 4~years in the future: (Model 1) just the FI, (Model 2) Model 1 + age and sex, and (Model 3) Model 2 + the baseline FPFP5 deficits. We use the nested structure to determine which variables provide additional predictive power. While the specific odds ratios have substantial inter-study differences (Supplemental Tables~S6 and S7), the overall patterns for prediction were very similar, as shown in Figure~\ref{fig:logreg} for HRS and ELSA. As indicated by the AUC, the FI is a good solo predictor of specific FPFP5 deficits (Model 1). Inclusion of sex and age (Model 2) did not have large effects. When we included the baseline FPFP5 variables (Model 3) we found that baseline status of specific deficits was  a  strong predictor of the followup deficit. This means that FPFP5 deficits have a stronger tendency to persist than to emerge. The  odds ratios varied considerably, however, with weakness being highly predictive of future weakness, versus weight loss being a relatively weak predictor. In general, none of the FPFP5 deficits were strong predictors of different FPFP5 deficits.

The other important possible outcome for each individual is that they die before the followup measurement. For time-to-death from the first measurement using Cox proportional hazard regression, Table~\ref{tab:cox} summarizes the same three nested models together with (Model 0): only age+sex. Harrell's C \cite{Harrell1982-qt} shows that the FI captures the majority of the predictive power when measured repeatedly (for HRS, every 2 years). This information overlaps with chronological age and specific deficits, and possibly sex, such that the hazard ratio for the FI per 0.1 changes considerably when other variables are included. For ELSA see Supplemental Table~S8. When the FI is measured just once (NHANES) there are also substantial gains to be made by including age (and potentially sex). 

We also see from Table~\ref{tab:cox} that the deficits in the FPFP5 increase the risk of death at most moderately (hazard ratios in last column of each study are small relative to age, which is by far the largest risk factor). When we included specific FPFP5 deficits (largest model), they had only moderate hazard ratios and made little difference in either of the model parameters: Harrell's C and the Bayesian Information Criterion (BIC). Regardless, these variables provide a more individualized prediction of risk factors for mortality that (partially) supplants the FI, as indicated by the shrinking hazard ratio when additional variables are included. 

\begin{table}
    \centering
    \begin{threeparttable}
    \caption{Cox survival hazard ratios with 95\% confidence intervals (and BIC) --- HRS (time-dependent) and NHANES (time-independent). HRS uses multiple FI measurements via start-stop formatting \cite{Moore2016-rh}; NHANES uses a single measurement.} \label{tab:cox}
    \begin{tabular}{l|llll|llll}
        Predictor\tnote{1} & HRS &  &  &  & NHANES &  &  &  \\
        or measure & Model 0 & Model 1 & Model 2 & Model 3 & Model 0 & Model 1 & Model 2 & Model 3 \\ \hline
        C & $0.69\down{0.69}\up{0.70}$ & $0.75\down{0.75}\up{0.76}$ &  $0.78\down{0.78}\up{0.79}$ & $0.79\down{0.79}\up{0.80}$ & $0.70\down{0.68}\up{0.71}$ & $0.65\down{0.63}\up{0.67}$ & $0.73\down{0.71}\up{0.75}$ & $0.74\down{0.73}\up{0.76}$ \\ 
        $\Delta$ BIC\tnote{2} & Ref. & -2703 & -4477 & -5077 & Ref. & 191 & -150 & -160 \\ 
        Female & $0.75\down{0.72}\up{0.78}$ &  & $0.62\down{0.59}\up{0.65}$ & $0.59\down{0.57}\up{0.62}$ &   $0.70\down{0.61}\up{0.80}$ & & $0.66\down{0.58}\up{0.76}$ & $0.62\down{0.54}\up{0.71}$ \\ 
        Age per 10 & $2.62\down{2.54}\up{2.69}$ & & $1.84\down{1.78}\up{1.89}$ & $1.70\down{1.64}\up{1.76}$  & $2.77\down{2.50}\up{3.07}$ & & $2.50\down{2.25}\up{2.77}$ & $2.46\down{2.21}\up{2.74}$ \\ 
        FI per 0.1 & & $1.50\down{1.48}\up{1.51}$  & $1.42\down{1.41}\up{1.44}$ & $1.31\down{1.29}\up{1.33}$ & & $1.40\down{1.34}\up{1.45}$ & $1.32\down{1.27}\up{1.37}$ & $1.20\down{1.11}\up{1.29}$ \\ 
        Weight loss &  &  &  & $1.65\down{1.55}\up{1.74}$ &  &  &  & $1.34\down{1.08}\up{1.65}$ \\ 
        Slow gait &  &  &  & $1.10\down{1.03}\up{1.18}$ &  &  &  & $1.38\down{1.10}\up{1.74}$ \\ 
        Weakness &  &  &  & $1.22\down{1.14}\up{1.30}$ &  &  &  & $1.25\down{1.02}\up{1.52}$ \\ 
        Exhaustion &  &  &  & $1.09\down{1.03}\up{1.16}$ &  &  &  & $0.90\down{0.68}\up{1.19}$ \\ 
        Low activity &  &  &  & $1.50\down{1.42}\up{1.59}$ &  &  &  & $1.40\down{1.14}\up{1.70}$ \\ \hline
    \end{tabular}
\begin{tablenotes}
\item[1] Predictors: Model 0: age and sex; Model 1: FI, Model 2: age, sex and FI; Model 3: age, sex, FI and FPFP5 status.
\item[2] Change in Bayesian information criterion (BIC), negative indicates a better model.
\end{tablenotes}
\end{threeparttable}
\end{table}

\subsection{The degree of FI is calibrated and predictive}
Our models indicate that the FI is a strong predictor of current and future FPFP5 deficits. We furthermore infer that risk from the FI is approximately additive with (independent from) risk from chronological age, the other major predictor. This makes understanding the calibration curves using the FI pertinent information for making predictions. A calibration curve compares observed outcome frequency to model prediction, varying the predictor variable, the FI ($f$). This includes future predictions, as well as inference.

\begin{figure*}[!ht]
    \centering
    \includegraphics[width=\textwidth]{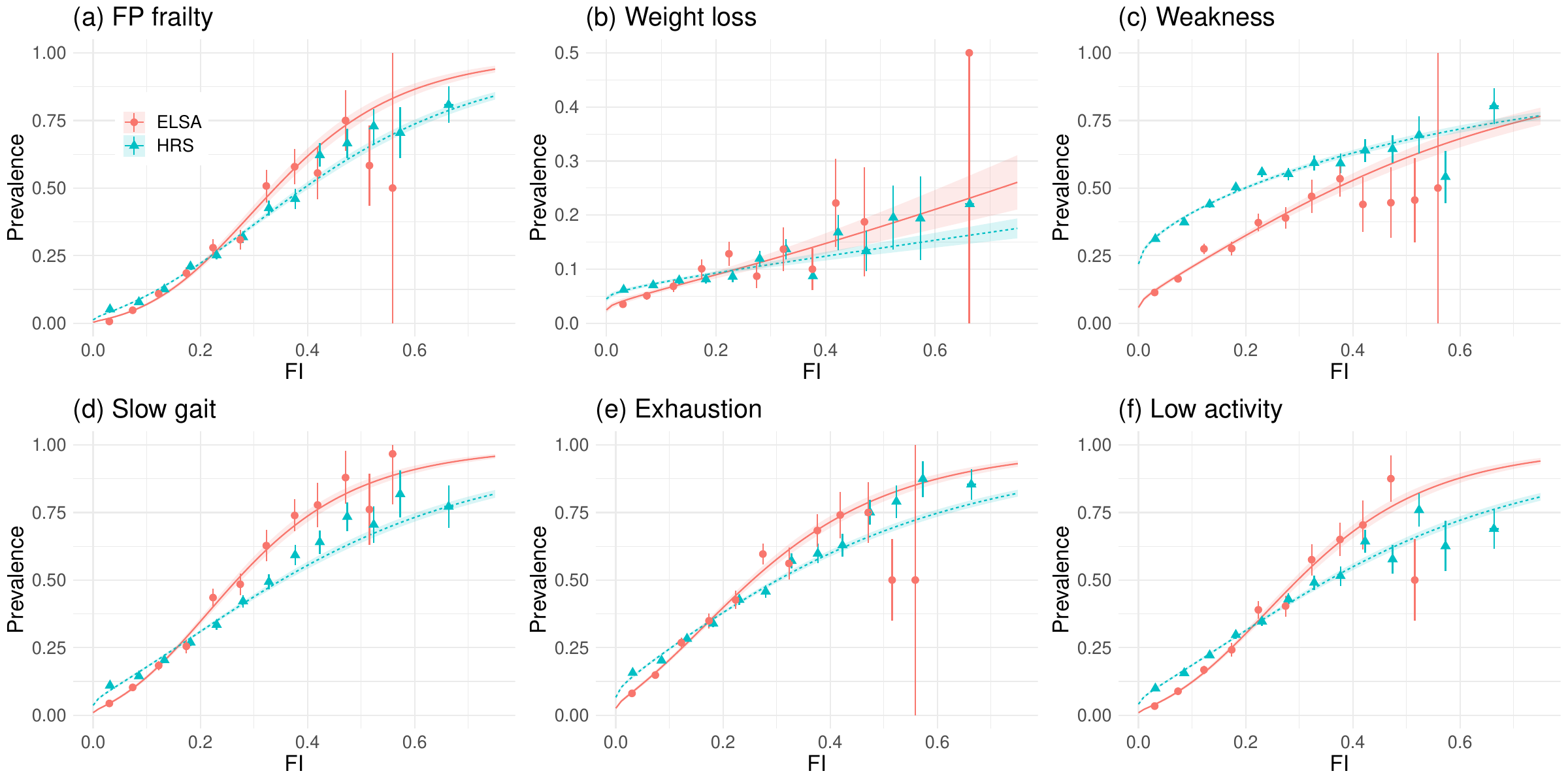} 
    \caption{\textbf{Future prediction calibration using the FI --- HRS and ELSA (longitudinal).} Prevalence increases proportionally to the FI, indicating the degree of frailty (FI) is predictive. Points are mean $\pm$ standard error, binned by FI. Lines are logit-square root regression fits with standard error, $\text{logit(prob.)}\propto\sqrt{f}$.}
    \label{fig:prognosticate}
\end{figure*}

We observe in Figure~\ref{fig:prognosticate} that the probability of future FPFP5 deficit increases non-linearly for each of the FPFP5 and FP with respect to the FI ($f$). 
We found that the model $\text{logit(prob.)}\propto\sqrt{f}$ fit all curves very well (for linear fits see Supplemental Section~S5). For most outcomes the observed frequencies (points) overlapped between the two studies, but the best fit lines differed substantially between HRS and ELSA, which may be due to population differences. A similar discrepancy is observed in the age calibration curves (Supplemental Figure~S4). 

The cross-sectional data was also used to demonstrate inference for NHANES. The curves look very similar to the prediction curves (Supplemental Figure~S3), ostensibly because of the strong association between current and future deficits (Figure~\ref{fig:logreg}). Weakness was notably steeper, but it was based off of self-reported weakness in NHANES whereas ELSA and HRS were an objective grip strength measurement. In general, the cross-sectional curves are sharper and steeper, consistent with the higher AUC observed earlier: this indicates that it is easier to infer current health than to predict future health.

\begin{figure*}[!ht]
    \centering
    \begin{subfigure}[t]{0.49\textwidth}
        \centering
        \includegraphics[width=\textwidth]{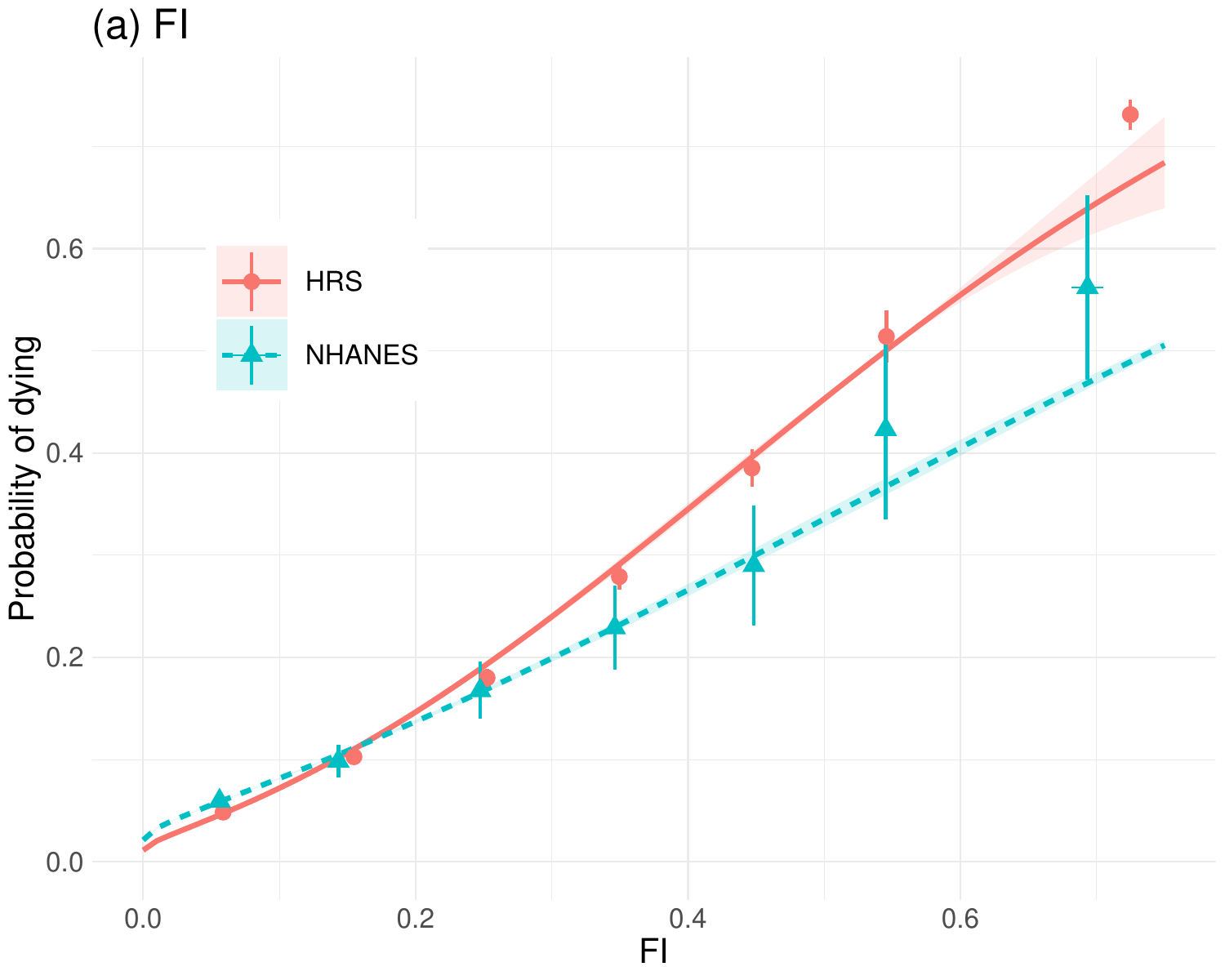} 
    \end{subfigure}
        \begin{subfigure}[t]{0.49\textwidth}
        \centering
        \includegraphics[width=\textwidth]{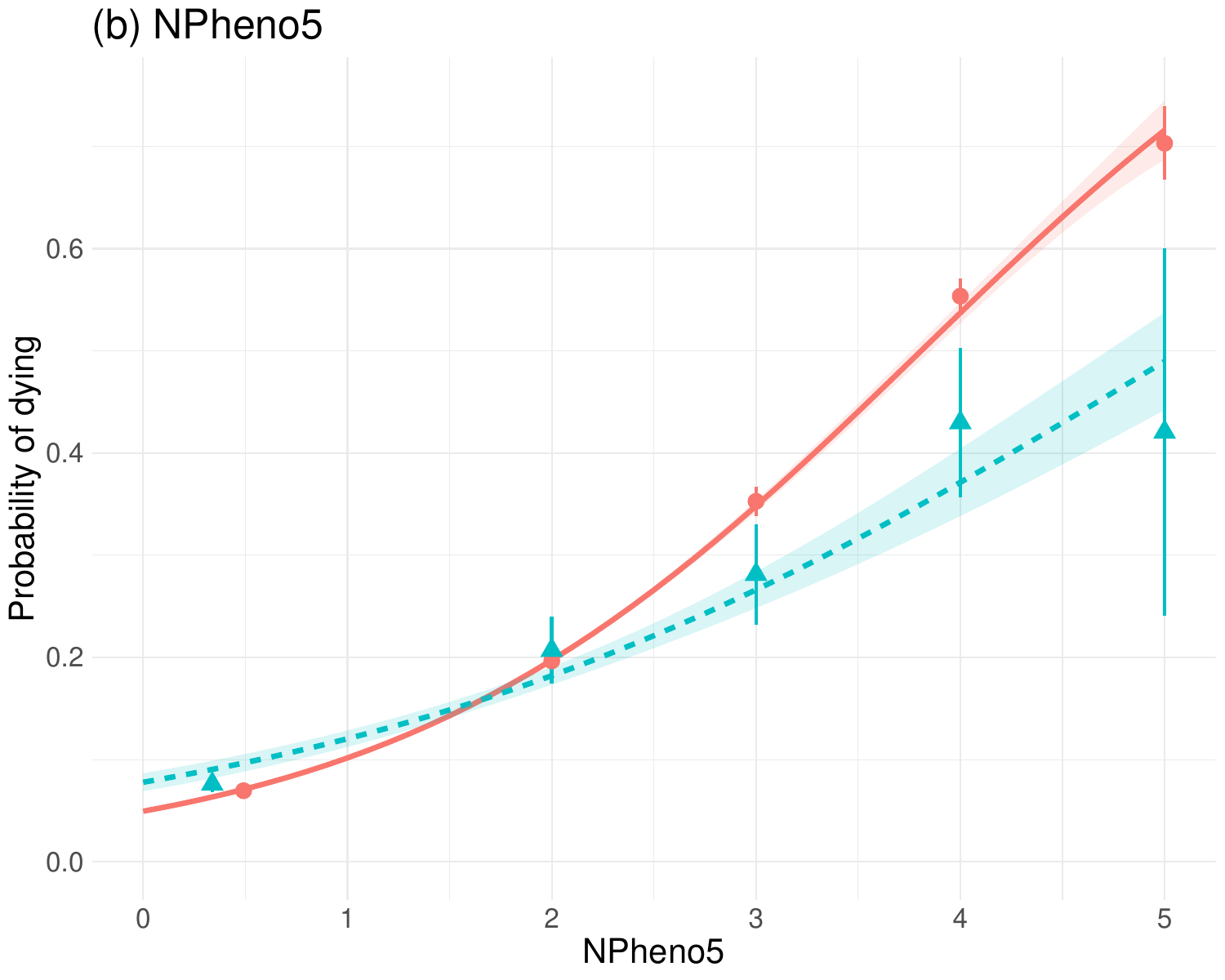}  
    \end{subfigure}
    \caption{\textbf{Probability of dying before followup (4 years).} The FI (\textbf{a}) and NFPFP5 (\textbf{b}) both have similar, weakly super-linear behaviour. Points are non-parametric Kaplan-Meier estimates with standard errors. Lines are logit-square root fit, $\text{logit(prob.)}\propto\sqrt{f}$ (to points). HRS uses two timepoints; NHANES uses one. ELSA showed the same functional form (supplemental).} \label{fig:death}
\end{figure*}

In contrast to previous claims \cite{Xue2021-zx}, we observe no evidence of a discontinuous jump in death probability when NFPFP5 reaches 5 (Figure~\ref{fig:death}b). Instead we observe that death probability increases smoothly and (weakly) super-linearly with increasing number of FPFP5 health deficits. The FI recapitulates this pattern. It is remarkable that the smooth square root-logit function fits across studies and outcomes, including cross-sectional, longitudinal, FPFP5 deficits and mortality. It also fit FPFP5 deficit or mortality prediction using chronological age excellently (Supplemental Figures~S4, S5 and S8). 

\FloatBarrier

\section{Discussion}
We performed future prediction and contemporaneous inference of the FPFP5 deficits: slow gait, weakness, low body weight/weight loss, low activity, and exhaustion; for older individuals age 60+. Inference within the cross-sectional study demonstrated that many deficits, including FP~frailty, can be estimated to a very high accuracy using an FI based solely on self-reported health. This has important implications for imputing missing data.  As for future prediction, across two longitudinal studies we observed that risk of future deficits depends on the degree of frailty, age and the history of the specific deficit. Some deficits were easier to predict, notably gait and low activity, in contrast to low body weight which we predicted only slightly better than a guess. The FI consistently out-performed the NFPFP5, even when the former was based solely on self-reported health with no direct knowledge of the prior FPFP5 deficits. Prediction between FPFP5 deficits was surprisingly weak, and they did not form a coherent cluster. These observations indicate that the FPFP5 are only a subset of a large number of related health deficit variables. The FI appears to be able to leverage this large set to improve prediction of FPFP deficits. This suggests that including more health deficits beyond the FPFP5 would improve FP performance.  
Frailty is strongly linked to physical health deficits \cite{Fried2004-gj, Dent2016-yh, Hoogendijk2019-pk,Kim2024-wx}. Consistent with this, we observed that both measures of frailty, the FI and NFPFP5, generally out-performed chronological age when predicting current and future FPFP5 deficits. Both of these measures purport to be sensitive to some form of frailty \cite{Dent2016-yh}, and they were moderate-to-strongly correlated  \cite{Thompson2018-ve, Beier2022-jo}. Frailty is defined by a NFPFP score of three or more by FP,\cite{Fried2001-mr} versus an FI of 0.2 or more\cite{Kim2024-wx}.  Although the FI and FP have only fair-to-moderate inter-measure kappa reproducibility for frailty classifications: in the range from 0.31-0.47 \cite{Thompson2018-ve,Zhu2016-ct,Beier2022-jo}, this may be explained by the similarly low intra-measure reproducibility of the graded FP, which is 0.41-0.45 \cite{Feenstra2021-bv} (for sequential measurements after 3~months). This indicates substantial overlap between frailty according to the FP versus the FI, with the latter generally performing better in the present study.
The NFPFP5 appears to be an inferior measure for inferring the physical component of frailty versus the FI, as indicated by the key physical deficits -- the FPFP5. The most likely reason for this is that we have generated FIs using primarily health deficits with strong physical components that likely share an etiology with the FPFP5, including physical activity and lower extremity mobility, or with strong physical dependencies related to the FPFP5, such as ADLs and IADLs. For example, gait speed depends heavily on walking ability, as do both of the deficits: difficulty walking one or several blocks. This was demonstrated when we clustered by similarity. Whereas health deficits normally cluster into domains,\cite{Pridham2023-yj,Gross2020-ee,Foote2024-ni} we observed that the FPFP5 did not form a coherent cluster, but rather were embedded in a large number of related health deficits (Figure~\ref{fig:clustering}). Where domains did appear, e.g. ADLs and IADLs, there were still many variables cutting across domains with strong similarities to the FPFP5. Furthermore, we saw that FPFP5 deficits were typically only modestly predictive between deficits, Figure~\ref{fig:logreg}, and that the NFPFP5 had weaker predictive power for FPFP5 deficits than the FI (Figure~\ref{fig:auc}). These signs indicate that there was nothing unique about the FPFP5 compared to most of the other health deficits used in this study. Adding together more, related, health deficits appears to improve performance without loss of specificity, explaining why the FI outperformed the NFPFP5. This is consistent with simulations of the Canadian Longitudinal Study on Aging that indicate a dramatic improvement in reliability, reproducibility and predictive power of the FI when the total number of deficit variables is increased from 5 to 25.\cite{Nguyen2021-be} Intuitively, more variables provides both a broader characterization of health and reduces sensitivity to binarization thresholds that: stifle dose-response relationships, such as those seen in grip strength \cite{Pham2023-vj,Cai2021-vh}, and are inaccurate due to individual variability, such as grip strength and gait speed within \cite{Mayhew2023-fq} and between populations \cite{Peel2013-an}. Contemporary FP theory does not limit expanding the variables used beyond the FPFP5, since it posits that physical frailty emerges from critical dysfunction across HPA stress response, musculoskeletal, and metabolic systems \cite{Fried2021-vh}, which affect a huge number of health variables beyond the FPFP5.\cite{Juster2010-kw} This suggests a missed opportunity in construct criteria for the FP \cite{Bandeen-Roche2020-za}. The NFPFP5 could be improved upon by including many more physical deficits related to the FPFP5, such as via clustering methods, to ensure specificity to the physical frailty, and then averaging them together as is done with the FI, e.g.\ a Fried physical frailty index. 
The existing FP methodology for summarizing FPFP5 deficits using NFPFP5 is vulnerable to biased results due to missing values, particularly from individuals who are too frail to measure. This makes the FP prone to informative missingness, such as when an individual's poor health is the underlying cause for a missing value \cite{Hardy2009-nl}. As with the FI,\cite{Pridham2022-ko} multiple imputation may provide a remedy. Imputation is the process by which missing values are inferred based on other information. Multiple imputation can be applied \textit{post hoc} and accounts for uncertainty in the imputed values by using multiple (stochastic) imputations. As we have demonstrated, the FI is a very strong predictor of FP and could therefore power the underlying inference engine needed for imputation. Furthermore, the FI can be constructed from questionnaire data which typically has much lower missingness since it doesn't require physical measurements. In contrast, the FP includes variables which conventionally must be measured by a health professional to ensure standardization and reliability, notably weight, gait speed, and grip strength.
We confirmed that informative missingness was present in our data when we observed that individuals missing FPFP5 variables had much higher mortality than those with all 5 measured. Failure to account for this effect leads to a biased population --- one which can be greatly deficient in frail individuals, e.g.\ 26.7\% vs 13.3\% prevalence \cite{Theou2015-uz}. The conclusions reached within this biased population may not translate well into a clinical setting where all individuals can be measured or their health inferred by the doctor. Multiple imputation of FP~frailty using the FI is a prospective \textit{post hoc} solution to this problem. Specific deficits could also be imputed, although our predictive ability varies for these, and one should include the deficit history e.g.\ impute gait using a model based on the current FI and the previous gait measurement. Bias in FP~frailty due to missing values is a concern, but one which has a simple prospective solution: imputation using the FI.

The probability of future and current FPFP5 deficits showed a proportional response in the calibration curves for the FI, NFPFP5 and age, as did mortality probability. These curves followed a simple, universal logit-square root functional form across studies and variables. This functional form is flexible enough that it can appear sub- or super-linear, depending on the position of the inflection point. Mortality probability was consistently super-linear with respect to both the FI and NFPFP5; as was: slow gait and FP~frailty with respect to the FI. This indicates that changes to the FI between the ranges of $0.1-0.4$ are particularly important for risk of these outcomes. This is consistent with observations that deficit risk increases dramatically starting near FI$\approx0.2$ due to a tipping point in loss of robustness and resilience with age and increasing FI\cite{Pridham2024-su}.
Across studies, our results broadly followed the same patterns, but the specific parameter estimates differed. 
Others have observed differences in ELSA and HRS responses given the same statistical level of disability \cite{Chan2012-xp}. Hence the same variables and functional forms apply across populations but calibration needs to be performed separately for each population of interest.

Our primary sources of error are differences in FI and FP definitions. These were primarily in NHANES versus the other two studies, notably in terms of weakness. It seems likely that the very different prediction curves for self-reported weakness (NHANES) vs grip-strength-based weakness (HRS and ELSA) are due to different definitions. Besides weakness we observed consistent overall behaviour as noted above, suggesting that the differences may affect specific risk values but are unlikely to affect qualitative risk factors. Our results are remarkably consistent given the underlying variation in definitions.

The FPFP5 are key physical health deficits that emerge in individuals who live with frailty. Being able to predict these deficits is useful for characterizing and forecasting individual health trajectories. We demonstrate that knowledge of an individual's age-related health state using the FI enhances our ability to infer their current FPFP5 state and predict future decline. Chronological age and specific deficit history are important additional considerations.

\section*{Acknowledgements}
The HRS (Health and Retirement Study) is sponsored by the National Institute on Aging (grant number NIA U01AG009740) and is conducted by the University of Michigan. ELSA is funded by the National Institute on Aging (R01AG017644), and by UK Government Departments coordinated by the National Institute for Health and Care Research (NIHR).

\section*{Author contributions statement}
ADR conceived the project. ADR and KR supervised. GP did the analysis and initial draft. All authors reviewed the manuscript.

\section*{Data availability}
All data used in the present study are publically available. ELSA \cite{elsa} is available from the UK Data Service \url{https://ukdataservice.ac.uk/}. HRS \cite{hrs} is available from \url{https://hrs.isr.umich.edu/data-products}. NHANES \cite{nhanes} is available from: \url{https://www.cdc.gov/nchs/nhanes/index.htm}.

\section*{Disclosures}
None directly related to this work. In the past three years KR has received honoraria for invited lectures, rounds and academic symposia on frailty from: Burnaby Family Practice, Chinese Medical Association, University of Nebraska-Omaha, the Australia New Zealand Society of Geriatric Medicine, the Atria Institute, University of British Columbia, McMaster University, and the Fraser Health Authority. KR is co-founder of Ardea Outcomes (DGI Clinical until 2021), which in the past 3 years has had contracts with pharma and device manufacturers (Danone, Hollister, INmune, Novartis, Takeda) on individualized outcome measurement.

\clearpage
\bibliography{ref}

\FloatBarrier

\clearpage

\renewcommand{\theequation}{S\arabic{equation}}
\renewcommand{\thefigure}{S\arabic{figure}}
\renewcommand{\thetable}{S\arabic{table}}
\renewcommand{\thesection}{S\arabic{section}}
\setcounter{equation}{0}  
\setcounter{figure}{0}  
\setcounter{table}{0}  
\setcounter{section}{0}  

\noindent\textbf{\LARGE Supplemental information for: Inferring and predicting Fried physical frailty phenotype deficits}

\section{FPFP5 deficits}

For NHANES we used the modified FP which uses self-reported weakness in place of grip strength \cite{Wilhelm-Leen2009-xe}. We computing the FPFP5 deficits following the previous work by Wilhelm-Leen \textit{et al.} (2009)\cite{Wilhelm-Leen2009-xe} with departures in parentheses. For weight loss we used $\text{BMI} \leq 22.5~kg/m^2$ (Wilhelm-Leen \textit{et al.} used $18.5~kg/m^2$ as the cutoff but we found it was too rare, with 1.5\% prevalence, and instead used the cut from Baniak \textit{et al.}, $22.5~kg/m^2$ \cite{Baniak2020-ef}, with 12.6\% prevalence). For weakness, if the individual responded as having \texttt{``no difficulty''} to the following question then they were encoded as non-deficit, otherwise they were encoded as deficit: \texttt{``how much difficulty (do you) have ... lifting or carrying something as heavy as 10 pounds [like a sack of potatoes or rice]?''} (PFQ060E). For slow gait we used the sex-specific lowest quintile (20\%) for the 20-foot walk (Wilhelm-Leen \textit{et al.} used the 8-foot walk). For exhaustion, if the individual responded as having \texttt{``no difficulty''} to the following question then they were encoded as non-deficit, otherwise they were encoded as deficit: \texttt{``how much difficulty (do you) have ... walking from one room to another on the same level?''} (PFQ060H). For low activity, if the individual responded as being \texttt{``less active''} to the following question then they were encoded as deficit: \texttt{``Compared with most (sex-specific people your) age, would you say that (you are) ...''} (PAQ520).

For ELSA, we computing the FPFP5 deficits following the previous work by Liljas \textit{et al.} (2017) \cite{Liljas2017-bp} with departures in parentheses. For weight loss we used either $\text{BMI} \leq 18.5~kg/m^2$ or a change of $\geq 10\%$ since the previous wave (4 years ago). For weakness we compared max of 3 attempts (both for dominant and non) then used the sex-specific lowest quintile (20\%) (Liljas adjusted for BMI but we saw no effect and did not; not shown). For slow gait we averaged together both attempts on 8~foot walk then applied the sex-specific lowest quintile (20\%) (Liljas adjusted for height but we saw the effect was weak and barely moved the cutoff, so we did not; not shown). A small number in each wave were bedbound or in wheelchairs and were encoded as deficit ($< 100$ per wave). Any individual who said they were unable to walk was encoded as deficit ($\sim 200$ per wave). For exhaustion, if the individual responded in affirmative to either of the following two CESD (Center for Epidemiological Studies Depression) questions then they were encoded as deficit: (i) \texttt{``Much of the time during the past week, have you felt that everything you did was an effort?''}, or (ii) \texttt{``Much of the time during the past week, could you not get going?''}. For low activity, if the individual reported that they performed moderate or vigorous activity at least 1-3 times per month then they were encoded as non-deficit, otherwise they were deficit (respondents are given examples of low, moderate and vigorous activity).

For HRS we sought to follow ELSA criteria as closely as possible. For weight loss we used either $\text{BMI} \leq 18.5~kg/m^2$ or a change of $\geq 10\%$ since the previous wave (2~years ago). For weakness and slow gait we used the sex-specific lowest quintiles (max grip strength, measured 2 times; 98.5~inch$\approx$8~foot walk). For exhaustion, if the individual responded in affirmative to either of the following two CESD questions then they were encoded as deficit, during the past week: (i) everything was an effort, or (ii) they could not get going. For low activity, if the individual reported that they performed moderate or vigorous activity at least 1-3 times per month then they were encoded as non-deficit, otherwise they were deficit (respondents are given examples of low, moderate and vigorous activity).
Individuals under age 65 were not measured for gait, this affected 2.3\% of our study population for feature selection and 3.7\% for survival prediction.

\section{FI variables}
For ELSA, our list of FI variables is based on previous work by Rogers \textit{et al.} \cite{Rogers2017-ke}. We modified their list by excluding depressive symptoms, as we had difficulty finding these 8~variables. We also extended their list to include Parkinson's, Alzheimer's and dementia diagnoses. The specific list is given in Table~\ref{tab:elsafi}. Ordinal variables were converted to a linear scale, this included self-reported general health, eyesight and hearing which are all on the scale: 0 (best), 1/4, 2/4, 3/4 or 4/4 (worst).

\begin{table}
    \caption{ELSA FI variables used} \label{tab:elsafi}
    \centering
    \begin{tabular}{lll}
        ~ & Description & Encoding \\ \hline
        1 & Difficulty walking 100 yards & 0: no, 1: yes \\ 
        2 & Difficulty sitting for about two hours & 0: no, 1: yes \\ 
        3 & Difficulty getting up from a chair after sitting for long periods & 0: no, 1: yes \\ 
        4 & Difficulty climbing several flights of stairs without resting & 0: no, 1: yes \\ 
        5 & Difficulty climbing one flight of stairs without resting & 0: no, 1: yes \\ 
        6 & Difficulty stooping kneeling or crouching & 0: no, 1: yes \\ 
        7 & Difficulty reaching or extending arms above shoulder level & 0: no, 1: yes \\ 
        8 & Difficulty pulling pushing large objects like a living room chair & 0: no, 1: yes \\ 
        9 & Difficulty lifting carrying over 10 lbs like a heavy bag of groceries & 0: no, 1: yes \\ 
        10 & Difficulty picking up a 5p coin from a table & 0: no, 1: yes \\ 
        11 & Difficulty dressing including putting on shoes and socks & 0: no, 1: yes \\ 
        12 & Difficulty walking across a room & 0: no, 1: yes \\ 
        13 & Difficulty bathing or showering & 0: no, 1: yes \\ 
        14 & Difficulty eating such as cutting up your food & 0: no, 1: yes \\ 
        15 & Difficulty getting in or out of bed & 0: no, 1: yes \\ 
        16 & Difficulty using the toilet including getting up or down & 0: no, 1: yes \\ 
        17 & Difficulty using a map to get around in a strange place & 0: no, 1: yes \\ 
        18 & Difficulty preparing a hot meal & 0: no, 1: yes \\ 
        19 & Difficulty shopping for groceries & 0: no, 1: yes \\ 
        20 & Difficulty making telephone calls & 0: no, 1: yes \\ 
        21 & Difficulty taking medications & 0: no, 1: yes \\ 
        22 & Difficulty doing work around the house or garden & 0: no, 1: yes \\ 
        23 & Difficulty managing money eg paying bills   keeping track ofexpenses & 0: no, 1: yes \\ 
        24 & Self-reported general health & 0: very good, ... , 4/4: very bad \\ 
        25 & Self-reported eyesight (corrected) & 0: excellent, ... , 4/4: very bad \\ 
        26 & Self-reported hearing (corrected) & 0: excellent, ... , 4/4: very bad \\ 
        27 & Chronic: lung disease diagnosis & 0: no, 1: yes \\ 
        28 & Chronic: asthma diagnosis & 0: no, 1: yes \\ 
        29 & Chronic: arthritis diagnosis & 0: no, 1: yes \\ 
        30 & Chronic: osteoporosis diagnosis & 0: no, 1: yes \\ 
        31 & Chronic: cancer diagnosis & 0: no, 1: yes \\ 
        32 & Chronic: Parkinson's diagnosis & 0: no, 1: yes \\ 
        33 & Chronic: psychiatric condition diagnosis & 0: no, 1: yes \\ 
        34 & Chronic: Alzheimer's diagnosis & 0: no, 1: yes \\ 
        35 & Chronic: dementia diagnosis & 0: no, 1: yes \\ 
        36 & CVD: high blood pressure diagnosis & 0: no, 1: yes \\ 
        37 & CVD: angina diagnosis & 0: no, 1: yes \\ 
        38 & CVD: heart attack & 0: no, 1: yes \\ 
        39 & CVD: congestive heart failure diagnosis & 0: no, 1: yes \\ 
        40 & CVD: heart murmur diagnosis & 0: no, 1: yes \\ 
        41 & CVD: abnormal heart rhythm & 0: no, 1: yes \\ 
        42 & CVD: diabetes or high blood sugar diagnosis & 0: no, 1: yes \\ 
        43 & CVD: stroke diagnosis & 0: no, 1: yes \\ \hline 
    \end{tabular}
\end{table}

For NHANES, we used the self-reported (``clinical'') FI variables from a previous study \cite{Pridham2023-yj}, which is, in turn based on an even earlier study \cite{Blodgett2017-jp}. The full list is provided in Table~\ref{tab:nhanesfi}. We used 26 variables. While 30+ is the standard, we note that the improvement is theoretically continuous and 26 should perform similarly well to 30.\cite{Pridham2023-yj}

\begin{table}
    \centering
        \caption{NHANES FI variables used} \label{tab:nhanesfi}
    \begin{tabular}{llll}
        ~ & Code & Description & Encoding \\ \hline
        1 & AUQ130 & General condition of hearing & 0: good, 1: otherwise \\ 
        2 & HUQ010 & General health condition & 0: excellent or good, 1: fair or poor \\ 
        3 & HUQ020 & Health now compared with 1 year ago & 0: better or same, 1: worse \\ 
        4 & KIQ046 & Leak urine during nonphysical activities & 0: no, 1: yes \\ 
        5 & OSQ010A & Broken or fractured a hip & 0: no, 1: yes \\ 
        6 & PFQ056 & Experience confusion/memory problems & 0: no, 1: yes \\ 
        7 & PFQ060A & Difficulty managing money & 0: no difficulty, 1: otherwise \\ 
        8 & PFQ060B & Difficulty walking a quarter mile & 0: no difficulty, 1: otherwise \\ 
        9 & PFQ060C & Difficulty walking 10 steps & 0: no difficulty, 1: otherwise \\ 
        10 & PFQ060D & Difficulty kneeling/crouching & 0: no difficulty, 1: otherwise \\ 
        11 & PFQ060F & Household chore difficulty & 0: no difficulty, 1: otherwise \\ 
        12 & PFQ060G & Difficulty preparing meals & 0: no difficulty, 1: otherwise \\ 
        13 & PFQ060I & Difficulty standing from armless chair & 0: no difficulty, 1: otherwise \\ 
        14 & PFQ060J & Difficulty getting in/out of bed & 0: no difficulty, 1: otherwise \\ 
        15 & PFQ060K & Difficulty using fork, knife, drinking from cup & 0: no difficulty, 1: otherwise \\ 
        16 & PFQ060L & Difficulty dressing self & 0: no difficulty, 1: otherwise \\ 
        17 & PFQ060M & Difficulty standing for long periods & 0: no difficulty, 1: otherwise \\ 
        18 & PFQ060N & Difficulty sitting for long periods & 0: no difficulty, 1: otherwise \\ 
        19 & PFQ060O & Difficulty reaching over head & 0: no difficulty, 1: otherwise \\ 
        20 & PFQ060P & Difficulty grasping/holding small objects & 0: no difficulty, 1: otherwise \\ 
        21 & PFQ060Q & Difficulty going to movies, events & 0: no difficulty, 1: otherwise \\ 
        22 & PFQ060R & Difficulty attending social events & 0: no difficulty, 1: otherwise \\ 
        23 & PFQ060S & Difficulty leasuring at home & 0: no difficulty, 1: otherwise \\ 
        24 & RDD030 & Coughing most days - over 3 mo period & 0: no, 1: yes \\ 
        25 & VIQ030 & General condition of eyesight & 0: excellent or good, 1: fair, poor or very poor \\ 
        26 & VIQ050C & Difficulty seeing steps/curbs-dim light & 0: no difficulty, 1: any difficulty \\  \hline
    \end{tabular}
\end{table}

We used the HRS FI developed elsewhere by Theou \textit{et al.} \cite{Theou2023-aw}. We excluded one variable: number of doctor visits in previous 2 years, since it was not obvious how to binarize. The specific variables used are reported in Table~\ref{tab:hrsfi}

\begin{table}
    \centering
        \caption{HRS FI variables used} \label{tab:hrsfi}
    \begin{tabular}{llll}
        ~ & Code & Description & Encoding \\ \hline
        1 & armsa & Difficulty reaching/extending arms up &0: no, 1: yes/any \\ 
        2 & arthre & Ever diagnosed with arthritis &0: no, 1: yes \\ 
        3 & batha & Difficulty bathing or showerng &0: no, 1: yes/any \\ 
        4 & bathh & Gets help bathing, showerng &0: no, 1: yes \\ 
        5 & beda & Difficulty getting in/out of bed &0: no, 1: yes/any \\ 
        6 & bede & Uses equipment to get in/out of bed &0: no, 1: yes \\ 
        7 & bedh & Gets help getting in/out of bed &0: no, 1: yes \\ 
        8 & cancre & Ever diagnosed with cancer &0: no, 1: yes \\ 
        9 & chaira & Difficulty getting up from chair &0: no, 1: yes/any \\ 
        10 & clim1a & Difficulty climbing one stair flight &0: no, 1: yes/any \\ 
        11 & climsa & Difficulty climbing several stair flight &0: no, 1: yes/any \\ 
        12 & diabe & Ever diagnosed with diabetes &0: no, 1: yes \\ 
        13 & dimea & Difficulty picking up a dime &0: no, 1: yes/any \\ 
        14 & dressa & Difficulty dressing &0: no, 1: yes/any \\ 
        15 & dressh & Gets help dressing &0: no, 1: yes \\ 
        16 & eata & Difficulty eating &0: no, 1: yes/any \\ 
        17 & eath & Gets help eating &0: no, 1: yes \\ 
        18 & hearte & Ever diagnosed with heart problems &0: no, 1: yes \\ 
        19 & hibpe & Ever diagnosed with high blood pressure &0: no, 1: yes \\ 
        20 & homcar & Received home health care within previous 2 years &0: no, 1: yes \\ 
        21 & hosp & Had a hospital stay within previous 2 years &0: no, 1: yes \\ 
        22 & lifta & Difficulty lifting/carrying 10lbs &0: no, 1: yes/any \\ 
        23 & lunge & Ever diagnosed with lung disease &0: no, 1: yes \\ 
        24 & moneya & Difficulty managing money &0: no, 1: yes/any \\ 
        25 & nhmliv & Living in nursing home at time of intervew &0: no, 1: yes \\ 
        26 & nrshom & Had a nursing home stay within previous 2 years &0: no, 1: yes \\ 
        27 & outpt & Had outpatient surgery within previous 2 years &0: no, 1: yes \\ 
        28 & phonea & Difficulty using the telephone &0: no, 1: yes/any \\ 
        29 & pusha & Difficulty pushing/pulling a large object &0: no, 1: yes/any \\ 
        30 & shlt & Self-reported health & 0: excellent-good, 1: fair-poor \\ 
        31 & shopa & Difficulty shoping for groceries &0: no, 1: yes/any \\ 
        32 & spcfac & Visited a specialized health facility within previous 2 years &0: no, 1: yes \\ 
        33 & stoopa & Difficulty stooping/kneeling/crouching &0: no, 1: yes/any \\ 
        34 & stroke & Ever diagnosed with a stroke &0: no, 1: yes \\ 
        35 & toilta & Difficulty using the toilet &0: no, 1: yes/any \\ 
        36 & toilth & Gets help using the toilet &0: no, 1: yes \\ 
        37 & walk1a & Difficulty walking one block &0: no, 1: yes/any \\ 
        38 & walkra & Difficulty  walking across rooms &0: no, 1: yes/any \\ 
        39 & walkre & Needs equipment to  walk across rooms &0: no, 1: yes \\ 
        40 & walkrh & Gets help walking across rooms &0: no, 1: yes \\ 
        41 & walksa & Difficulty walking several blocks &0: no, 1: yes/any \\ \hline
    \end{tabular}
\end{table}

\FloatBarrier

\section{Additional results}
\begin{figure*}[!ht]
    \centering
    \begin{subfigure}[t]{0.5\textwidth}
        \centering
        \includegraphics[width=\textwidth]{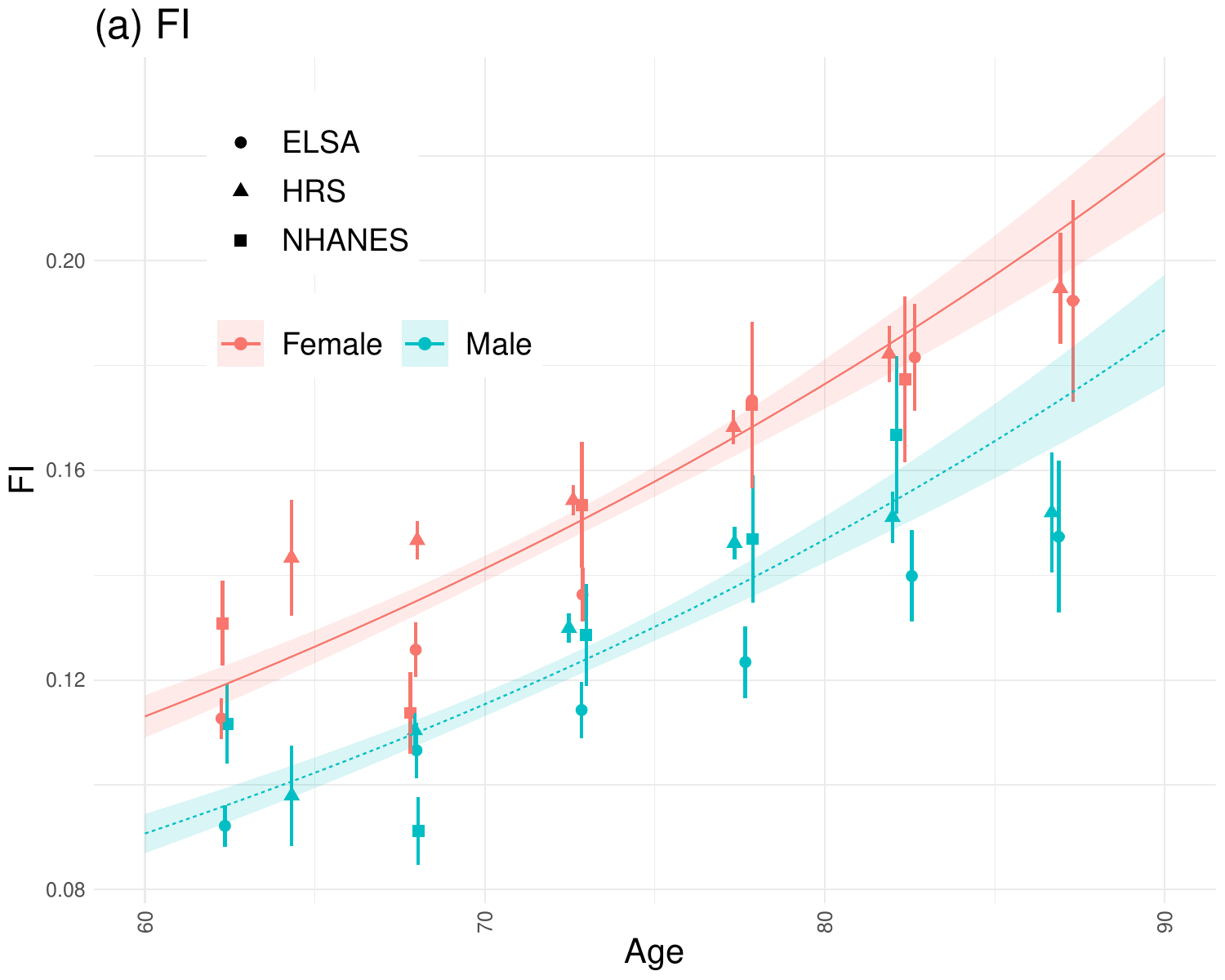} 
    \end{subfigure}%
    ~ 
    \begin{subfigure}[t]{0.5\textwidth}
        \centering
        \includegraphics[width=\textwidth]{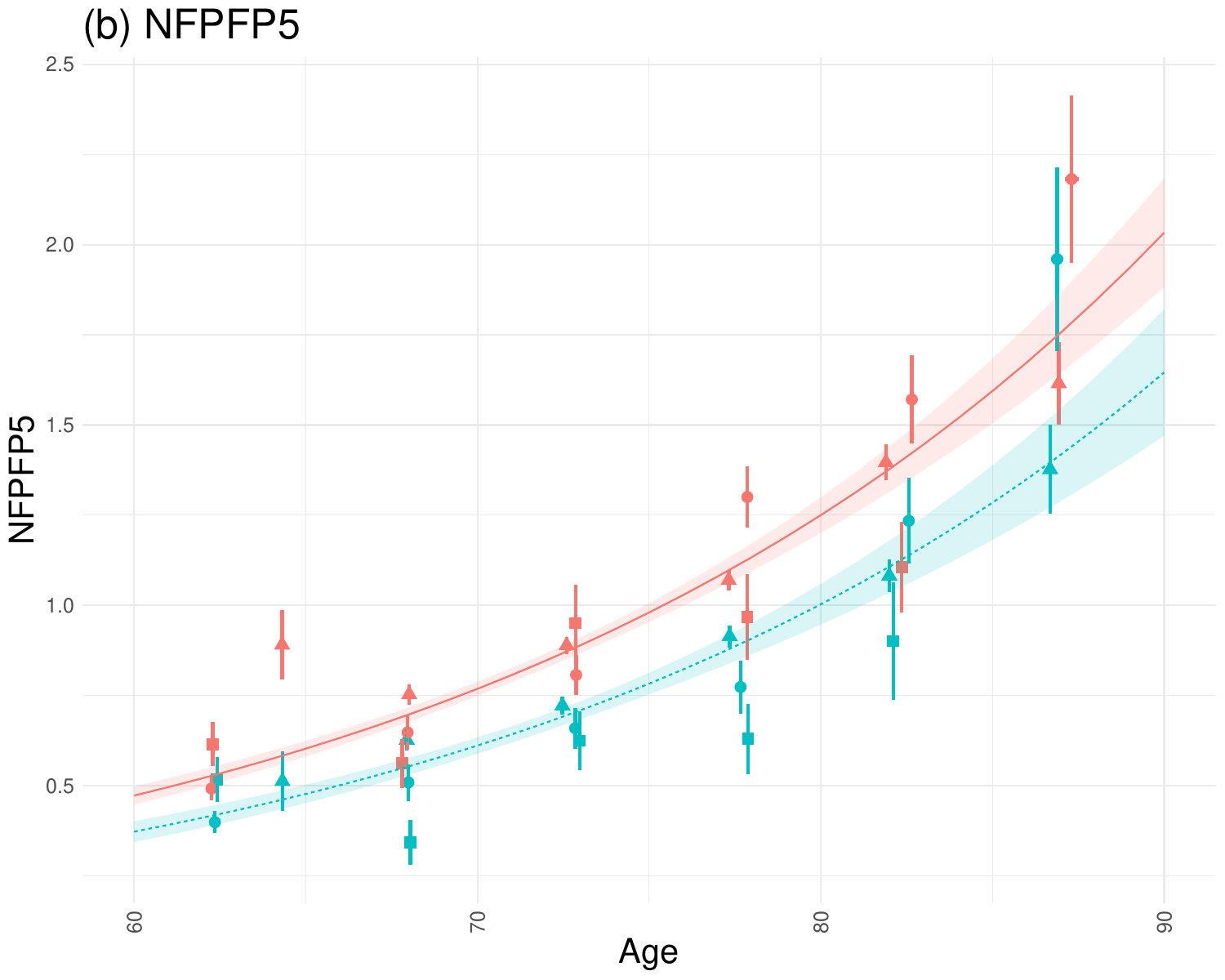} 
    \end{subfigure}
    \caption{\textbf{The FI and NFPFP5 both increase exponentially with age} (lines) with males always lower (dashed blue). Points are data binned by 5 year intervals. Lines are weighted least-squares exponential model fit, $\ln{(y)}\propto \text{Age}$. Error bars are standard errors.} \label{fig:age}
\end{figure*}

Both the FI and NFPFP5 grow exponentially with age, with strong sex-effects, Figure~\ref{fig:age} (males always lower).

\begin{figure*}[!ht]
    \centering
        \includegraphics[width=0.5\textwidth]{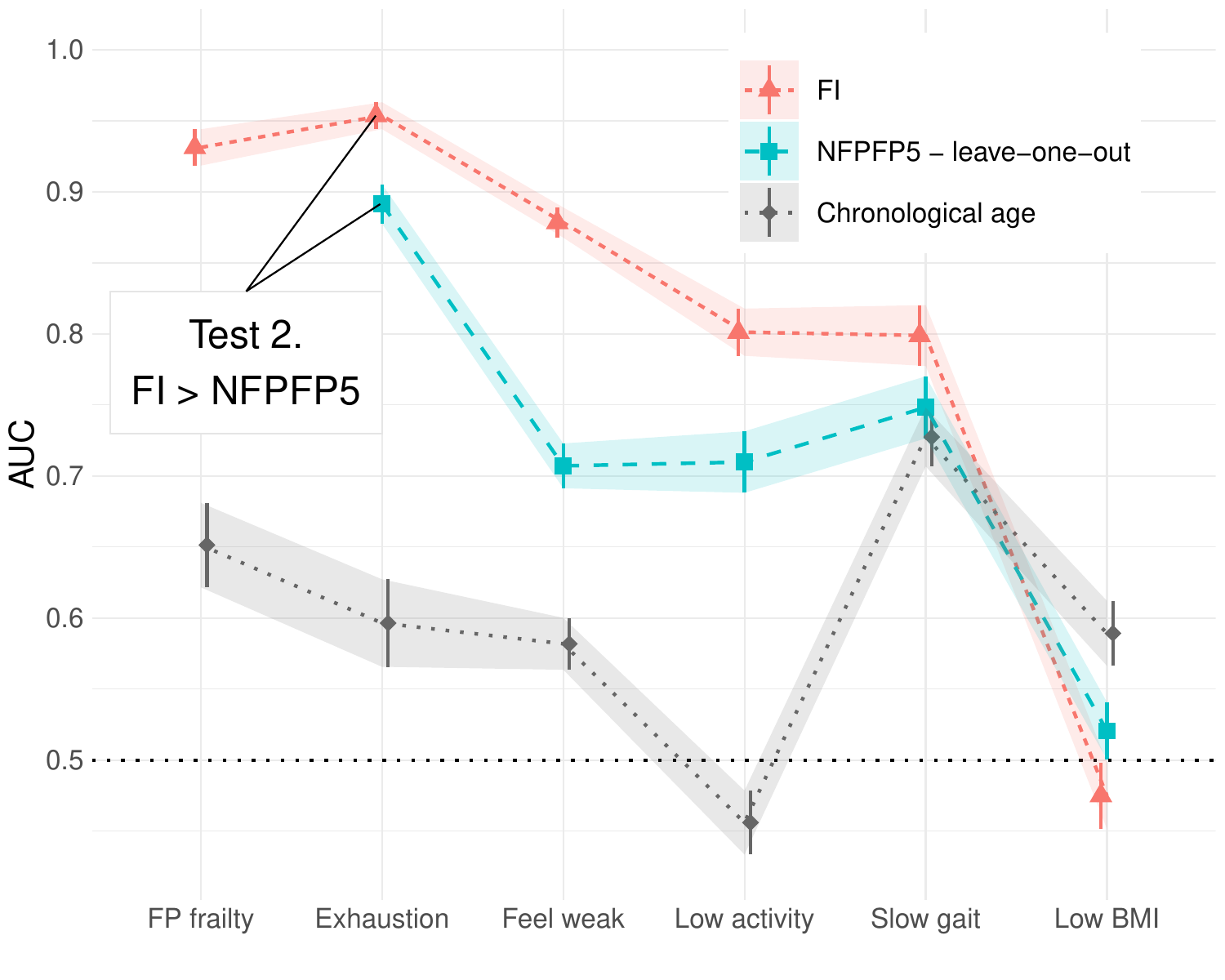} 
    \caption{\textbf{The FI better infers contemporaneous FPFP5 deficits than the NFPFP5 (NHANES)}. Predictive power for predicting current FPFP5 deficits using the current FI versus (leave-one-out) NFPFP5 (NHANES). The FI is the superior predictor. NFPFP5 is the number of FPFP5 deficits excluding the current outcome (leave-one-out; not applicable for FP~frailty). Note: the definitions for low activity and exhaustion are quite different from the longitudinal datasets (see Table~1). Error bars are standard errors.} \label{fig:auc_nhanes}
\end{figure*}

For the cross-sectional NHANES data, only Test 2 is non-trivial: leave-one-out NFPFP5 vs the FI. We see in Figure~\ref{fig:auc_nhanes} that
leave-one-out NFPFP5 performs poorly.

\begin{figure*}
    \centering
    \includegraphics[width=\textwidth]{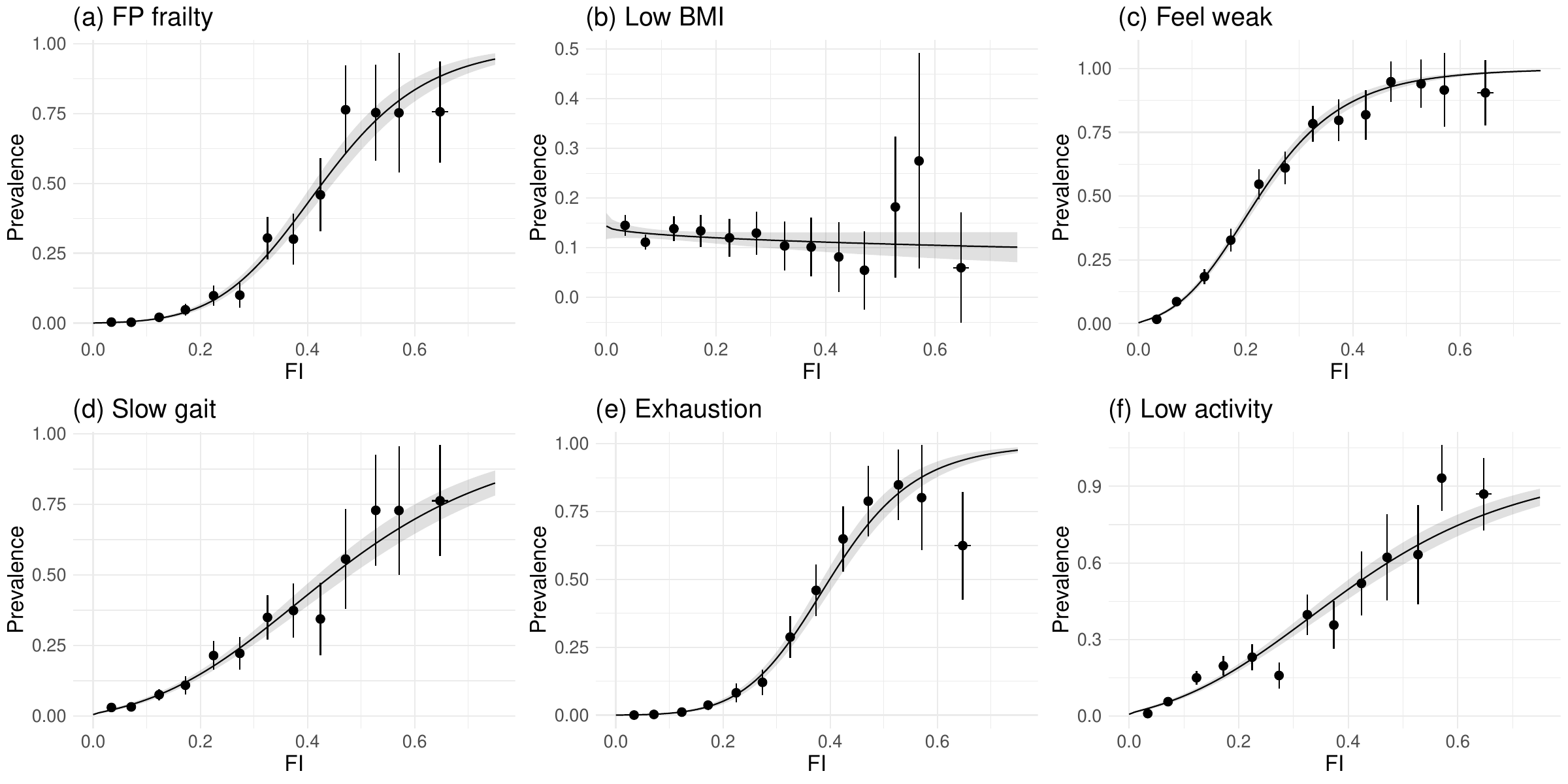} 
    \caption{\textbf{Inference calibration using the FI --- NHANES (cross-sectional).} The current condition of an individual can be inferred from their FI score, which shows a graded response. Points are mean $\pm$ standard error, binned by FI. Lines are sqrt logistic regression fits with standard error, $\text{logit(prob.)}\propto\sqrt{f}$.}
    \label{fig:calibration_nhanes}
\end{figure*}

If we instead look at cross-sectional data we have the inference (within-wave) calibration curves, Figure~\ref{fig:calibration_nhanes}. We observe that the inference curves look very similar to the future prediction curves, as expected since the current deficit and future deficit tend to be very strongly associated (as we saw in Figure~3). Weakness had a much stronger response in the cross-sectional data, but was also based on a different measurement than HRS and ELSA (self-reported functional limitations rather than direct grip strength measurement). In general, the cross-sectional curves are sharper and steeper, consistent with the higher AUC observed earlier: it is easier to infer current health than predict future health.

The (linear) Pearson correlation matrix for the key outcomes is reported in Table~\ref{tab:pearson}, while the (non-parametric) Spearman correlation matrix is reported in Table~\ref{tab:spearman}. The relationship between age and the FI or NFPFP5 were both non-linear (Figure~\ref{fig:age}), whereas the relationship between the FI and NFPFP5 was linear (not shown). Hence we expect the Spearman correlation will be most accurate for the former and Pearson for the latter.

\begin{table}
    \caption{Pearson correlation matrix ($n=100$ bootstraps)} \label{tab:pearson}
    \centering
    \begin{tabular}{lllll}
        Dataset & Variable & FI & FI /w FPFP5 & NFPFP5 \\ \hline
        HRS & Age & $0.12\down{0.10}\up{0.14}$ & $0.15\down{0.13}\up{0.17}$ & $0.20\down{0.18}\up{0.22}$ \\ 
         & FI & ~ & $0.99\down{0.99}\up{0.99}$ & $0.52\down{0.50}\up{0.53}$ \\ 
         & FI /w FPFP5 & ~ & ~ & $0.65\down{0.63}\up{0.66}$ \\ 
        ELSA & Age & $0.21\down{0.17}\up{0.24}$ & $0.25\down{0.21}\up{0.28}$ & $0.32\down{0.29}\up{0.36}$ \\ 
         & FI & ~ & $0.99\down{0.98}\up{0.99}$ & $0.58\down{0.55}\up{0.61}$ \\ 
         & FI /w FPFP5 & ~ & ~ & $0.71\down{0.68}\up{0.73}$ \\ 
        NHANES & Age & $0.21\down{0.15}\up{0.26}$ & $0.22\down{0.16}\up{0.27}$ & $0.19\down{0.14}\up{0.24}$ \\ 
         & FI & ~ & $0.98\down{0.98}\up{0.99}$ & $0.70\down{0.67}\up{0.74}$ \\ 
         & FI /w FPFP5 & ~ & ~ & $0.82\down{0.80}\up{0.84}$ \\ \hline
    \end{tabular}
\end{table}

\begin{table}
    \caption{Spearman correlation matrix ($n=100$ bootstraps)} \label{tab:spearman}
    \centering
    \begin{tabular}{lllll}
        Dataset & Variable & FI & FI /w FPFP5 & NFPFP5 \\ \hline
        HRS & Age & $0.15\down{0.13}\up{0.17}$ & $0.17\down{0.15}\up{0.19}$ & $0.19\down{0.17}\up{0.21}$ \\ 
         & FI & ~ & $0.98\down{0.98}\up{0.98}$ & $0.45\down{0.43}\up{0.46}$ \\ 
         & FI /w FPFP5 & ~ & ~ & $0.60\down{0.59}\up{0.62}$ \\ 
        ELSA & Age & $0.25\down{0.21}\up{0.28}$ & $0.28\down{0.25}\up{0.31}$ & $0.29\down{0.25}\up{0.32}$ \\ 
         & FI & ~ & $0.98\down{0.98}\up{0.98}$ & $0.49\down{0.46}\up{0.52}$ \\ 
         & FI /w FPFP5 & ~ & ~ & $0.63\down{0.61}\up{0.66}$ \\ 
        NHANES & Age & $0.21\down{0.16}\up{0.26}$ & $0.23\down{0.18}\up{0.28}$ & $0.17\down{0.12}\up{0.23}$ \\ 
         & FI & ~ & $0.96\down{0.95}\up{0.97}$ & $0.56\down{0.51}\up{0.60}$ \\ 
         & FI /w FPFP5 & ~ & ~ & $0.73\down{0.71}\up{0.76}$ \\ \hline
    \end{tabular}
\end{table}

Figure~3 demonstrated that HRS and ELSA broadly had the same prediction patterns for (linear) logistic regression prediction. The specific odds ratios are reported in Table~\ref{tab:glm_elsa} for ELSA and Table~\ref{tab:glm_hrs} for HRS. The odds ratios are close but generally do not agree within confidence intervals between the two different studies. Note that although the logit-square root fit best for the FI and NFPFP5, the linear model also fit reasonably well (Supplemental Section~\ref{sec:fit_diag}).

\begin{table}
    \centering
    \begin{threeparttable}  
    \caption{Logistic regression odds ratios with 95\% confidence intervals (and AUC) --- ELSA } \label{tab:glm_elsa}
    \begin{tabular}{llllllllll}
        Outcome & AUC\tnote{*} & FI per 0.1 & Age per 10 & Female & Weight loss & Weakness & Slow gait & Exhaustion & Low activity \\ \hline
        FP frailty & $0.830\down{0.811}\up{0.849}$ & $3.62\down{3.58}\up{3.70}$ & ~ & ~ & ~ & ~ & ~ & ~ & ~ \\ 
        FP frailty & $0.891\down{0.876}\up{0.907}$ & $3.91\down{3.90}\up{3.93}$ & $5.01\down{3.79}\up{8.63}$ & $0.71\down{0.69}\up{0.75}$ & ~ & ~ & ~ & ~ & ~ \\ 
        FP frailty & $0.913\down{0.898}\up{0.929}$ & $2.65\down{2.60}\up{2.75}$ & $4.20\down{0.62}\up{181.81}$ & $0.63\down{0.52}\up{0.91}$ & $2.59\down{2.14}\up{3.78}$ & $2.12\down{1.75}\up{3.08}$ & $2.31\down{1.90}\up{3.36}$ & $1.63\down{1.34}\up{2.38}$ & $3.13\down{2.58}\up{4.56}$ \\ 
        Weight loss & $0.634\down{0.589}\up{0.678}$ & $1.51\down{1.50}\up{1.54}$ & ~ & ~ & ~ & ~ & ~ & ~ & ~ \\ 
        Weight loss & $0.657\down{0.616}\up{0.698}$ & $1.37\down{1.36}\up{1.37}$ & $1.61\down{1.33}\up{2.33}$ & $1.50\down{1.47}\up{1.55}$ & ~ & ~ & ~ & ~ & ~ \\ 
        Weight loss & $0.667\down{0.627}\up{0.707}$ & $1.25\down{1.24}\up{1.28}$ & $1.56\down{0.52}\up{13.39}$ & $1.46\down{1.31}\up{1.81}$ & $2.09\down{1.87}\up{2.59}$ & $0.89\down{0.80}\up{1.10}$ & $1.36\down{1.22}\up{1.69}$ & $0.98\down{0.88}\up{1.22}$ & $1.40\down{1.26}\up{1.74}$ \\ 
        Weakness & $0.676\down{0.654}\up{0.698}$ & $1.83\down{1.82}\up{1.87}$ & ~ & ~ & ~ & ~ & ~ & ~ & ~ \\ 
        Weakness & $0.784\down{0.763}\up{0.806}$ & $1.71\down{1.70}\up{1.71}$ & $3.56\down{3.01}\up{4.92}$ & $0.65\down{0.64}\up{0.67}$ & ~ & ~ & ~ & ~ & ~ \\ 
        Weakness & $0.830\down{0.812}\up{0.849}$ & $1.41\down{1.39}\up{1.46}$ & $2.92\down{0.46}\up{107.61}$ & $0.67\down{0.55}\up{0.95}$ & $1.05\down{0.88}\up{1.51}$ & $8.10\down{6.74}\up{11.62}$ & $1.20\down{1.00}\up{1.72}$ & $1.19\down{0.99}\up{1.70}$ & $1.24\down{1.03}\up{1.78}$ \\ 
        Slow gait & $0.794\down{0.772}\up{0.815}$ & $3.25\down{3.21}\up{3.32}$ & ~ & ~ & ~ & ~ & ~ & ~ & ~ \\ 
        Slow gait & $0.857\down{0.840}\up{0.874}$ & $3.25\down{3.24}\up{3.27}$ & $3.75\down{2.72}\up{7.07}$ & $0.64\down{0.62}\up{0.68}$ & ~ & ~ & ~ & ~ & ~ \\ 
        Slow gait & $0.874\down{0.860}\up{0.888}$ & $2.34\down{2.30}\up{2.41}$ & $3.29\down{0.66}\up{76.53}$ & $0.65\down{0.56}\up{0.89}$ & $1.44\down{1.22}\up{1.97}$ & $1.25\down{1.06}\up{1.71}$ & $3.66\down{3.12}\up{5.02}$ & $1.06\down{0.90}\up{1.45}$ & $2.07\down{1.77}\up{2.84}$ \\ 
        Exhaustion & $0.751\down{0.729}\up{0.773}$ & $2.62\down{2.61}\up{2.66}$ & ~ & ~ & ~ & ~ & ~ & ~ & ~ \\ 
        Exhaustion & $0.754\down{0.735}\up{0.773}$ & $2.47\down{2.47}\up{2.48}$ & $1.33\down{1.21}\up{1.62}$ & $1.32\down{1.31}\up{1.35}$ & ~ & ~ & ~ & ~ & ~ \\ 
        Exhaustion & $0.787\down{0.769}\up{0.804}$ & $2.00\down{1.99}\up{2.03}$ & $1.33\down{0.64}\up{5.63}$ & $1.21\down{1.13}\up{1.40}$ & $1.46\down{1.36}\up{1.69}$ & $0.91\down{0.84}\up{1.05}$ & $1.22\down{1.13}\up{1.41}$ & $3.37\down{3.13}\up{3.89}$ & $1.30\down{1.21}\up{1.51}$ \\ 
        Low activity & $0.791\down{0.769}\up{0.812}$ & $3.09\down{3.07}\up{3.13}$ & ~ & ~ & ~ & ~ & ~ & ~ & ~ \\ 
        Low activity & $0.827\down{0.808}\up{0.846}$ & $2.82\down{2.82}\up{2.83}$ & $2.64\down{2.31}\up{3.44}$ & $1.33\down{1.31}\up{1.36}$ & ~ & ~ & ~ & ~ & ~ \\ 
        Low activity & $0.852\down{0.830}\up{0.873}$ & $2.13\down{2.11}\up{2.17}$ & $2.41\down{0.98}\up{13.89}$ & $1.18\down{1.08}\up{1.41}$ & $1.56\down{1.42}\up{1.85}$ & $1.14\down{1.05}\up{1.36}$ & $1.42\down{1.30}\up{1.69}$ & $1.43\down{1.31}\up{1.70}$ & $4.66\down{4.26}\up{5.55}$ \\ \hline
    \end{tabular}
\begin{tablenotes}
\item[*] cross-validated estimate (out-of-sample).
\end{tablenotes}
\end{threeparttable}
\end{table}

\begin{table}
    \centering
    \begin{threeparttable}
    \caption{Logistic regression odds ratios with 95\% confidence intervals (and AUC) --- HRS}\label{tab:glm_hrs}
    \begin{tabular}{llllllllll}
        Outcome & AUC\tnote{*} & FI per 0.1 & Age per 10 & Female & Weight loss & Weakness & Slow gait & Exhaustion & Low activity \\ \hline
        FP frailty & $0.756\down{0.742}\up{0.771}$ & $2.32\down{2.30}\up{2.34}$ & ~ & ~ & ~ & ~ & ~ & ~ & ~ \\ 
        FP frailty & $0.786\down{0.772}\up{0.800}$ & $2.29\down{2.28}\up{2.29}$ & $2.78\down{2.43}\up{3.62}$ & $0.92\down{0.91}\up{0.95}$ & ~ & ~ & ~ & ~ & ~ \\ 
        FP frailty & $0.835\down{0.824}\up{0.847}$ & $1.69\down{1.67}\up{1.71}$ & $2.36\down{1.09}\up{10.60}$ & $0.87\down{0.80}\up{1.01}$ & $1.61\down{1.49}\up{1.87}$ & $2.54\down{2.35}\up{2.95}$ & $1.86\down{1.72}\up{2.16}$ & $2.56\down{2.37}\up{2.98}$ & $2.49\down{2.30}\up{2.89}$ \\ 
        Weight loss & $0.574\down{0.549}\up{0.599}$ & $1.26\down{1.25}\up{1.27}$ & ~ & ~ & ~ & ~ & ~ & ~ & ~ \\ 
        Weight loss & $0.606\down{0.585}\up{0.626}$ & $1.22\down{1.22}\up{1.22}$ & $1.38\down{1.27}\up{1.62}$ & $1.51\down{1.50}\up{1.54}$ & ~ & ~ & ~ & ~ & ~ \\ 
        Weight loss & $0.651\down{0.629}\up{0.673}$ & $1.13\down{1.13}\up{1.15}$ & $1.25\down{0.68}\up{4.11}$ & $1.41\down{1.33}\up{1.59}$ & $3.40\down{3.20}\up{3.83}$ & $1.45\down{1.37}\up{1.64}$ & $1.20\down{1.13}\up{1.35}$ & $1.17\down{1.10}\up{1.32}$ & $1.12\down{1.05}\up{1.26}$ \\ 
        Weakness & $0.619\down{0.607}\up{0.631}$ & $1.44\down{1.44}\up{1.46}$ & ~ & ~ & ~ & ~ & ~ & ~ & ~ \\ 
        Weakness & $0.713\down{0.703}\up{0.724}$ & $1.38\down{1.38}\up{1.38}$ & $3.87\down{3.59}\up{4.49}$ & $1.04\down{1.04}\up{1.06}$ & ~ & ~ & ~ & ~ & ~ \\ 
        Weakness & $0.800\down{0.789}\up{0.811}$ & $1.19\down{1.19}\up{1.20}$ & $2.99\down{1.89}\up{7.35}$ & $1.00\down{0.96}\up{1.10}$ & $1.11\down{1.06}\up{1.22}$ & $9.36\down{8.94}\up{10.24}$ & $1.26\down{1.20}\up{1.37}$ & $1.05\down{1.00}\up{1.15}$ & $1.15\down{1.10}\up{1.26}$ \\ 
        Slow gait & $0.709\down{0.694}\up{0.725}$ & $2.00\down{1.98}\up{2.03}$ & ~ & ~ & ~ & ~ & ~ & ~ & ~ \\ 
        Slow gait & $0.747\down{0.735}\up{0.759}$ & $2.02\down{2.01}\up{2.02}$ & $2.35\down{2.01}\up{3.19}$ & $0.61\down{0.60}\up{0.63}$ & ~ & ~ & ~ & ~ & ~ \\ 
        Slow gait & $0.780\down{0.767}\up{0.794}$ & $1.61\down{1.60}\up{1.64}$ & $2.10\down{0.90}\up{11.05}$ & $0.62\down{0.57}\up{0.73}$ & $1.27\down{1.17}\up{1.50}$ & $1.55\down{1.42}\up{1.83}$ & $3.20\down{2.94}\up{3.78}$ & $1.58\down{1.45}\up{1.86}$ & $1.51\down{1.38}\up{1.78}$ \\ 
        Exhaustion & $0.688\down{0.675}\up{0.700}$ & $1.87\down{1.87}\up{1.88}$ & ~ & ~ & ~ & ~ & ~ & ~ & ~ \\ 
        Exhaustion & $0.688\down{0.676}\up{0.700}$ & $1.86\down{1.86}\up{1.87}$ & $1.07\down{1.01}\up{1.19}$ & $1.02\down{1.02}\up{1.03}$ & ~ & ~ & ~ & ~ & ~ \\ 
        Exhaustion & $0.774\down{0.763}\up{0.785}$ & $1.47\down{1.46}\up{1.48}$ & $1.13\down{0.78}\up{2.33}$ & $1.03\down{0.99}\up{1.11}$ & $1.05\down{1.01}\up{1.13}$ & $1.01\down{0.97}\up{1.09}$ & $1.16\down{1.12}\up{1.25}$ & $5.74\down{5.53}\up{6.17}$ & $1.24\down{1.19}\up{1.33}$ \\ 
        Low activity & $0.707\down{0.695}\up{0.720}$ & $1.95\down{1.95}\up{1.96}$ & ~ & ~ & ~ & ~ & ~ & ~ & ~ \\ 
        Low activity & $0.719\down{0.707}\up{0.732}$ & $1.89\down{1.89}\up{1.89}$ & $1.51\down{1.42}\up{1.71}$ & $1.47\down{1.46}\up{1.49}$ & ~ & ~ & ~ & ~ & ~ \\ 
        Low activity & $0.781\down{0.769}\up{0.793}$ & $1.52\down{1.51}\up{1.53}$ & $1.48\down{0.96}\up{3.46}$ & $1.29\down{1.24}\up{1.41}$ & $1.16\down{1.11}\up{1.26}$ & $1.27\down{1.22}\up{1.39}$ & $1.12\down{1.08}\up{1.22}$ & $1.24\down{1.19}\up{1.35}$ & $6.54\down{6.26}\up{7.12}$ \\ \hline
    \end{tabular}
\begin{tablenotes}
\item[*] 10-fold, 10-repeat cross-validation estimate (out-of-sample).
\end{tablenotes}
\end{threeparttable}
\end{table}

\begin{figure*}[!ht]
    \centering
    \includegraphics[width=\textwidth]{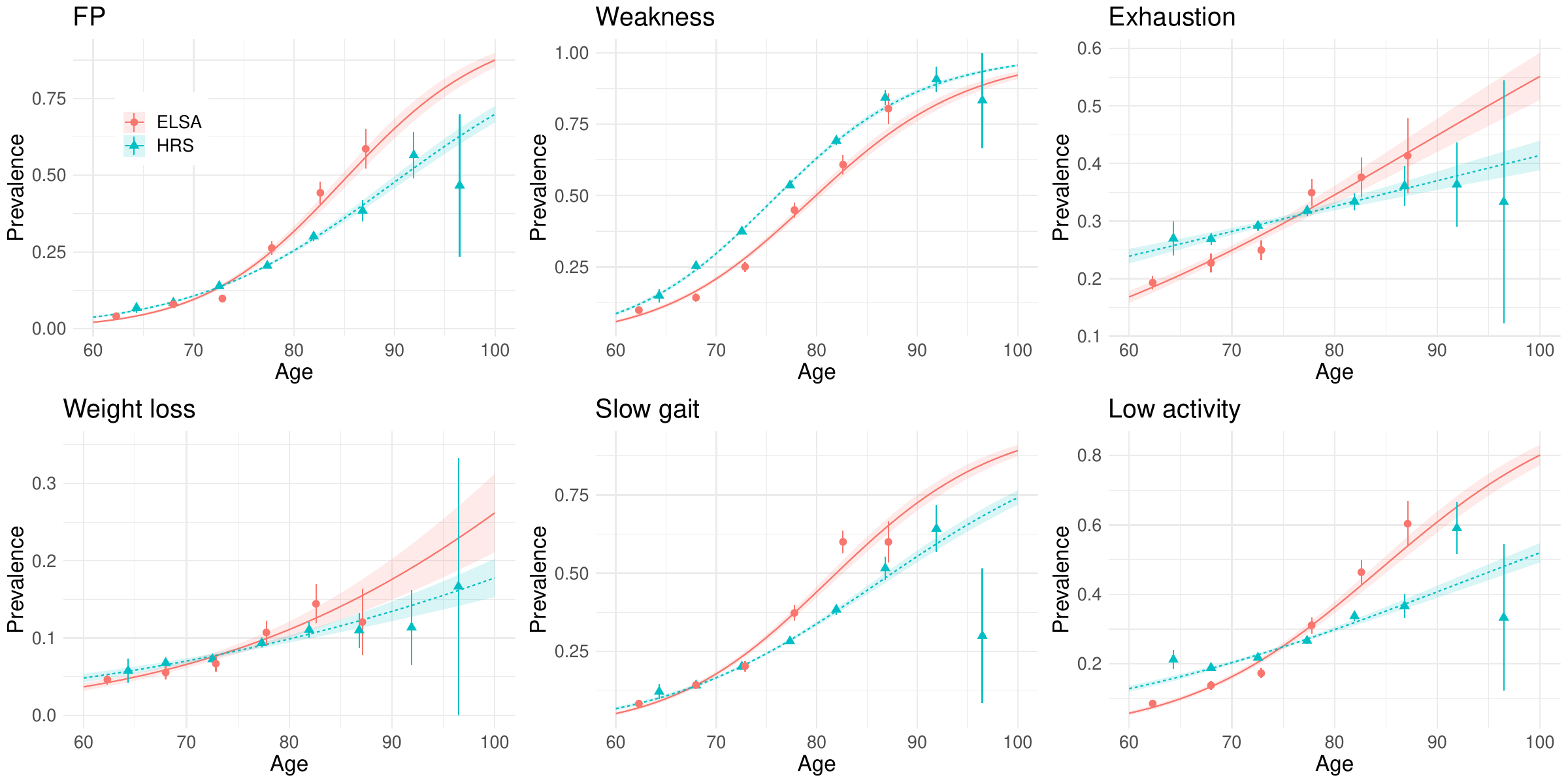} 
    \caption{Future prediction using chronological age. HRS and ELSA (longitudinal). Lines are sqrt logistic regression fits. The future condition of an individual at followup can be inferred from their current FI score. }
    \label{fig:prognosticate_age}
\end{figure*}

\begin{figure*}
    \centering
    \includegraphics[width=0.5\textwidth]{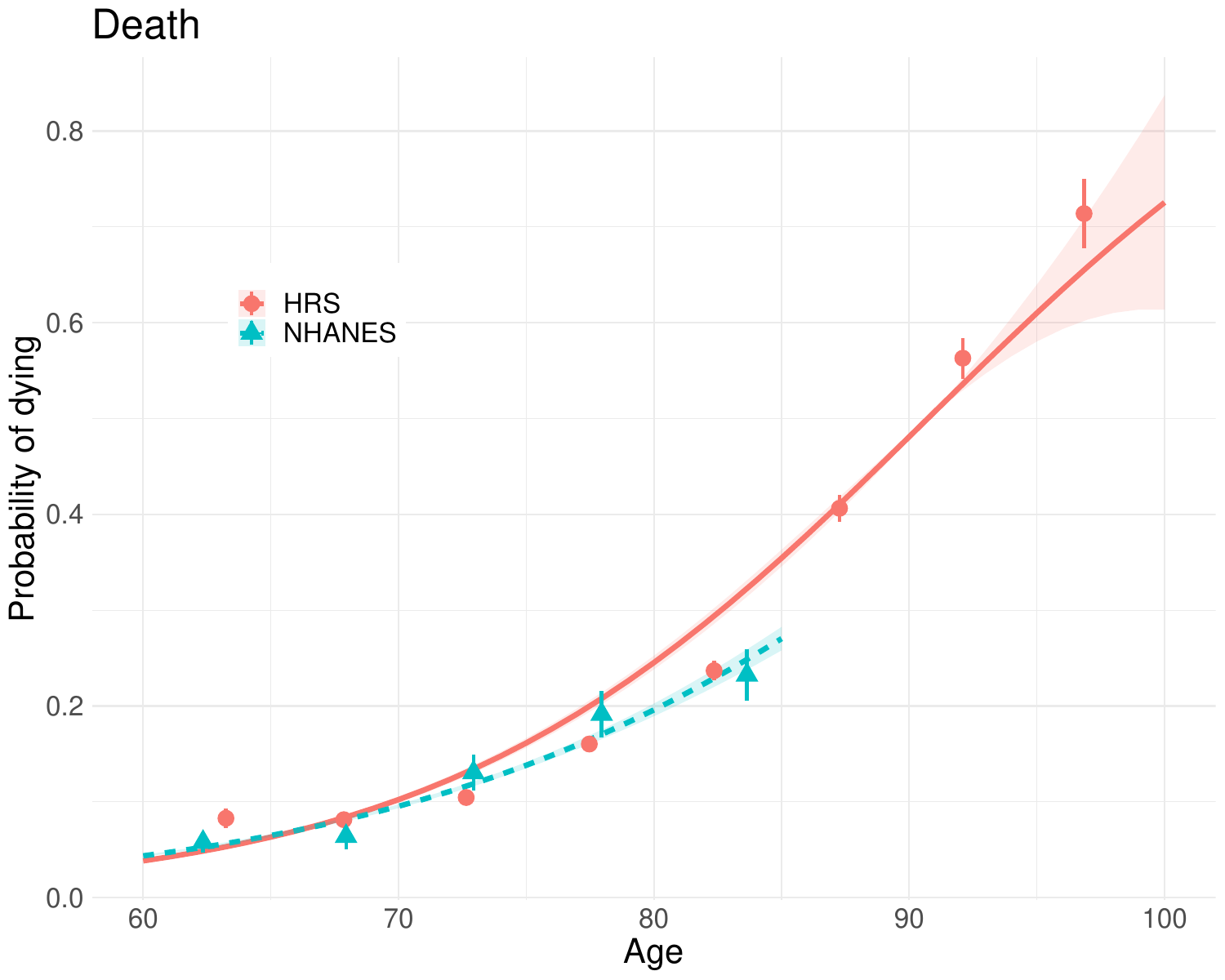}
    \caption{Probability of dying before followup (4 years) --- age. Points are non-parametric Kaplain-Meier estimates with standard errors. Line is the logit-square root fit, $\text{logit(prob.)}\propto\sqrt{\text{Age}}$ (to points).}
    \label{fig:age_death}
\end{figure*}

The future prediction calibration curves for chronological age are provided for the FPFP5 deficits in Figure~\ref{fig:prognosticate_age} and for survival in Figure~\ref{fig:age_death}. The logit-square root fits observed FPFP5 decline frequencies and mortality excellently. We observed that there are significant inter-study differences between the calibration curves, indicating that the differences we observed in the main text are not necessarily related to FI reproducibility.

\subsection{ELSA survival} \label{sec:s_elsa}
\begin{figure*}[!ht]
    \centering
    \includegraphics[width=0.5\textwidth]{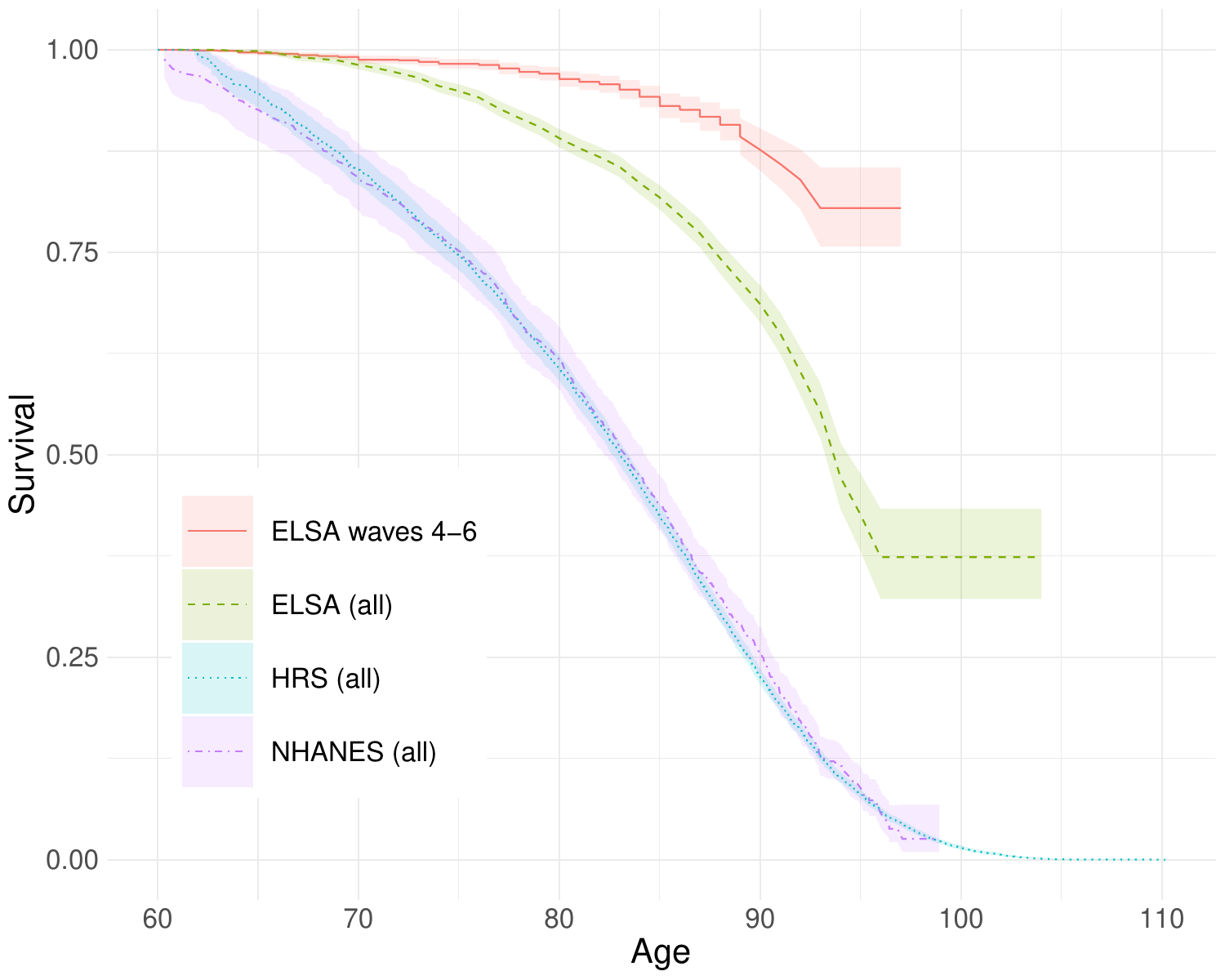} 

    \caption{ELSA survival had abnormally low event rate. Whereas HRS (dotted blue line) and NHANES (dot-dashed blue line) strongly overlapped, the overall ELSA population survival (solid red line) was much higher. When we looked at the interval we used in the present study the survival was even higher (waves 4-6, dashed green line).} \label{fig:s_elsa}
\end{figure*}

We did not include our ELSA survival results in the main text because of the very low hazard observed, especially between waves 4 and 6. This is illustrated by Figure~\ref{fig:s_elsa}. HRS and NHANES survival overlap perfectly but ELSA has much higher survival, especially between waves 4 to 6, indicating an anomalously low hazard. We are therefore reluctant to put emphasis on the ELSA survival results but nevertheless include them here in the supplemental. Broadly, the ELSA results qualitatively agree with the HRS and NHANES results but with a notably lower mortality rate.

\begin{figure*}[!ht]
    \centering
    \begin{subfigure}[t]{0.49\textwidth}
        \centering
        \includegraphics[width=\textwidth]{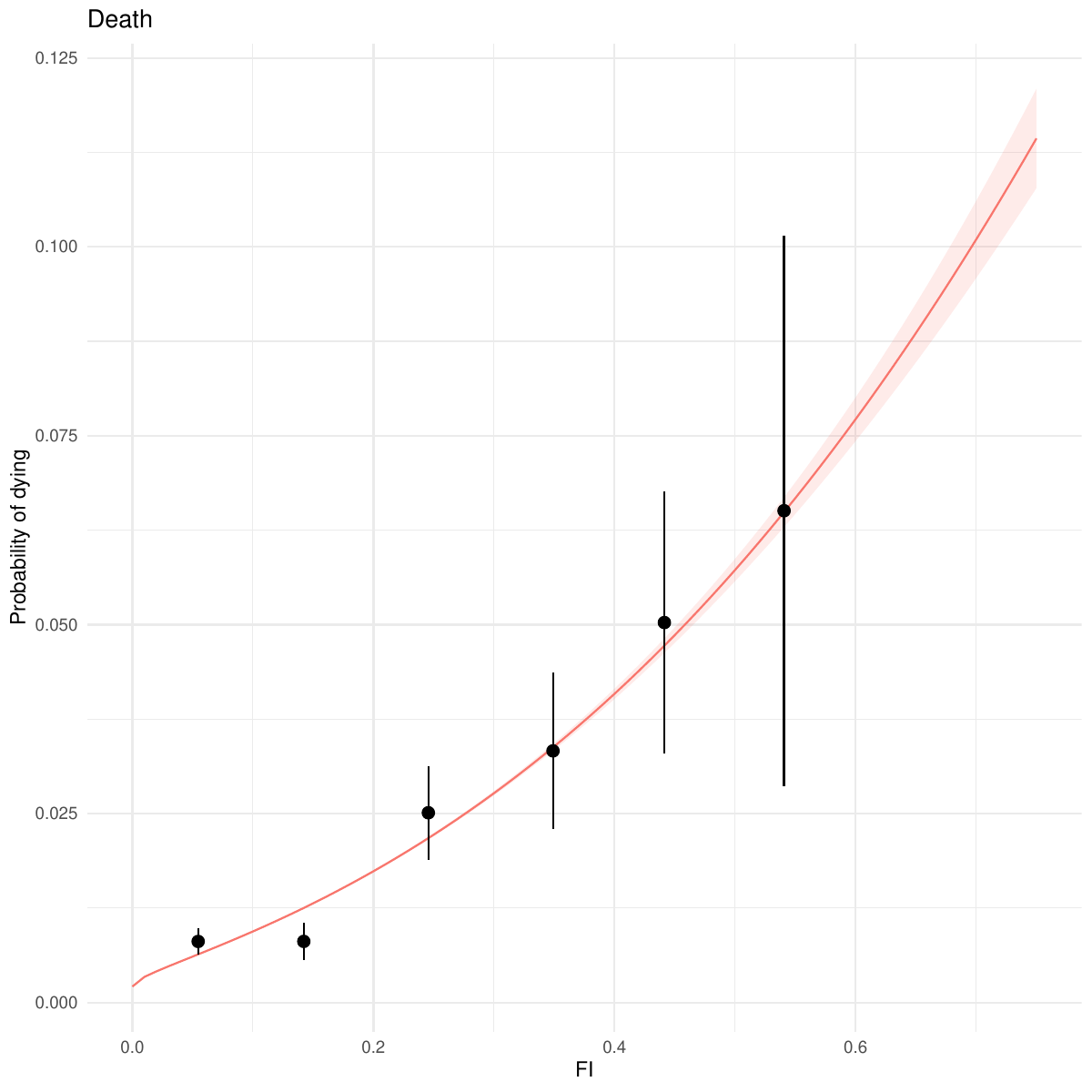} 
        \caption{FI.}
    \end{subfigure}
        \begin{subfigure}[t]{0.49\textwidth}
        \centering
        \includegraphics[width=\textwidth]{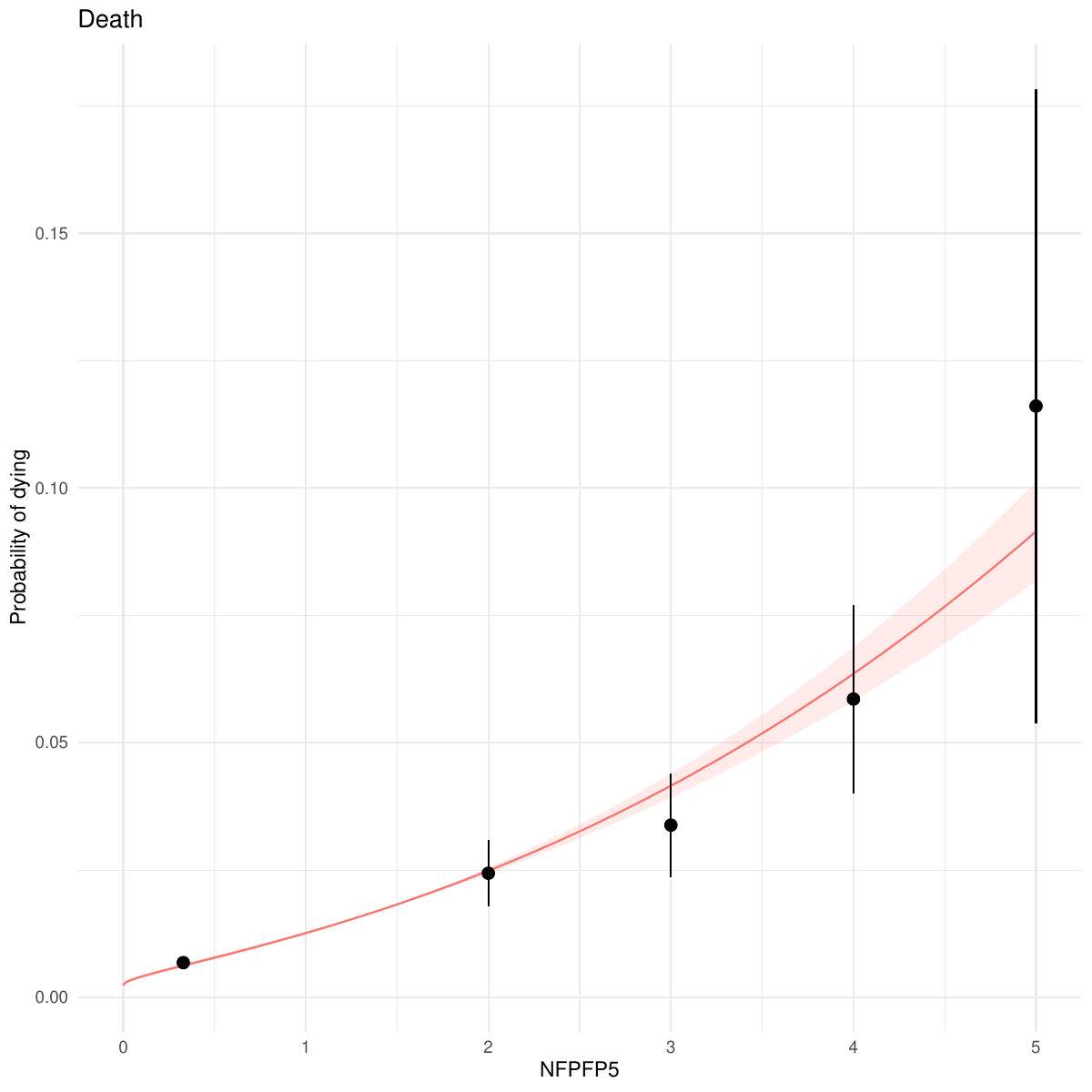}  
        \caption{NFPFP5.}
    \end{subfigure}
    \caption{Probability of dying before followup (4 years) --- ELSA. The FI (\textbf{a}) and NFPFP5 (\textbf{b}) both have similar, weakly super-linear behaviour. Points are non-parametric Kaplain-Meier estimates with standard errors. Line is the logit-square root fit, $\text{logit(prob.)}\propto\sqrt{f}$ (to points).} \label{fig:elsa_death}
\end{figure*}

\begin{figure*}[!ht]
    \centering
        \includegraphics[width=0.5\textwidth]{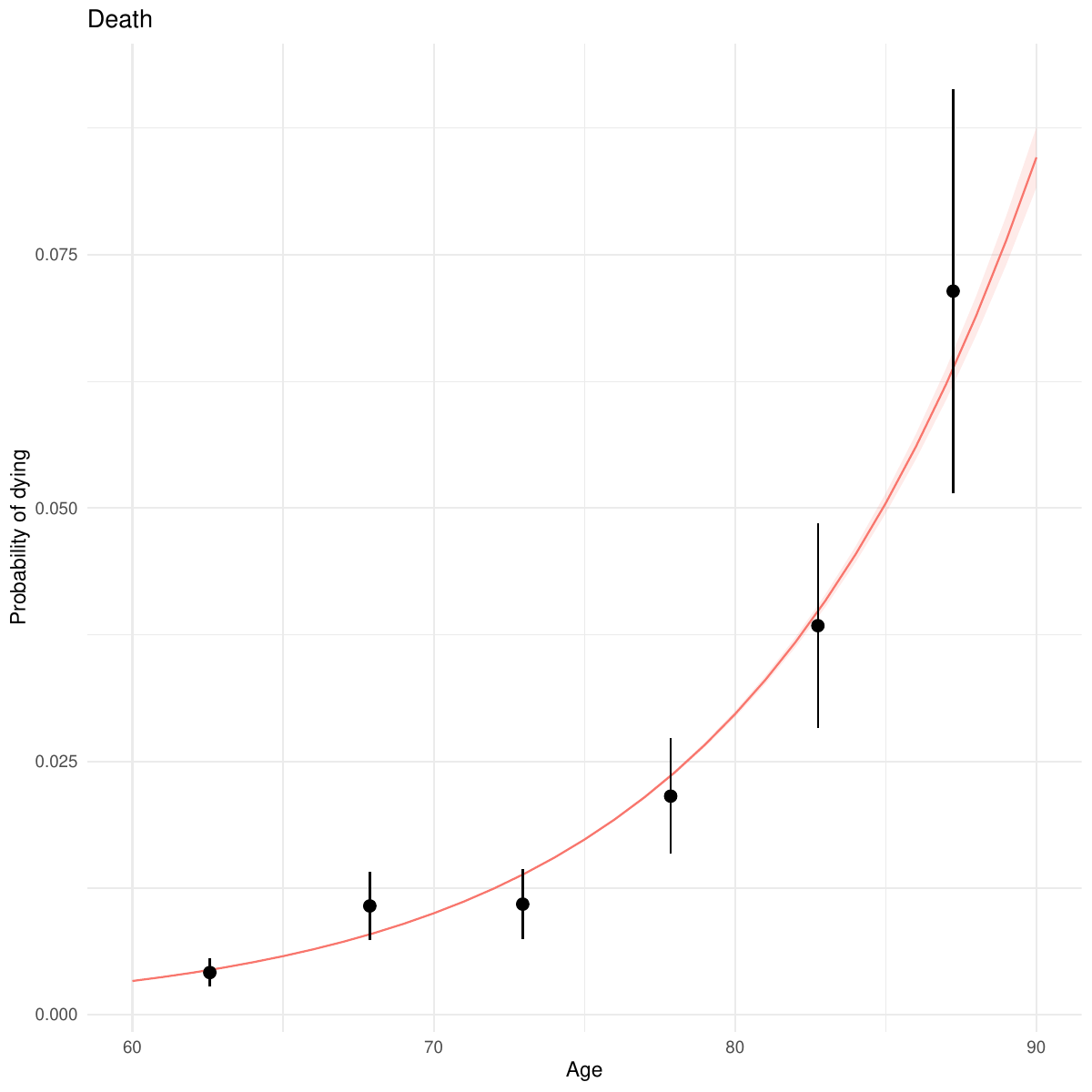} 
    \caption{Probability of dying before followup (4 years) --- ELSA, age. Points are non-parametric Kaplain-Meier estimates with standard errors. Line is the logit-square root fit, $\text{logit(prob.)}\propto\sqrt{\text{Age}}$ (to points).} \label{fig:elsa_age_death}
\end{figure*}

As in the main text, both the FI and NFPFP5 demonstrated a logit-square root relationship with the probability of dying before followup,
Figure~\ref{fig:elsa_death}. Again, we see no evidence of a sudden jump at NFPFP5 = 5. These observations also apply to the age-dependence, Figure~\ref{fig:elsa_age_death}.

Finally we have the full logistic regression models reported in Table~\ref{tab:elsa_cox}. Recall that we considered 4 models, mostly nested: (Model 0) age and sex alone, (Model 1) the FI alone, (Model 2) Model 1 + age and sex, and (Model 3) Model 2 + the specific FPFP5 deficits. The values are all comparable to HRS and NHANES (Table~3). While weight loss shows a much larger point estimate for ELSA, the confidence interval still overlaps with both HRS and NHANES.

\begin{table}
    \centering
    \caption{Cox survival hazard ratios with 95\% confidence intervals (and BIC) --- ELSA (time-dependent).} \label{tab:elsa_cox}
\begin{threeparttable}
    \begin{tabular}{l|llll}
        Predictor/Measure & Model 0 & Model 1 & Model 2 & Model 3 \\ \hline
        C & $0.73\down{0.68}\up{0.77}$ & $0.68\down{0.63}\up{0.73}$ &  $0.77\down{0.73}\up{0.81}$ & $0.80\down{0.76}\up{0.84}$ \\ 
        BIC\tnote{1} & Ref. &  33 & -28 & -28 \\ 
        Female & $0.52\down{0.36}\up{0.75}$ & & $0.44\down{0.31}\up{0.64}$ & $0.41\down{0.28}\up{0.60}$ \\ 
        Age per 10 & $2.80\down{2.22}\up{3.54}$ & & $2.38\down{1.88}\up{3.02}$ & $2.05\down{1.58}\up{2.66}$ \\ 
        FI per 0.1 & & $1.56\down{1.40}\up{1.74}$ & $1.47\down{1.30}\up{1.66}$ & $1.19\down{1.00}\up{1.41}$ \\ 
        Weight loss & ~ & ~ & ~ & $2.55\down{1.53}\up{4.26}$ \\ 
        Weakness & ~ & ~ & ~ & $1.23\down{0.81}\up{1.86}$ \\ 
        Slow gait & ~ & ~ & ~ & $1.27\down{0.78}\up{2.08}$ \\ 
        Exhaustion & ~ & ~ & ~ & $1.59\down{1.04}\up{2.42}$ \\ 
        Low activity & ~ & ~ & ~ & $1.21\down{0.76}\up{1.94}$ \\ \hline
    \end{tabular}
\begin{tablenotes}
\item[1] Change in Bayesian information criterion (BIC), negative indicates a better model.
\end{tablenotes}
\end{threeparttable}
\end{table}

\subsection{Health deficit similarity and clustering}
\begin{figure*}[!ht]
        \includegraphics[width=\textwidth]{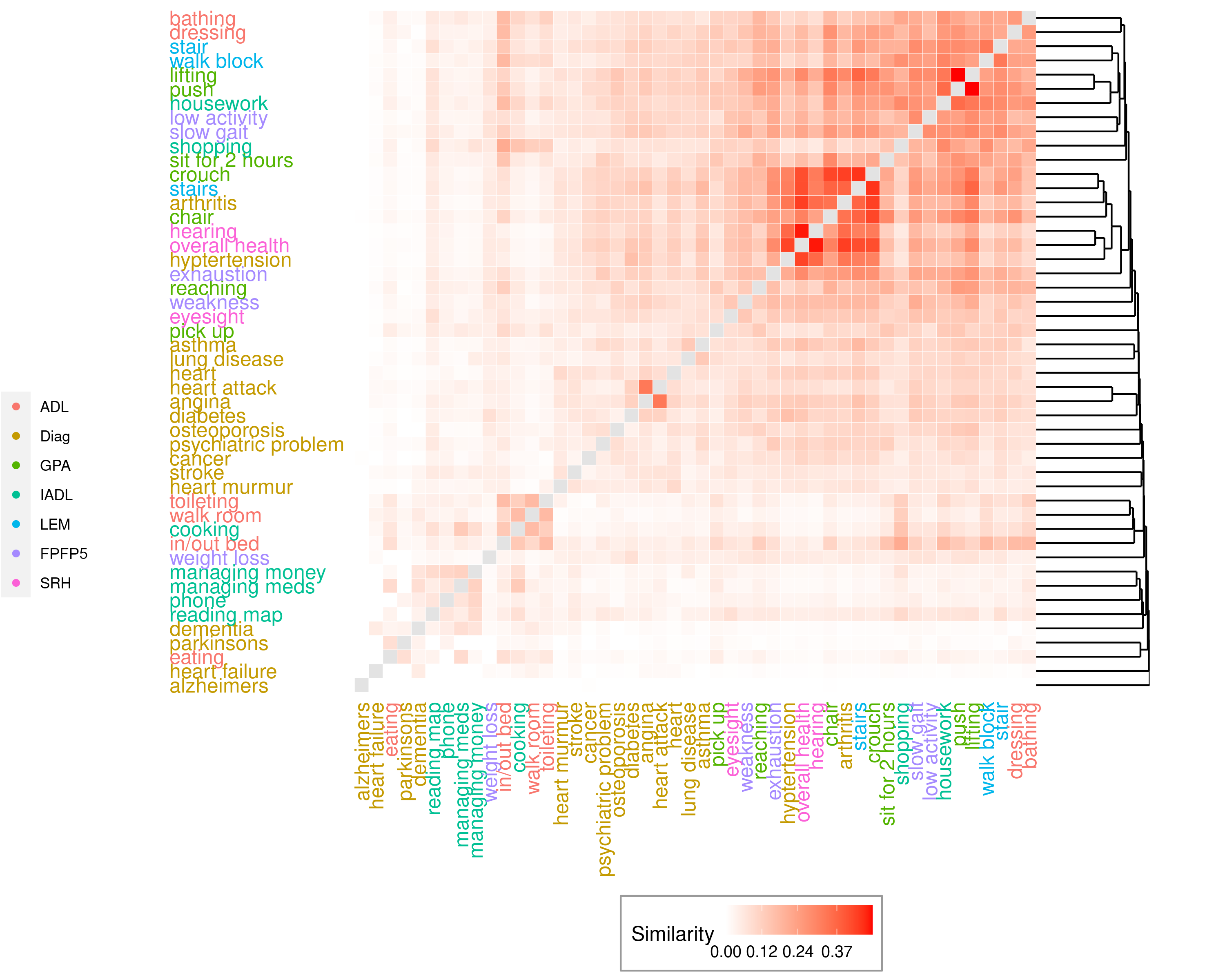} 
    \caption{\textbf{There was no distinct FPFP5 cluster among the health deficits (purple labels)}. ELSA. As with HRS (Figure~2), the FI ostensibly performs better than the NFPFP5 because there is no FPFP5 cluster, instead there is a huge number of closely related deficits to the FPFP5 that can help better describe the FPFP5 health state by including them in the summary metric. Red tiles are similar deficits whereas white are different, as measured by the Jaccard similarity coefficient ($1-\text{Jaccard distance}$). ADL: activity of daily living, Diag: medical diagnosis, GPA: general physical activity, HU: health care utilization, IADL: instrumental ADL, LEM: lower extremity mobility, SRH: self-reported health.} 
 \label{fig:elsa_clustering}
\end{figure*}

\begin{figure*}[!ht]
        \includegraphics[width=\textwidth]{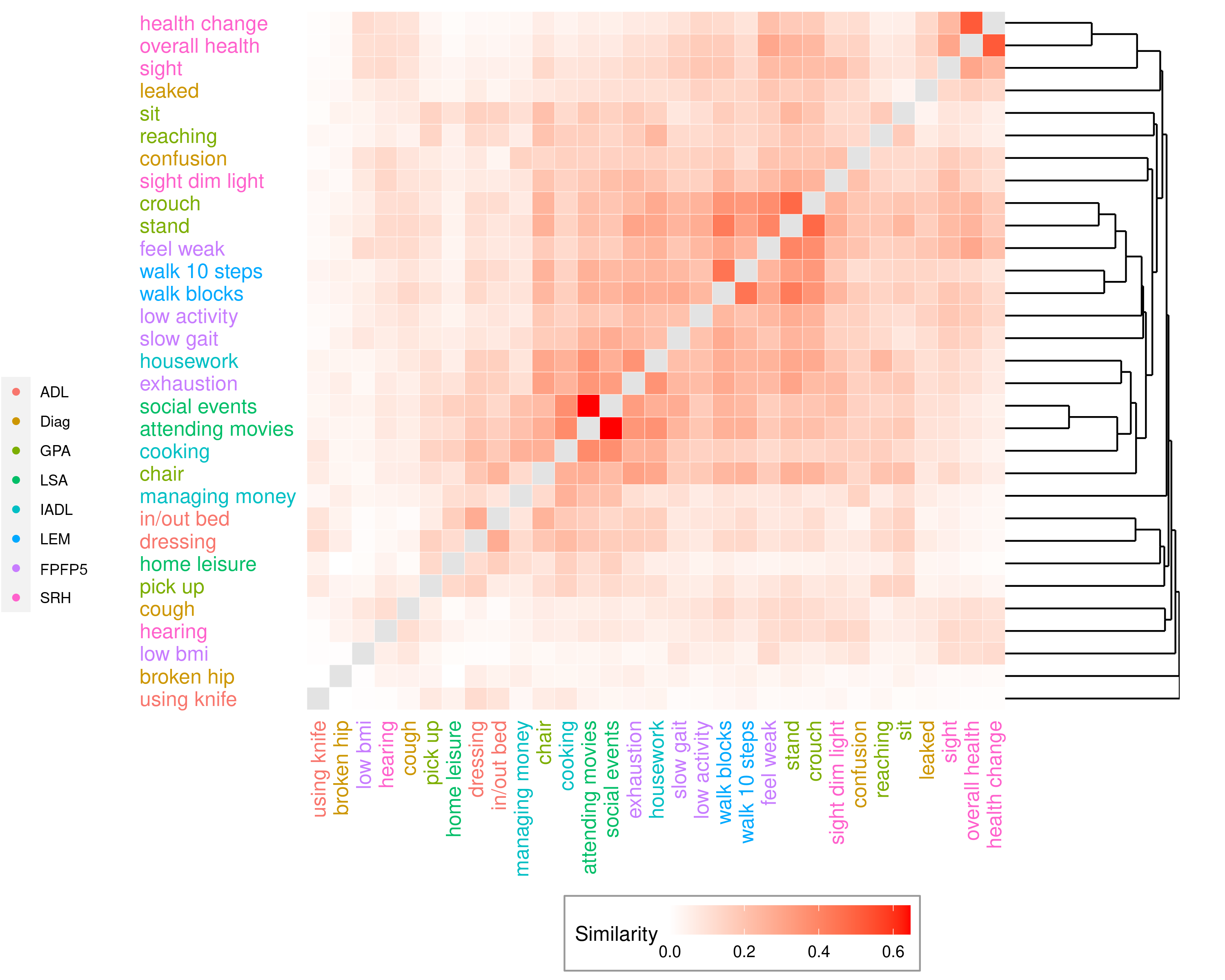} 
    \caption{\textbf{There was no distinct FPFP5 cluster among the health deficits (purple labels)}. NHANES. As with HRS (Figure~2), the FI ostensibly performs better than the NFPFP5 because there is no FPFP5 cluster, instead there is a huge number of closely related deficits to the FPFP5 that can help better describe the FPFP5 health state by including them in the summary metric. Red tiles are similar deficits whereas white are different, as measured by the Jaccard similarity coefficient ($1-\text{Jaccard distance}$). ADL: activity of daily living, Diag: medical diagnosis, GPA: general physical activity, HU: health care utilization, IADL: instrumental ADL, LEM: lower extremity mobility, SRH: self-reported health.} \label{fig:nhanes_clustering}
\end{figure*}

In Figure~2 we analyzed the clustering of FPFP5 deficits relative to the deficits used to generate the FI using the HRS dataset. Here we provide clustering for the other two datasets as well as a complete exposition of the underlying methods. ELSA clustering is shown in Figure~\ref{fig:elsa_clustering} and NHANES in Figure~\ref{fig:nhanes_clustering}. Our key observations hold across all three datasets: while clusters do exist, the FPFP5 do not form a unique cluster but rather are 5 deficits among many closely related, and likely redundant, health deficits. This explains why the overall average (FI) is a better predictor than the specific average (NFPFP5). 

Hierarchical clustering was performed using \texttt{hclust}, a default package in \texttt{R} \cite{R_Core_Team2021-uq}. We used the \texttt{average} method and the \texttt{binary} distance measure, which is the Jaccard distance; we then plot the Jaccard similarity together with the clustering dendrogram. The Jaccard similarity is the number of times both variables are 1 divided by the number of times at least one is 1 (the distance is 1 minus the similarity). (We binarized the ELSA self-reported deficits at $0.49$.) When plotting we classified (coloured) variable types according to: activities of daily living (ADLs), instrumental activities of daily living (IADLs), diagnoses (Diag), general physical activity (GPA), health care utilization (HU), lower extreme mobility (LEM), the FPFP5, or self-reported health (SRH) (list based on Kuo \textit{et al} \cite{Kuo2006-pl}).

\FloatBarrier

\section{Sex effects}
Sex effects were small and did not change our central messages. Both sexes showed similar patterns in feature selection, the calibration curves and survival prediction. Data between the sexes overlapped within confidence intervals. Females might be a bit easier to predict in general. Some significant differences were observed in the calibration curves, although these could likely be accounted for by including sex as a binarized covariate since they appear to simply shift the position of the inflection point. Given the lack of substantive difference and indication that sex as a covariate appears sufficient, we neglect to perform the computationally intensive logistic regression models by sex.

\begin{figure*}[!ht]
    \centering
        \includegraphics[width=\textwidth]{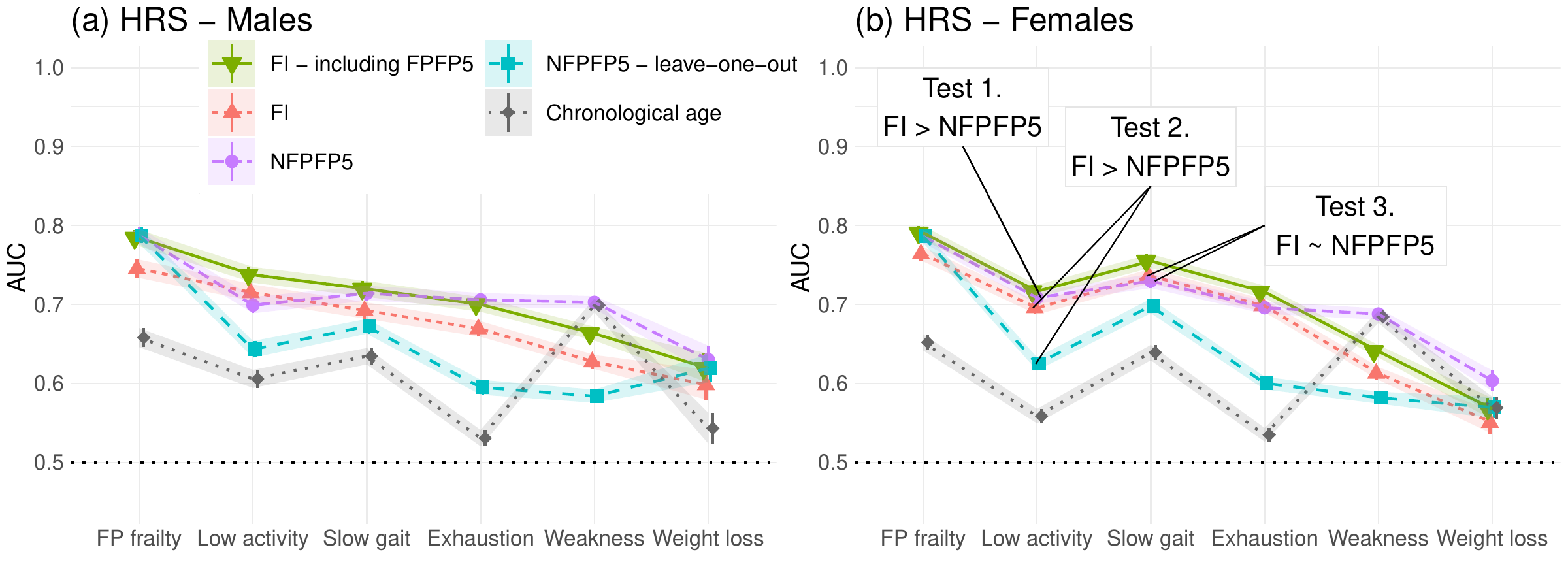} 
    \caption{\textbf{The FI predicts future FPFP5 deficits better than NFPFP5.} HRS, sex-stratified. The curves are similar between sexes. Visually, individual points appear to agree between sexes within error. The AUC is the probability that a metric will correctly rank positive individuals as higher than negative individuals \cite{Hanley1982-pm} (0.5 is guess, dotted line; 1 is oracle). Note: the x-axes are sorted by AUC. Leave-one-out excludes the outcome deficit from the predictor (doesn't affect FP~frailty prediction). Error bars are standard errors.} \label{fig:auc_hrs_sex}
\end{figure*}

\begin{figure*}[!ht]
    \centering
        \includegraphics[width=\textwidth]{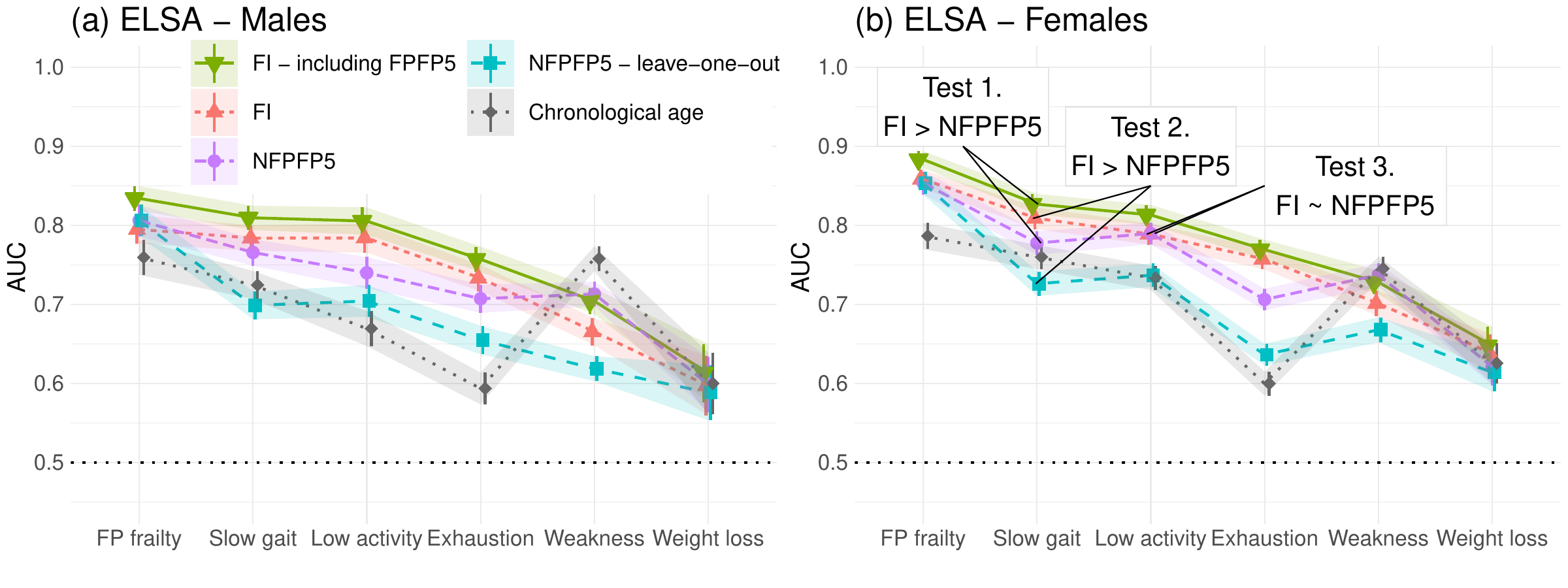} 
    \caption{\textbf{The FI predicts future FPFP5 deficits better than NFPFP5.} ELSA, sex-stratified. The curves are similar between sexes. Visually, individual points appear to agree between sexes within error. The AUC is notably higher for females when predicting FP~frailty, however, by $\sim0.05$. The AUC is the probability that a metric will correctly rank positive individuals as higher than negative individuals \cite{Hanley1982-pm} (0.5 is guess, dotted line; 1 is oracle). Note: the x-axes are sorted by AUC. Leave-one-out excludes the outcome deficit from the predictor (doesn't affect FP~frailty prediction). Error bars are standard errors.} \label{fig:auc_elsa_sex}
\end{figure*}

\begin{figure*}[!ht]
    \centering
        \includegraphics[width=\textwidth]{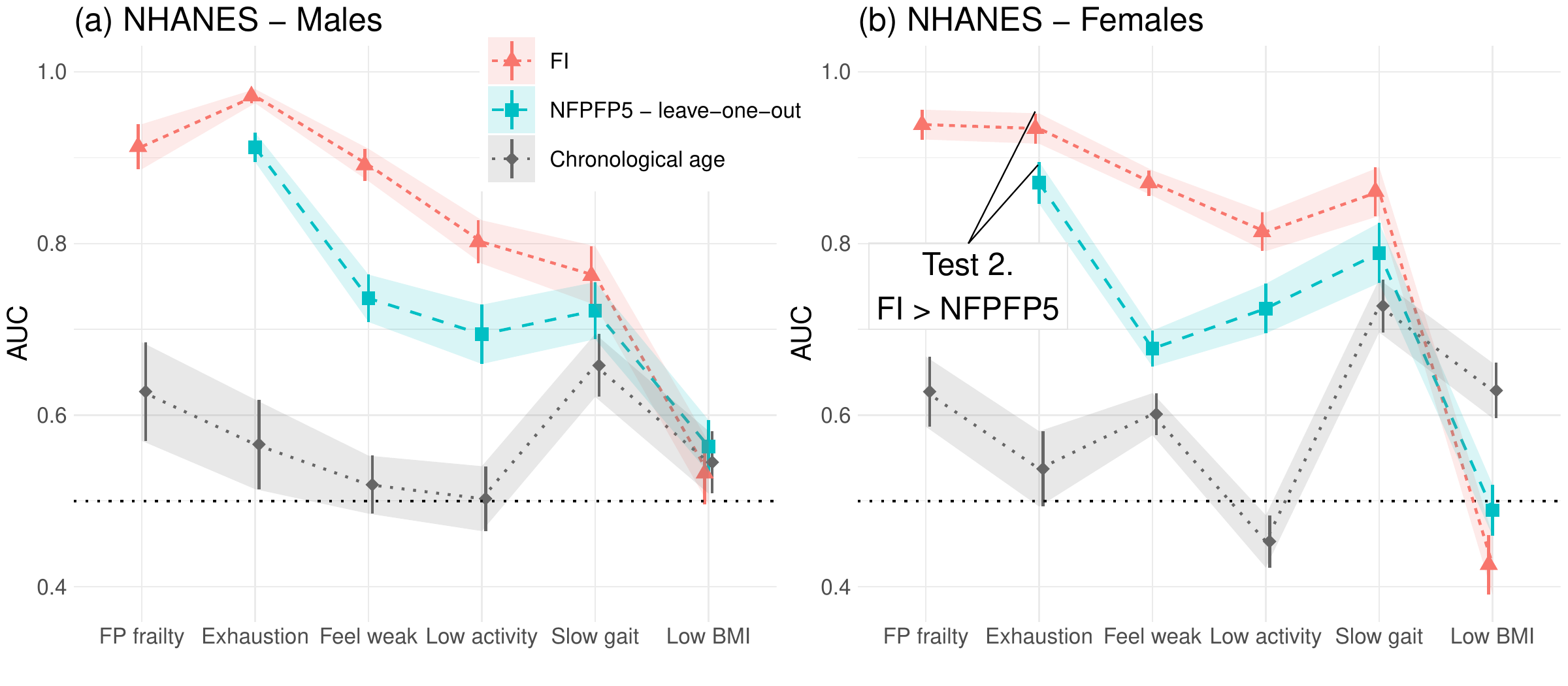} 
    \caption{\textbf{The FI better infers current FPFP5 deficits than the NFPFP5}. NHANES, sex-stratified. Sex effects were on the order of the error bars, with similar curves. Predictive power for predicting current FPFP5 deficits using the current FI versus (leave-one-out) NFPFP5 (NHANES). NFPFP5 is the number of FPFP5 deficits excluding the current outcome (leave-one-out; not applicable for FP~frailty). Note: the definitions for low activity and exhaustion are quite different from the longitudinal datasets (Table~1). Error bars are standard errors.} \label{fig:auc_nhanes_sex}
\end{figure*}

The feature selection curves are provided in Figures: \ref{fig:auc_hrs_sex} (HRS), \ref{fig:auc_elsa_sex} (ELSA), and \ref{fig:auc_nhanes_sex} (NHANES). There are differences between the sexes, but if you look closely they typically are of the same order as the error bars. Female points do seem to be more likely to be higher than male points, although the differences are typically on the order of the error bars. Females may be easier to predict but the difference is small.

\begin{figure*}[!ht]
    \centering
    \includegraphics[width=\textwidth]{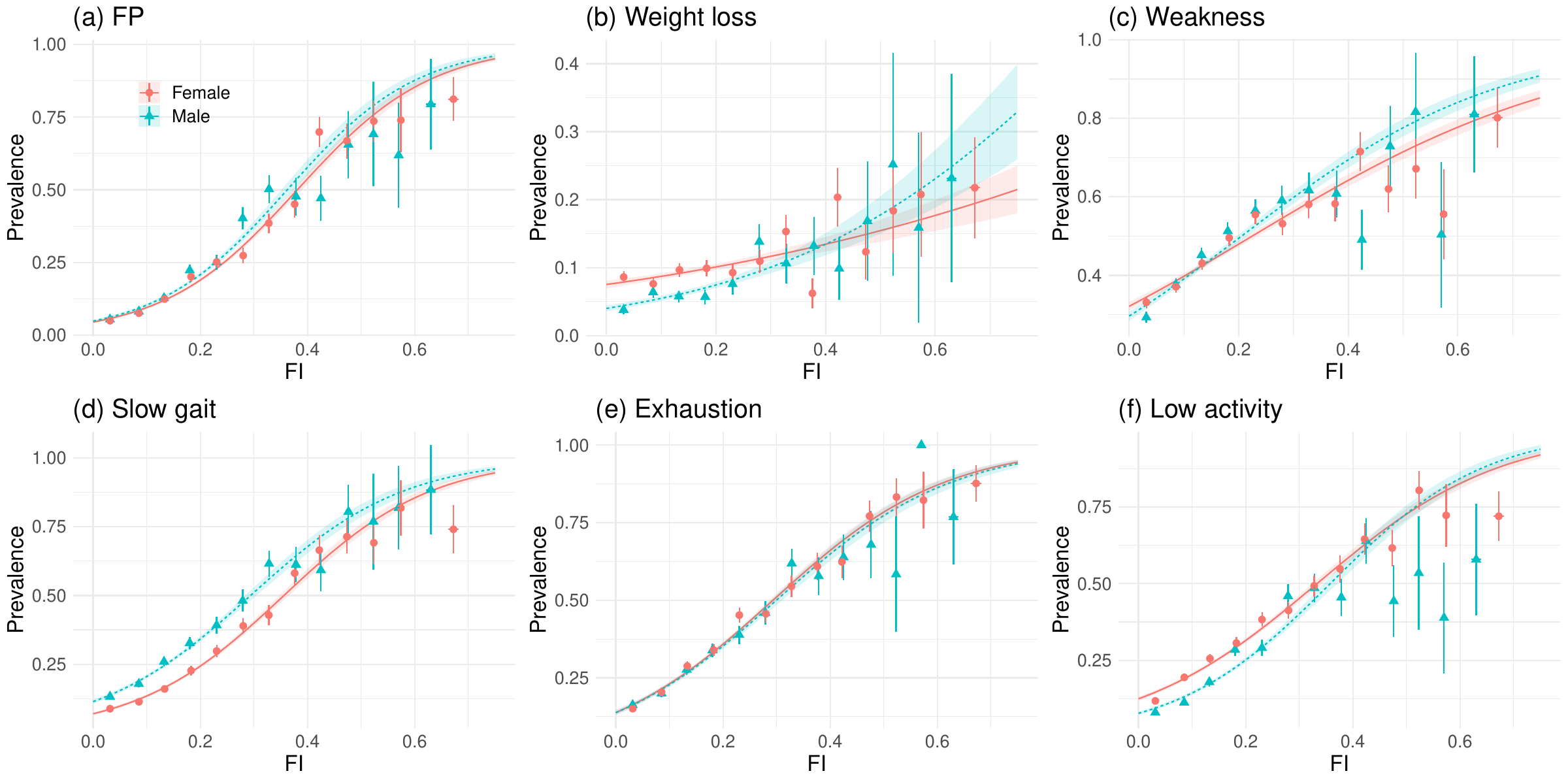} 
    \caption{\textbf{Future prediction calibration using the FI. HRS (longitudinal)} --- sex-stratified. While the prevalence of slow gait is noticeably lower among females, the curves have the same functional form across all FPFP5 deficits. The curve can be shifted by including sex as a covariate. Points are mean $\pm$ standard error, binned by FI. Lines are sqrt logistic regression fits, $\text{logit(prob.)}\propto\sqrt{f}$.}
    \label{fig:calibration_hrs_sex}
\end{figure*}

\begin{figure*}[!ht]
    \centering
    \includegraphics[width=\textwidth]{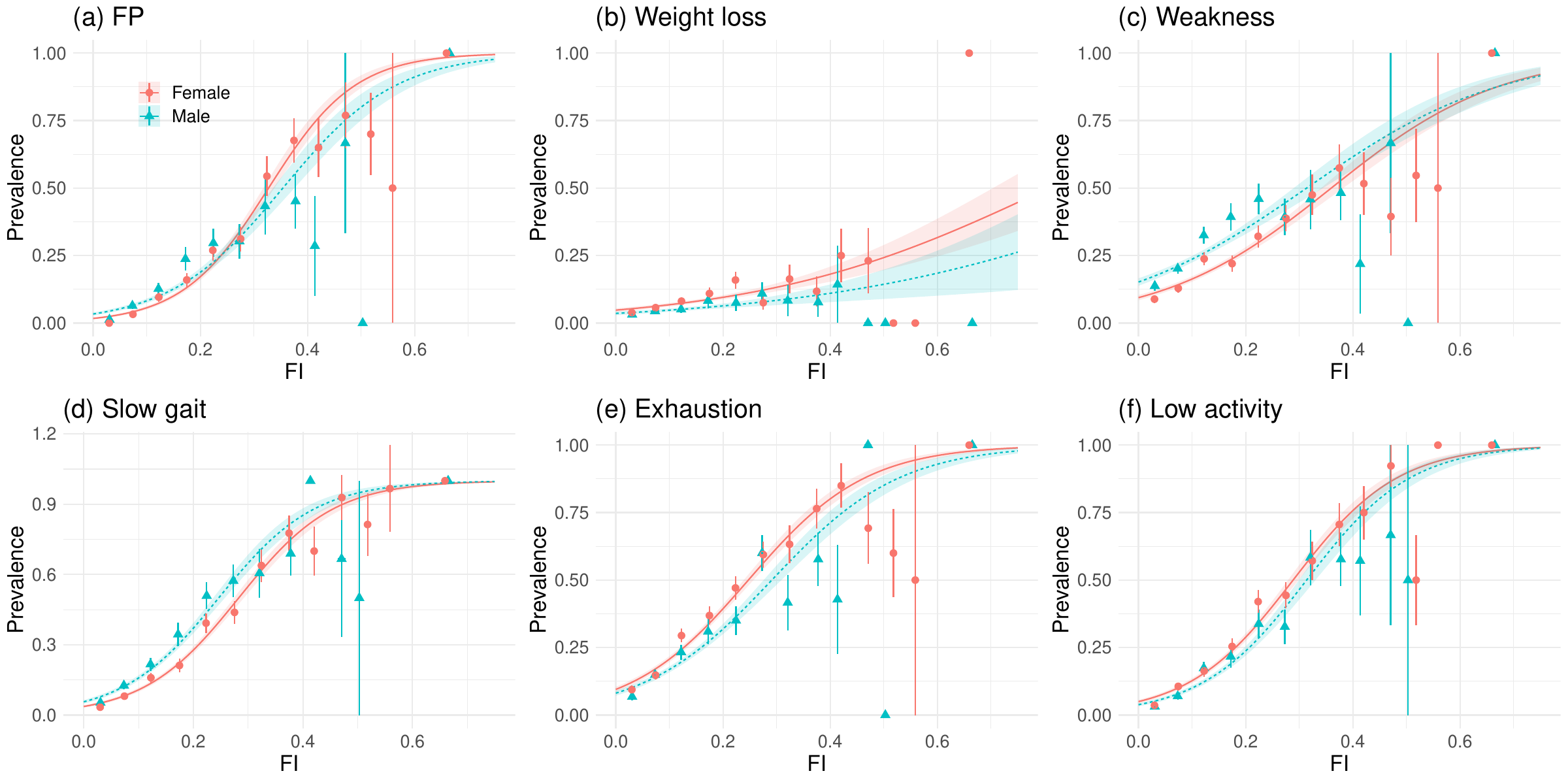} 
    \caption{\textbf{Future prediction calibration using the FI. ELSA (longitudinal)} --- sex-stratified. As with HRS, the prevalence of slow gait was lower among females, although the curves have the same functional form across all FPFP5 deficits. The curve can be shifted by including sex as a covariate. Points are mean $\pm$ standard error, binned by FI. Lines are sqrt logistic regression fits, $\text{logit(prob.)}\propto\sqrt{f}$.}
    \label{fig:calibration_elsa_sex}
\end{figure*}

\begin{figure*}[!ht]
    \centering
    \includegraphics[width=\textwidth]{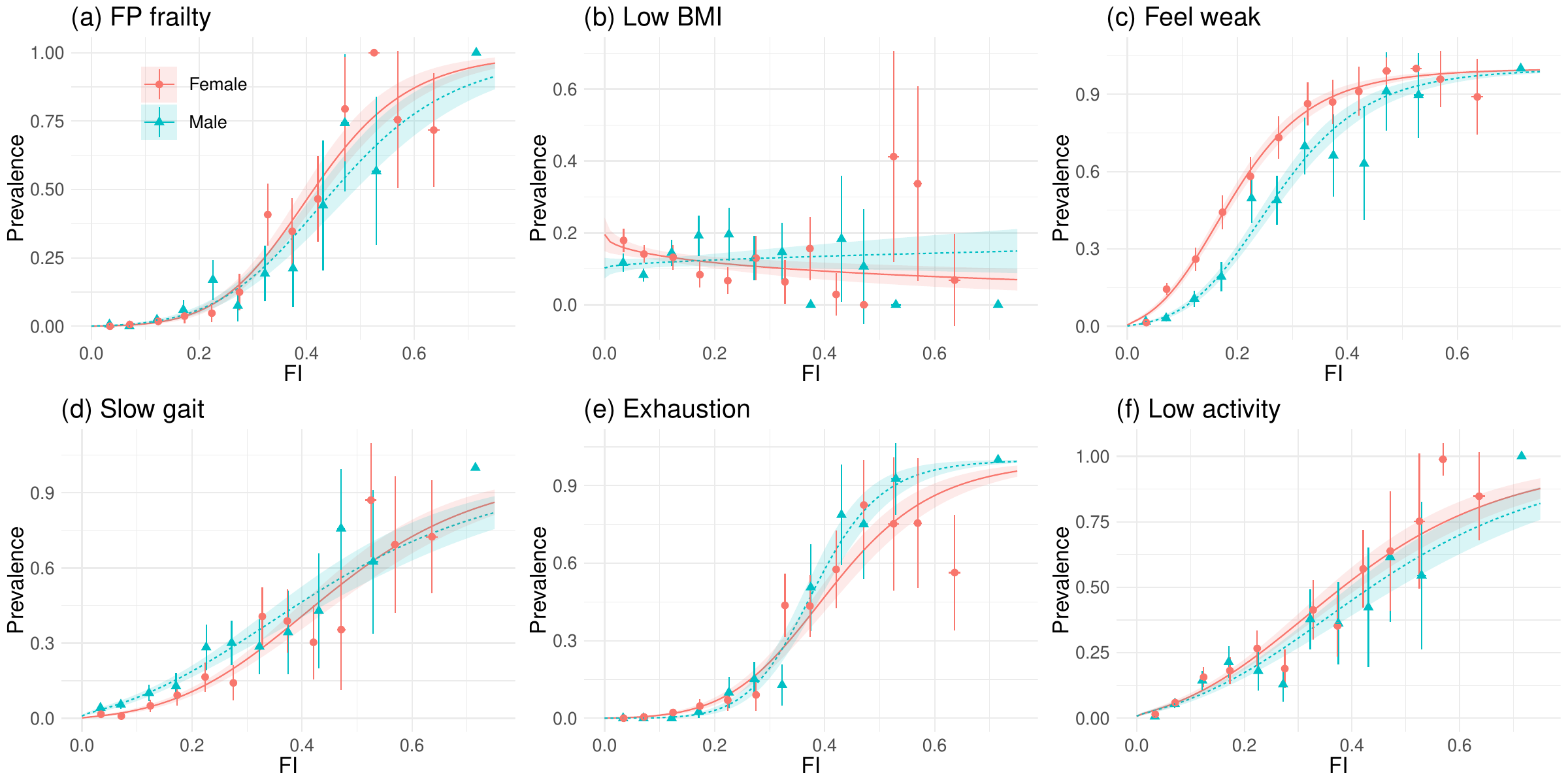} 
    \caption{\textbf{Inference calibration using the FI. NHANES (cross-sectional)} --- sex-stratified. While the prevalence of feeling weak is noticeably higher among females, the curves have the same functional form across all FPFP5 deficits. The curve can be shifted by including sex as a covariate. Points are mean $\pm$ standard error, binned by FI. Lines are sqrt logistic regression fits, $\text{logit(prob.)}\propto\sqrt{f}$.}
    \label{fig:calibration_nhanes_sex}
\end{figure*}

The calibration curves are provided in Figures: \ref{fig:calibration_hrs_sex} (HRS), \ref{fig:calibration_elsa_sex} (ELSA), and \ref{fig:calibration_nhanes_sex} (NHANES). There are differences between the sexes, although they are not very large on the natural scale. Visually, shifting the curves would reconcile most of these differences. Adding sex as a covariate in logistic regression would shift $x$. Hence we expect that these differences will be mostly captured by including sex as a covariate. 

\begin{table}
    \centering
        \caption{Cox survival hazard ratios by sex with 95\% confidence intervals (and BIC) --- HRS (time-dependent).} \label{tab:hrs_cox_sex}
\begin{threeparttable}
    \begin{tabular}{l|llll|llll}
        Predictor or  & HRS  & (males) &  &  & HRS & (females)  &  &  \\ 
        measure & Model 0 & Model 1 & Model 2 & Model 3 & Model 0 & Model 1 & Model 2 & Model 3 \\ \hline
        C & $0.68\down{0.67}\up{0.69}$ & $0.75\down{0.74}\up{0.76}$ & $0.77\down{0.76}\up{0.77}$ & $0.78\down{0.77}\up{0.79}$ & $0.70\down{0.69}\up{0.71}$ & $0.77\down{0.76}\up{0.78}$ & $0.79\down{0.79}\up{0.80}$ & $0.80\down{0.79}\up{0.81}$ \\ 
        $\Delta$BIC\tnote{1} & Ref. & -1328 & -1983 & -2265 & Ref. & -1671 & -2486 & -2772 \\ 
        Age per 10 & $2.55\down{2.43}\up{2.67}$ &  & $1.89\down{1.80}\up{1.98}$ & $1.75\down{1.66}\up{1.84}$ & $2.67\down{2.57}\up{2.77}$ &  & $1.80\down{1.73}\up{1.87}$ & $1.66\down{1.59}\up{1.74}$ \\ 
        FI per 0.1 &  & $1.50\down{1.48}\up{1.52}$ & $1.42\down{1.40}\up{1.44}$ & $1.29\down{1.26}\up{1.32}$ &  & $1.53\down{1.51}\up{1.54}$ & $1.42\down{1.41}\up{1.44}$ & $1.32\down{1.30}\up{1.35}$ \\ 
        Low BMI &  &  &  & $1.70\down{1.56}\up{1.86}$ &  &  &  & $1.62\down{1.50}\up{1.74}$ \\ 
        Slow &  &  &  & $1.26\down{1.15}\up{1.38}$ &  &  &  & $1.17\down{1.07}\up{1.28}$ \\ 
        Weakness &  &  &  & $1.08\down{0.98}\up{1.18}$ &  &  &  & $1.11\down{1.01}\up{1.23}$ \\ 
        Exhaustion &  &  &  & $1.14\down{1.05}\up{1.23}$ &  &  &  & $1.05\down{0.97}\up{1.14}$ \\ 
        Low activity &  &  &  & $1.54\down{1.42}\up{1.67}$ &  &  &  & $1.47\down{1.36}\up{1.59}$ \\ 
    \end{tabular}
\begin{tablenotes}
\item[1] Change in Bayesian information criterion (BIC), negative indicates a better model.
\end{tablenotes}
\end{threeparttable}
\end{table}

\begin{table}
    \centering
        \caption{Cox survival hazard ratios by sex with 95\% confidence intervals (and BIC) --- NHANES (time-independent).} \label{tab:nhanes_cox_sex}
\begin{threeparttable}
    \begin{tabular}{l|llll|llll}
        Predictor or & NHANES &  (males) & ~ & ~ & NHANES& (females) &  ~  & ~ \\ 
        measure & Model 0 & Model 1 & Model 2 & Model 3 & Model 0 & Model 1 & Model 2 & Model 3 \\ \hline
        C & $0.67\down{0.65}\up{0.70}$ & $0.65\down{0.62}\up{0.67}$ & $0.71\down{0.68}\up{0.73}$ & $0.72\down{0.70}\up{0.75}$ & $0.71\down{0.68}\up{0.73}$ & $0.67\down{0.64}\up{0.69}$ & $0.75\down{0.72}\up{0.77}$ & $0.75\down{0.73}\up{0.78}$ \\ 
        $\Delta$BIC\tnote{1} & Ref. & 73 & -55 & -56 & Ref. & 92 & -97 & -82 \\ 
        Age per 10 & $2.56\down{2.22}\up{2.94}$ & ~ & $2.29\down{1.98}\up{2.63}$ & $2.26\down{1.95}\up{2.61}$ & $3.02\down{2.60}\up{3.51}$ & ~ & $2.79\down{2.40}\up{3.24}$ & $2.71\down{2.31}\up{3.18}$ \\ 
        FI per 0.1 & ~ & $1.38\down{1.31}\up{1.45}$ & $1.27\down{1.21}\up{1.35}$ & $1.18\down{1.07}\up{1.31}$ & ~ & $1.45\down{1.37}\up{1.53}$ & $1.38\down{1.31}\up{1.47}$ & $1.25\down{1.12}\up{1.40}$ \\ 
        Low BMI & ~ & ~ & ~ & $1.49\down{1.13}\up{1.97}$ & ~ & ~ & ~ & $1.20\down{0.89}\up{1.63}$ \\ 
        Slow & ~ & ~ & ~ & $1.41\down{1.04}\up{1.92}$ & ~ & ~ & ~ & $1.33\down{0.94}\up{1.87}$ \\ 
        Weakness & ~ & ~ & ~ & $1.03\down{0.76}\up{1.40}$ & ~ & ~ & ~ & $1.37\down{1.05}\up{1.79}$ \\ 
        Exhaustion & ~ & ~ & ~ & $0.97\down{0.63}\up{1.49}$ & ~ & ~ & ~ & $0.90\down{0.63}\up{1.29}$ \\ 
        Low activity & ~ & ~ & ~ & $1.61\down{1.22}\up{2.12}$ & ~ & ~ & ~ & $1.19\down{0.88}\up{1.59}$ \\ \hline
    \end{tabular}
\begin{tablenotes}
\item[1] Change in Bayesian information criterion (BIC), negative indicates a better model.
\end{tablenotes}
\end{threeparttable}
\end{table}

\FloatBarrier

\section{Fit diagnostics} \label{sec:fit_diag}
We report the fit coefficients for two parametric models: logistic regression for FPFP5 deficits and Cox time-to-event regression for survival. In both cases we assumed linearity of the predictor variables  because we believe that they're easier to work with compared to using the square root. In this section we demonstrate that the linear assumption is a fair approximation. For comparison we'll include the square root which generally fits better visually.

\begin{figure*}[!ht]
    \centering
    \includegraphics[width=\textwidth]{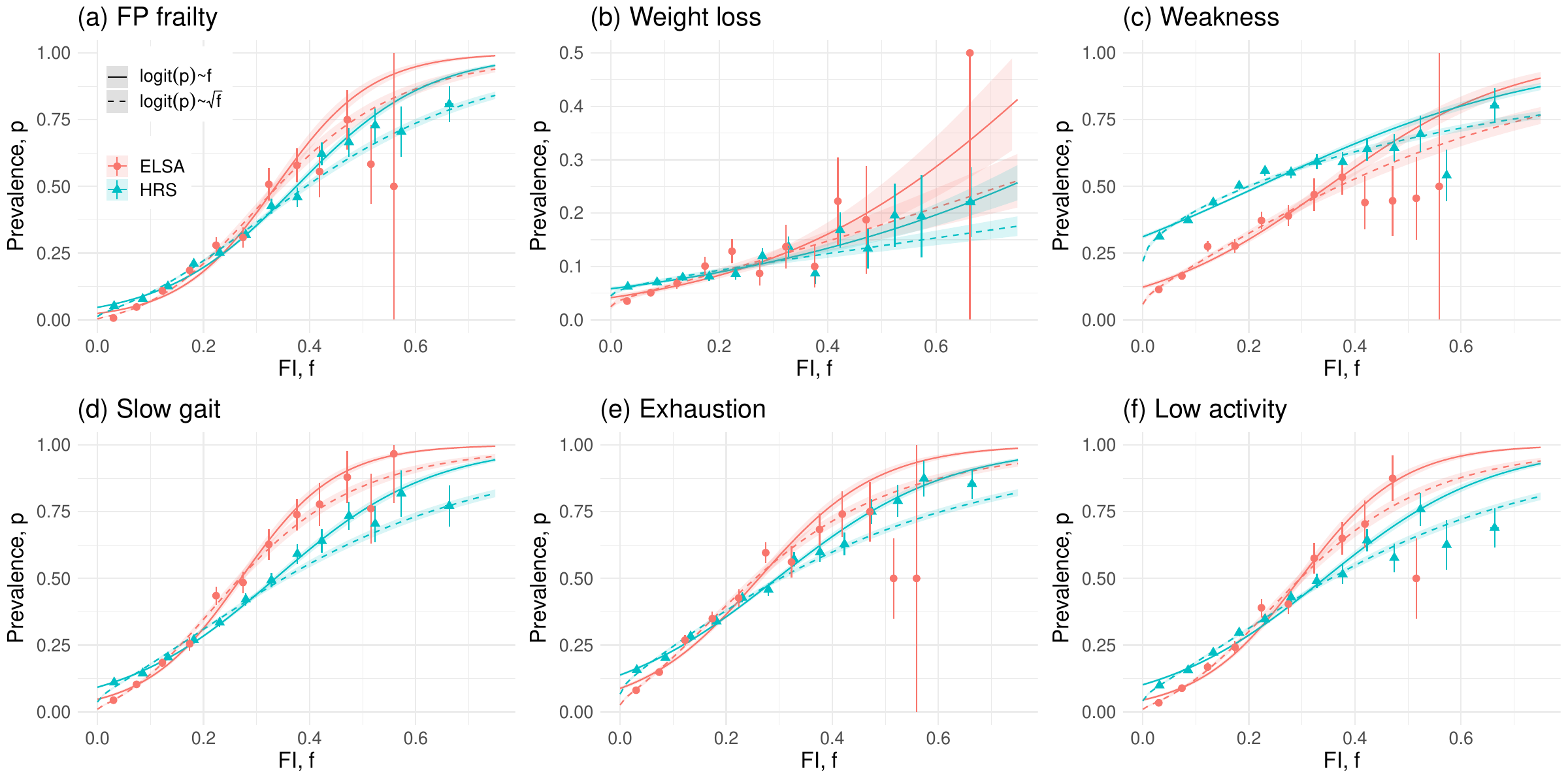} 
    \caption{A linear logistic assumption is a good approximation (solid lines), but a square root is a better fit (dashed lines) --- HRS and ELSA (longitudinal). Points are the binned future frequencies with standard errors, lines are the logistic regression fits.} \label{fig:lin_logit}
\end{figure*}

\begin{figure*}[!ht]
    \centering
    \includegraphics[width=\textwidth]{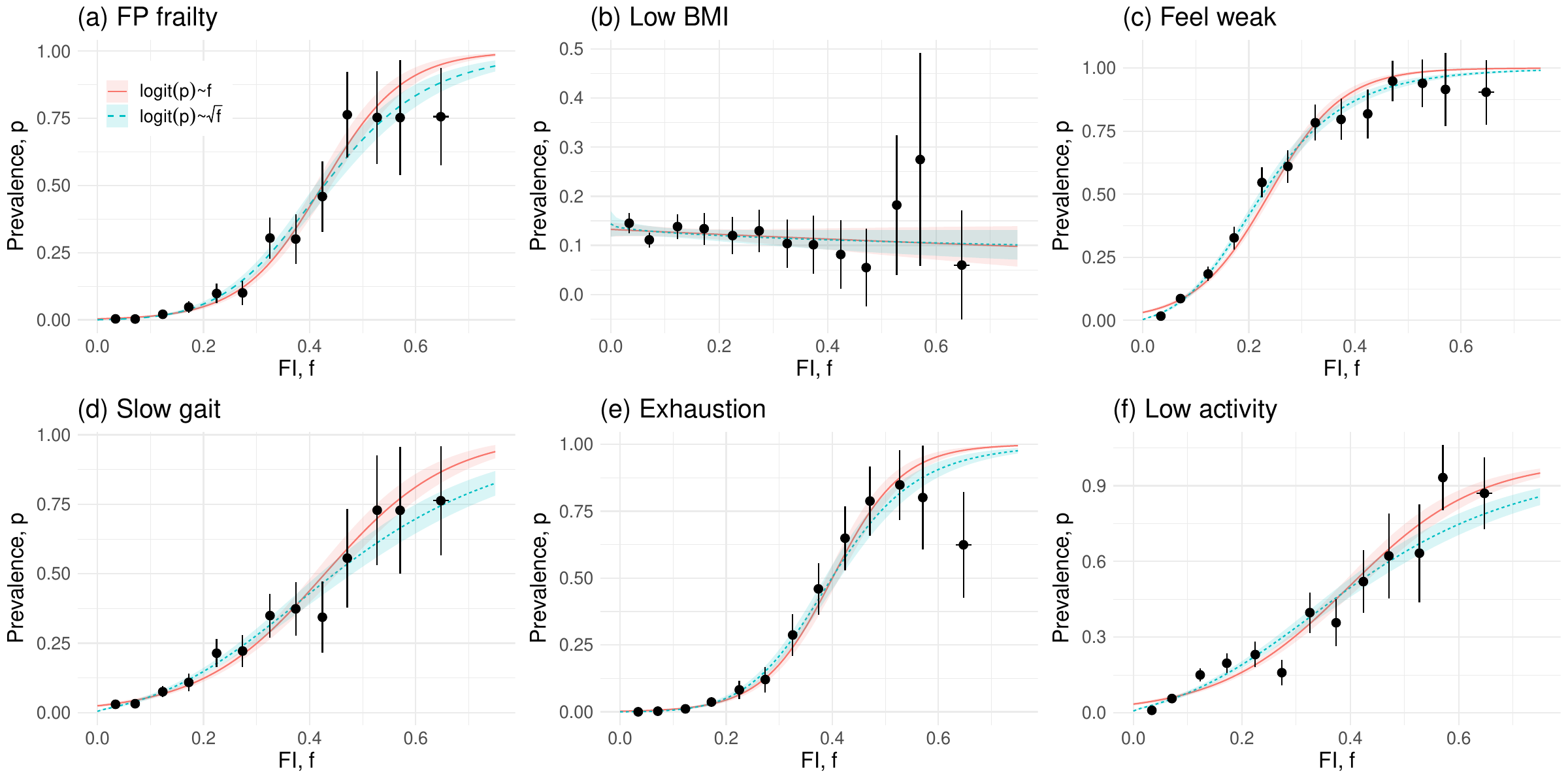} 
    \caption{A linear logistic assumption is a good approximation (solid lines), and fits the data well, as does a square root (dashed lines) --- NHANES (cross-sectional). Points are the binned frequencies with standard errors, lines are the logistic regression fits. The linear fit (solid red lines) looks good, the square root (dashed blue lines) looks about the same. } \label{fig:lin_logit_nhanes}
\end{figure*}

The linear logistic regression assumption is tested for the FI in Figure~\ref{fig:lin_logit} (HRS and ELSA) and Figure~\ref{fig:lin_logit_nhanes} (NHANES). In all cases the linear model is a good approximation, although the square root typically visually fits noticeably better.

\begin{figure*}[!ht]
    \centering
    \includegraphics[width=0.67\textwidth]{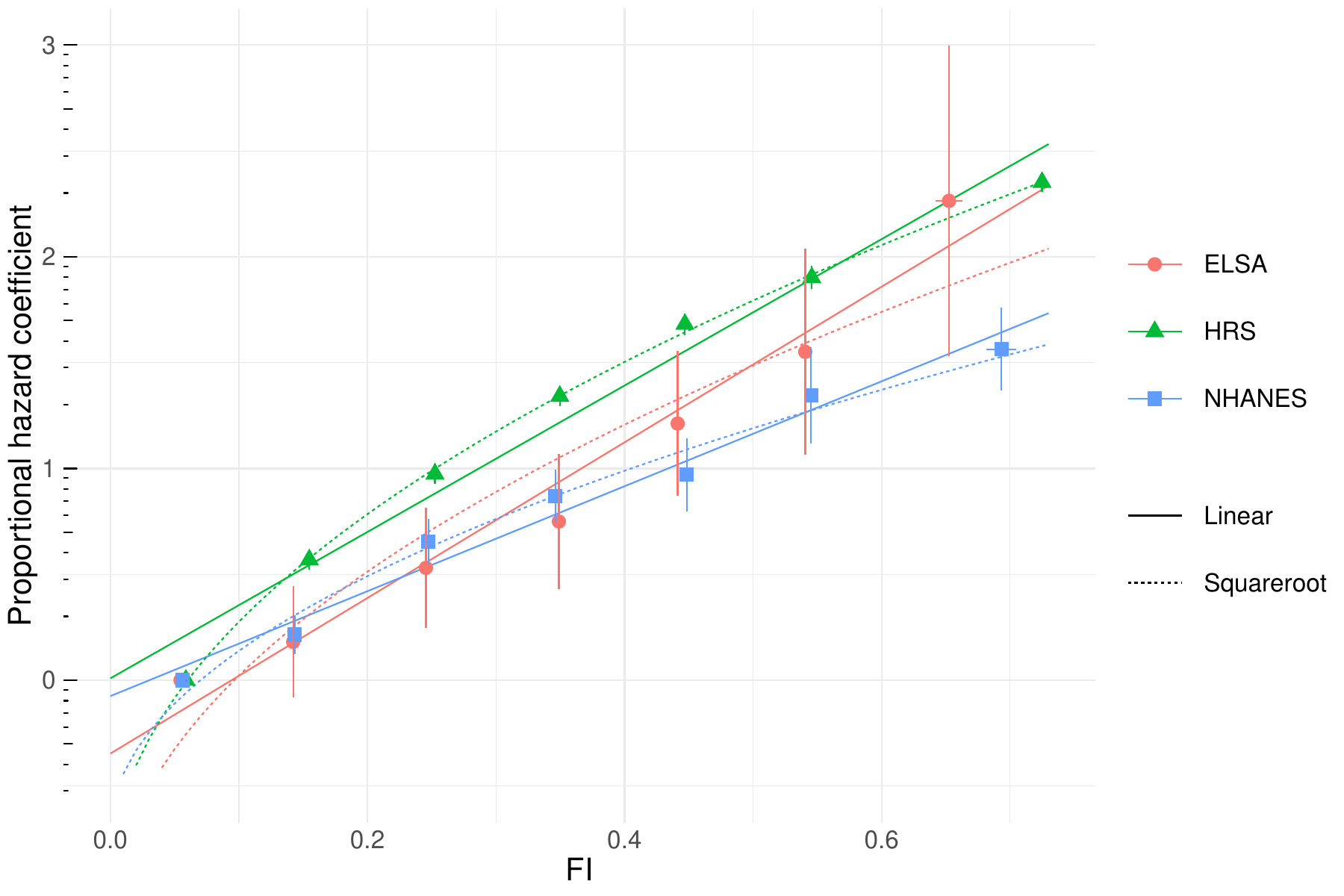} 
    \caption{A linear Cox proportional hazard assumption is fair approximation, but a square root is much more realistic. Points are the binned fits with standard errors, lines are linear regression fits to the points. For HRS and NHANES: the linear assumption is fine (solid lines) but the square root fits excellently (dashed lines). For ELSA the linear fits better, although the error bars are very large (likely due to the low number of events, as discussed in Section~\ref{sec:s_elsa}).} \label{fig:cox_ph}
\end{figure*}

The Cox proportional hazard assumption is tested for the FI in Figure~\ref{fig:cox_ph}. We again see that the linear is qualitatively correct and fits reasonably well, but for HRS and NHANES the true curve is clearly sub-linear. The square root of the FI visually fits much better.

\end{document}


\flushbottom
\maketitle
%
%
\thispagestyle{empty}


\section*{Extended Results}

\begin{figure*}[!ht]
    \centering
    \includegraphics[width=\textwidth]{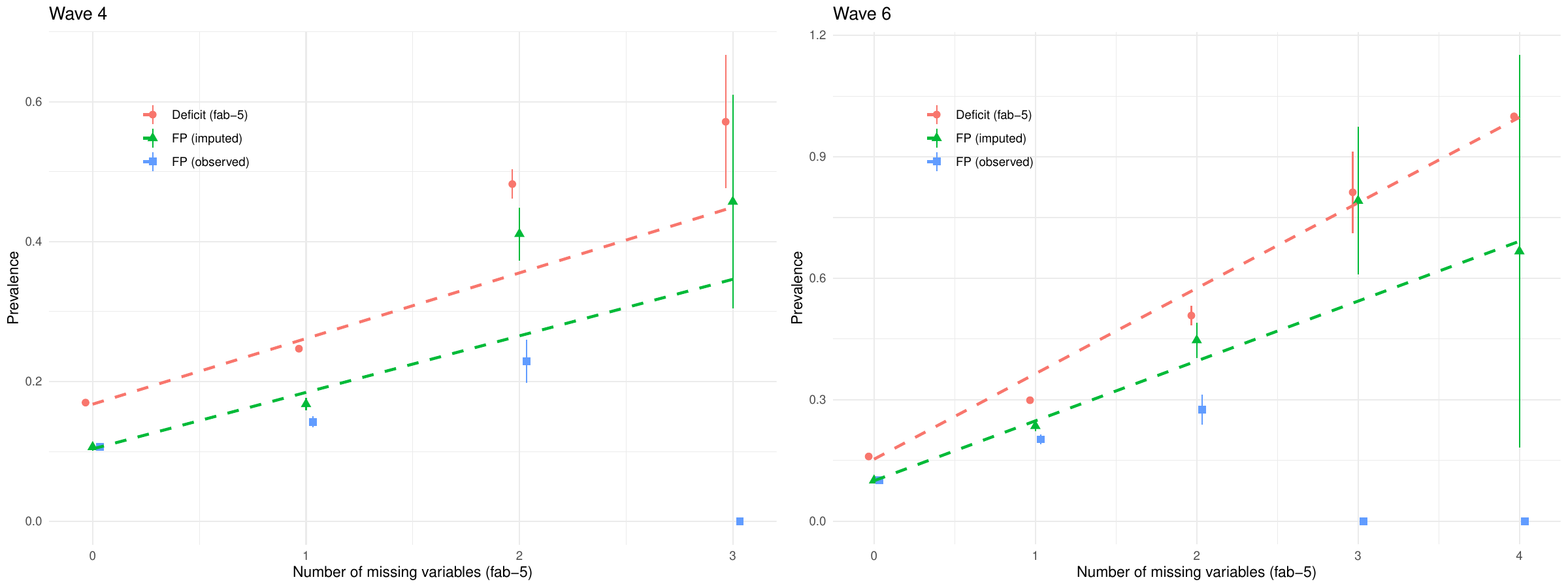}
    \caption{\textbf{Failure to impute would lead to a biased population (ELSA)}. Individuals missing measurements of the fab-5 were much more likely to have deficits (red points), increasing linearly with missingness. The imputed values showed a commensurate change in FP prevalence, suggesting they capture this effect (green triangles). The unimputed FP prevalence also shows increasing prevalence with increasing missingness, but clearly has reduced prevalence due to ignorance e.g. once 3 are missing --- i.e. 2 measured --- it is impossible to have 3+ deficits and the prevalence goes to 0.} \label{fig:missingness_elsa}
\end{figure*}

\begin{figure*}[!ht]{0.5\textwidth}
        \centering
        \includegraphics[width=\textwidth]{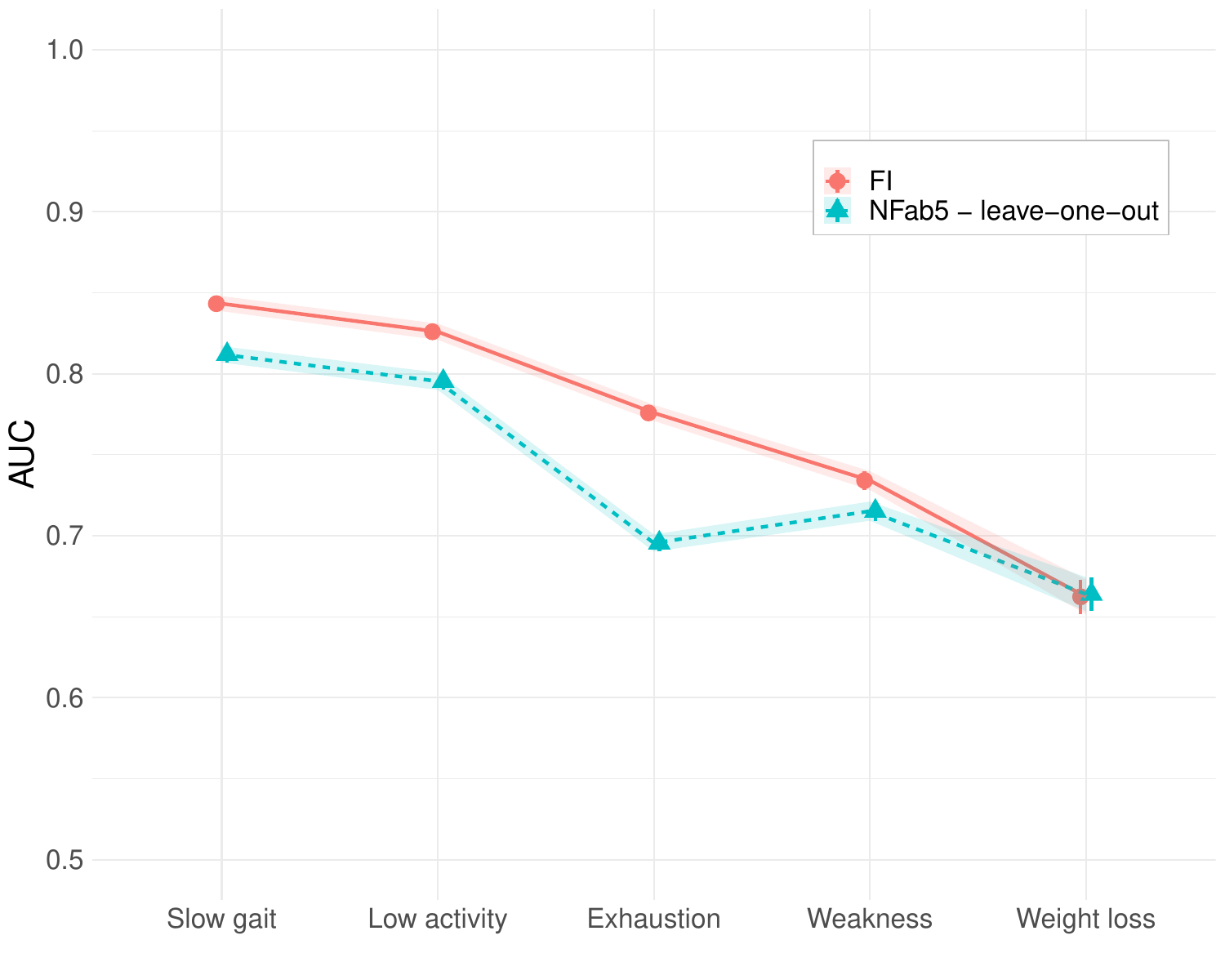} 
        \caption{AUC for current health deficits (cross-sectional). The FI includes only questionnaire data whereas the NFab5 includes the 4 deficits not being predicted (leave-one-out). The FI out-performs the NFab5 in 4/5 and ties for weight loss.}
\end{figure*}

\begin{figure*}[!ht]
    \centering
    \begin{subfigure}[t]{0.5\textwidth}
        \centering
        \includegraphics[width=\textwidth]{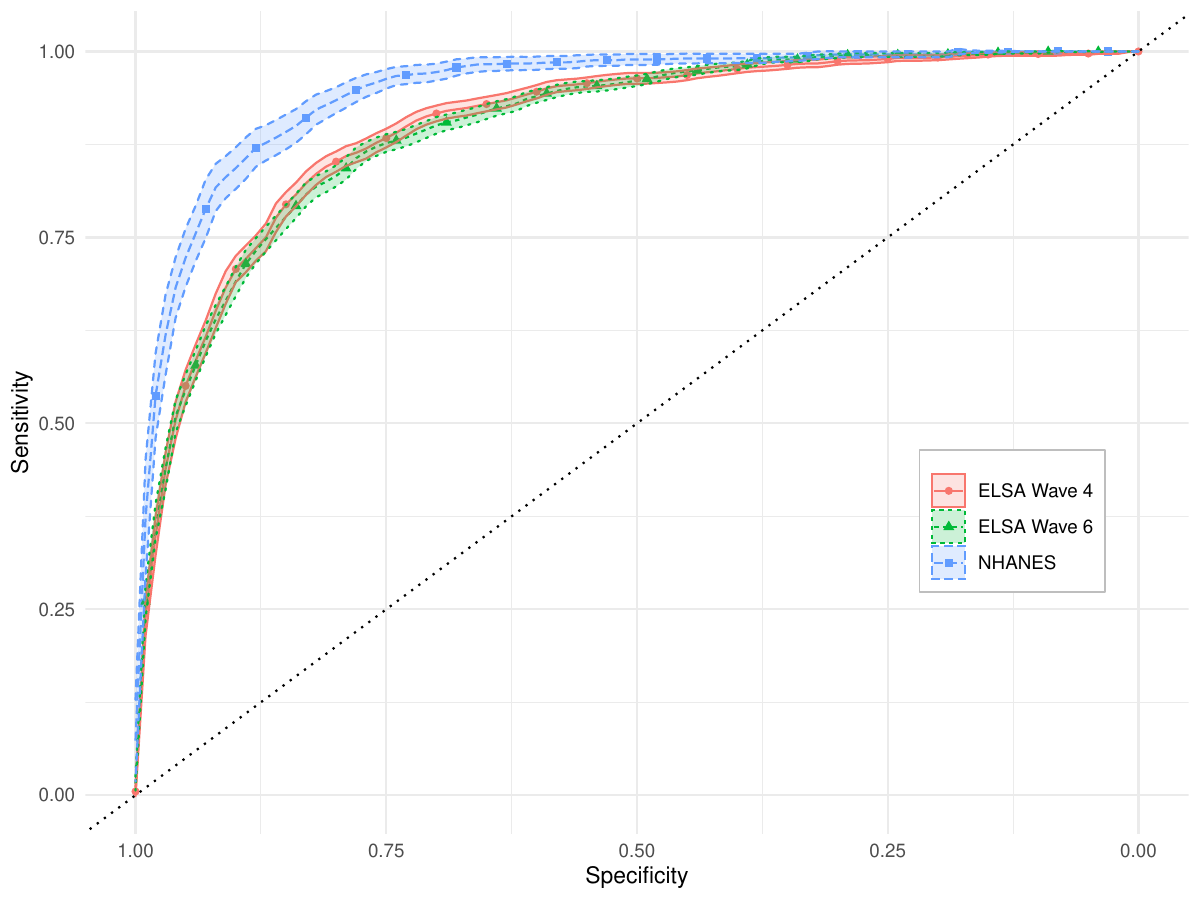} 
        \caption{ROC curve for predicting FP using the FI. The FI is an excellent predictor of the FP. AUC: $0.899\pm 0.006$ (ELSA wave 4), $0.897\pm 0.006$ (ELSA wave 6), and $0.946\pm 0.008$ (NHANES).}
    \end{subfigure}%
    ~ 
    \begin{subfigure}[t]{0.5\textwidth}
        \centering
        \includegraphics[width=\textwidth]{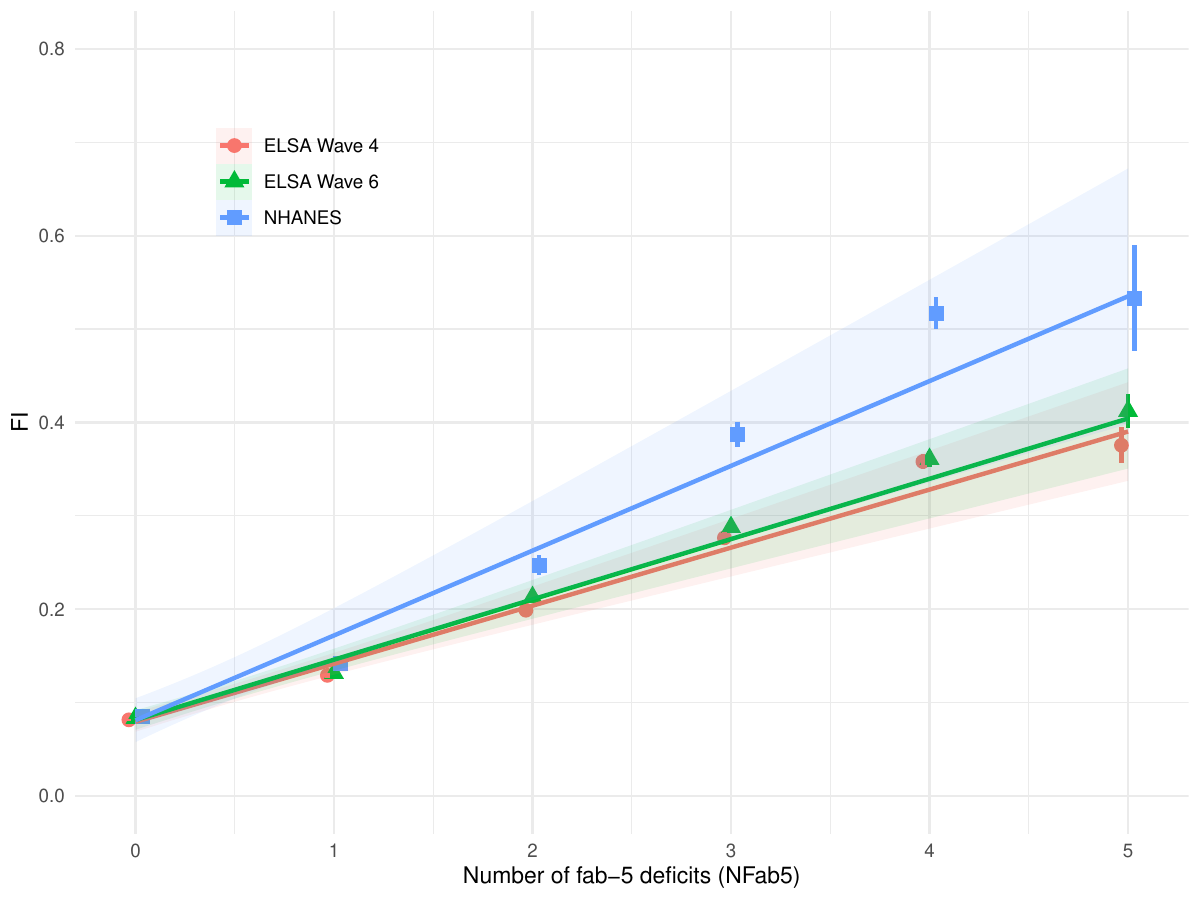} 
        \caption{The FI responds linearly to the number of fab-5 deficits (NFab5) across datasets. This suggests that the FI is a good predictor of the NFab5. Points are mean $\pm$ standard error.}
    \end{subfigure}
    \caption{\textbf{The FI and the NFab5 have a strong mutual association}. The FI is an excellent predictor of the FP (NFab5 $\geq 3$) (a) and has a linear relationship with the average number of fab-5 deficits (b).} \label{fig:fi_vs_fp}
\end{figure*}

\begin{figure*}[!ht]
    \centering
    \begin{subfigure}[t]{0.5\textwidth}
        \centering
        \includegraphics[width=\textwidth]{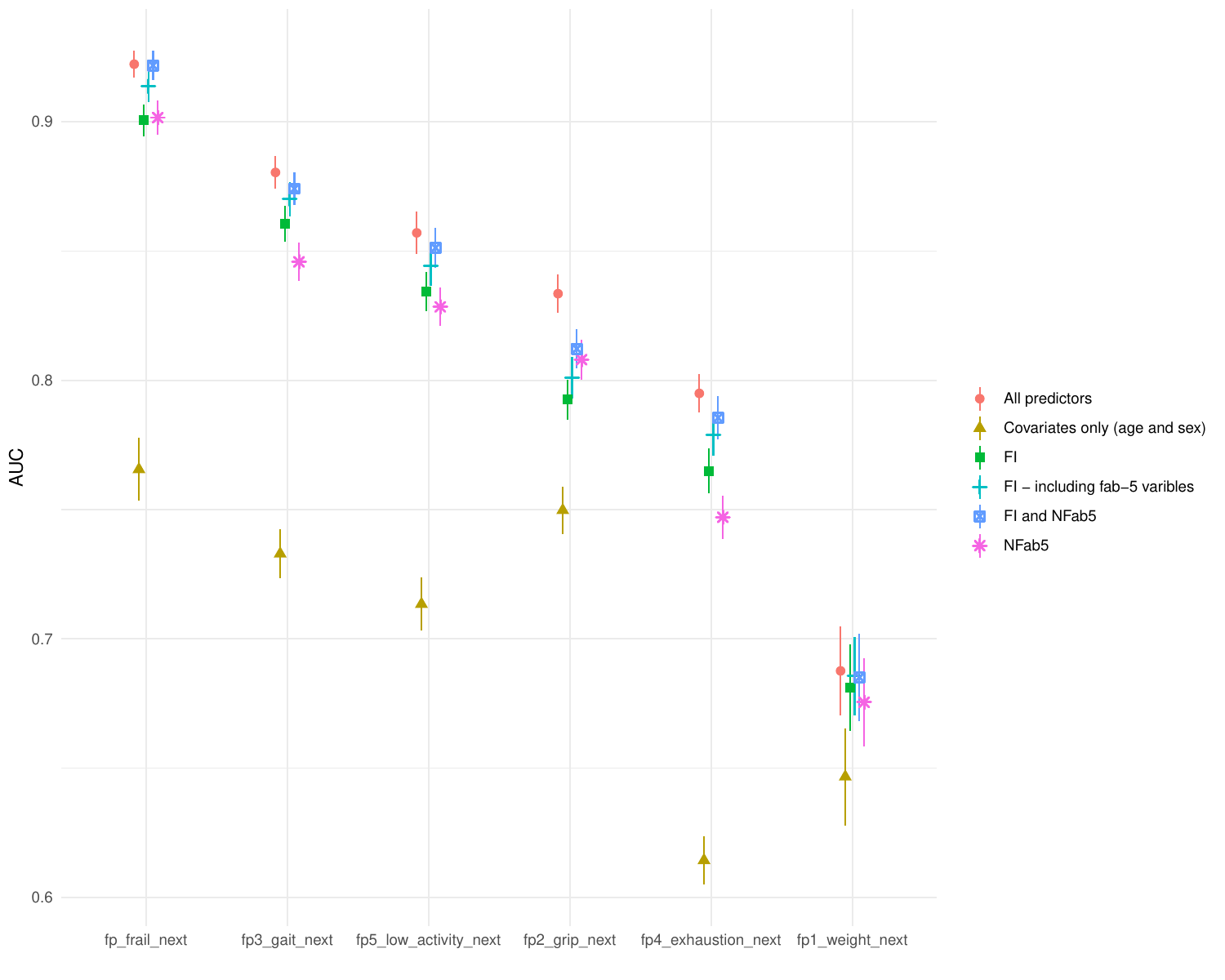} 
        \caption{AUC for predicting future health deficits, GLM (ELSA). We see that the FI (green squares) and NFab5 (pink stars) perform similarly well, much better than simply age and sex alone (tan triangles). Together the FI and NFab5 predict slightly better (dark blue x-box), but no better than the modified FI which includes the fab-5 deficits in its definition (light blue cross).}
    \end{subfigure}%
    ~ 
    \begin{subfigure}[t]{0.5\textwidth}
        \centering
        \includegraphics[width=\textwidth]{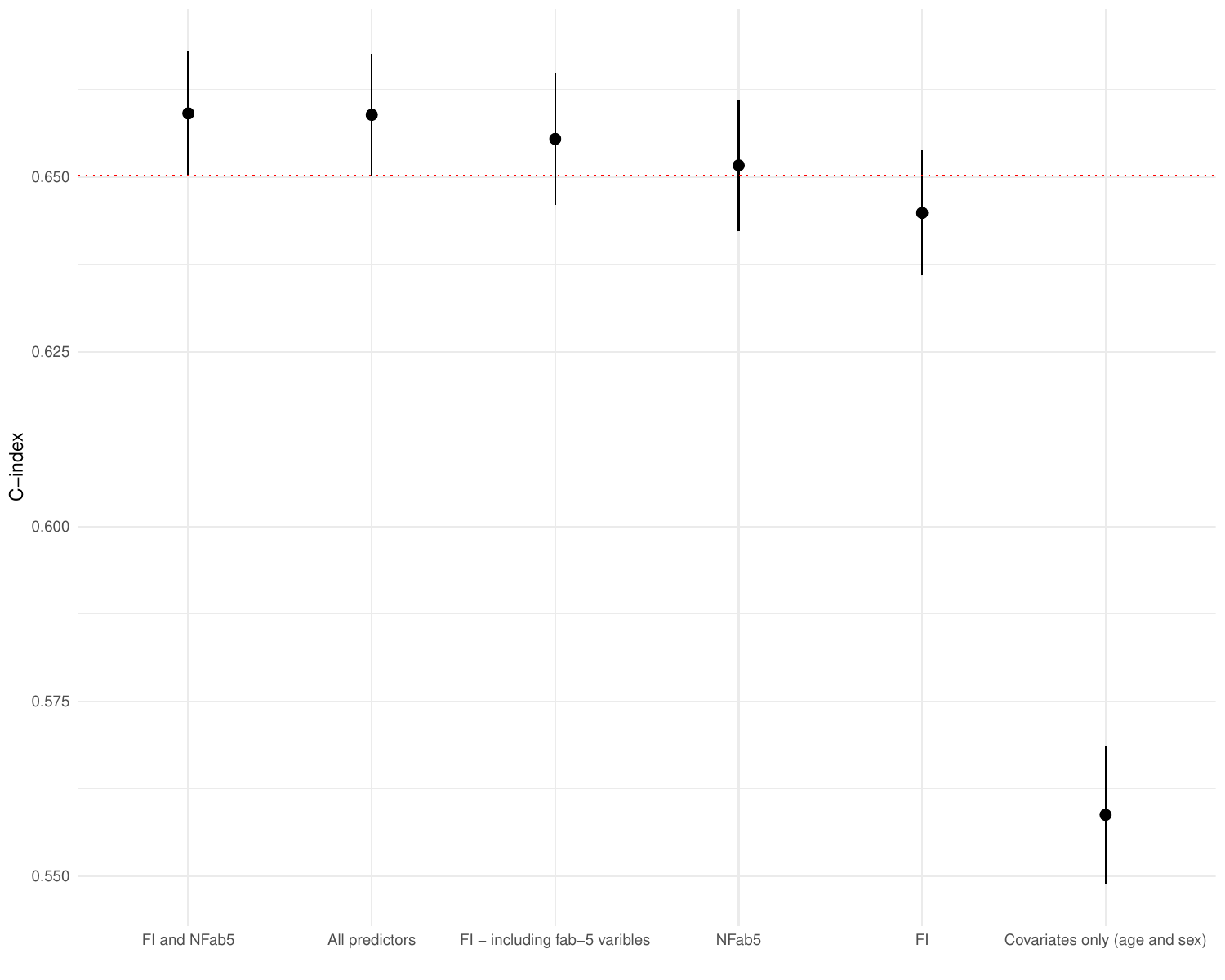} 
        \caption{Concordance (C)--index for survival. Higher is better: 0.5 is a guess and 1 is perfect. Each model performs as well as each other, within error (red line). Only the null model (age and sex) performed worse.}
    \end{subfigure}
    \caption{\textbf{The FI and NFab5 contain little complementary information regarding future health trajectory.} Using GLMs we trained models to predict future adverse outcomes comparing models incorporating both the FI and NFab5 to those which consider just one. While including both measures gave a small performance boost, an equivalent boost was observed if the fab-5 deficits were simply included in the definition of the FI. Hence the FI includes similar information to the NFab5 and can subsume the NFab5 if domain-specific information about the fab-5 is included.} \label{fig:glm}
\end{figure*}

While we observed that the FI performed better with access to the same level of information, we also observed that the FI without fab-5 knowledge and the NFab5 share a great deal of information. We can make definite how much useful information is shared between the FI and NFab5 by considering how well they predict adverse outcomes together versus alone. We used logistic and Cox regression to predict adverse outcomes at followup using: the FI alone, the NFab5 alone, the FI and NFab5 together, the FI including fab-5 variables and all of the predictors including the FI, NFab5 and the fab-5 themselves. Each model included age and sex as covariates. In Figure~\ref{fig:glm} we present the predictive power of each model. We observe that the null model performed reasonably well, but much worse than either the FI or NFab5 --- which performed equally-well. Combining the FI and NFab5 led to a small, but significant, improvement. Consistent with our initial observations, simply including the fab-5 as part of the FI yielded equivalent performance to the model which included both the FI and NFab5. We saw a minimal improvement by including all 5 of the binary fab-5 variables in the predictive models (except grip strength, which may capture persistent individual differences in strength). To summarize, the FI and NFab5 performed equally-well but slightly better if used together. This implies a small amount of complementary information. As indicated by the performance of the modified FI, this complementary information can be easily incorporated into the FI by including fab-5 variables when generating the FI, suggesting that the FI subsumes the NFab5. Finally, individual fab-5 deficits are minimally useful for prediction, implying they offer little additional information over the FI and NFab5 i.e. the aggregate forms performed as well as the specific. This is an indication that most of the deficit information is captured by the FI (or NFab5), consistent with health changes being driven by an underlying phenomenon such as frailty.

\begin{figure*}[!ht]
    \centering
    \begin{subfigure}[t]{0.5\textwidth}
        \centering
        \includegraphics[width=\textwidth]{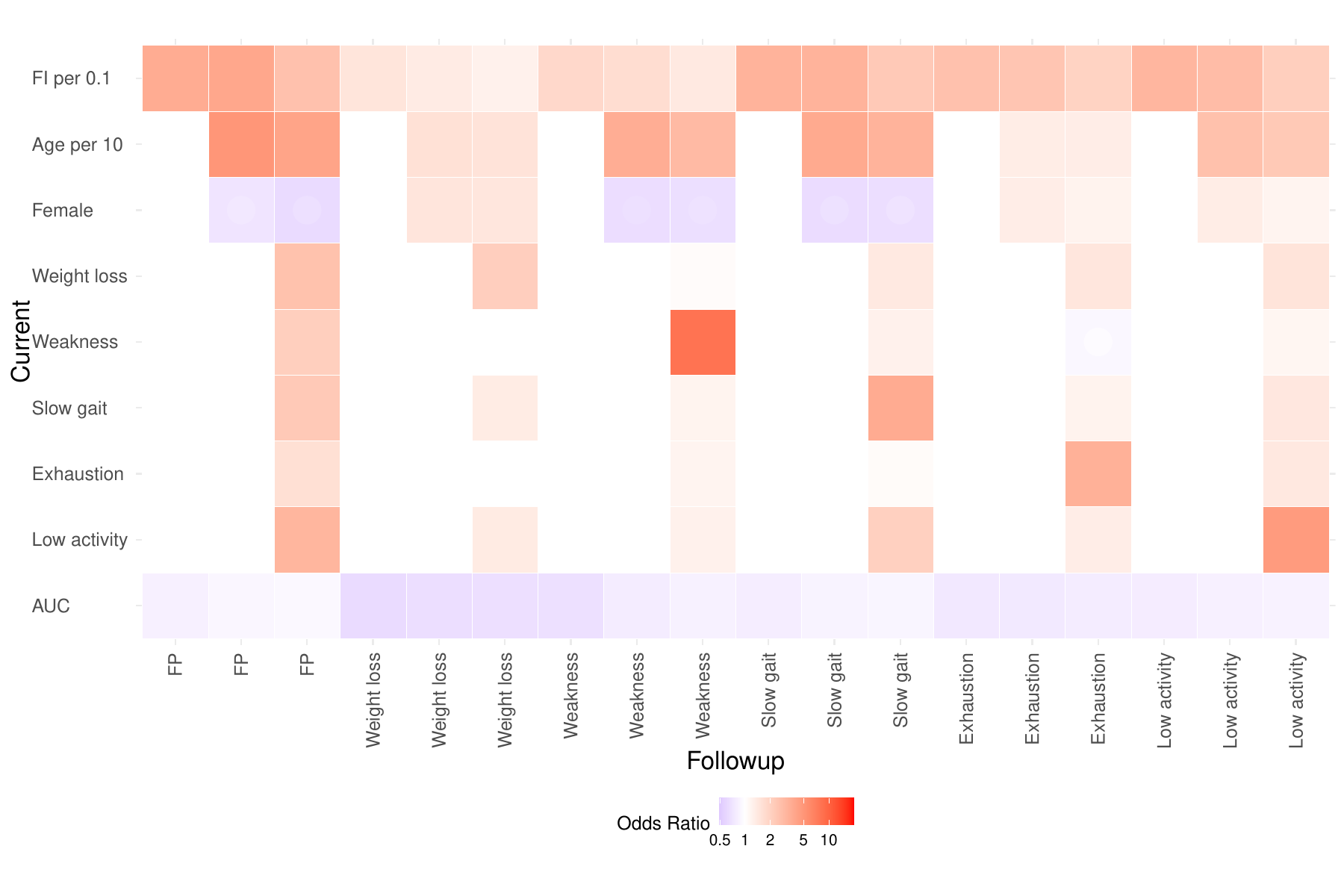} 
        \caption{Calibration curves for FI and number of fab-5 deficits (NFab5). Both the FI and NFab5 are linearly related to the the followup NFab5.}
    \end{subfigure}%
    ~ 
    \begin{subfigure}[t]{0.5\textwidth}
        \centering
        \includegraphics[width=\textwidth]{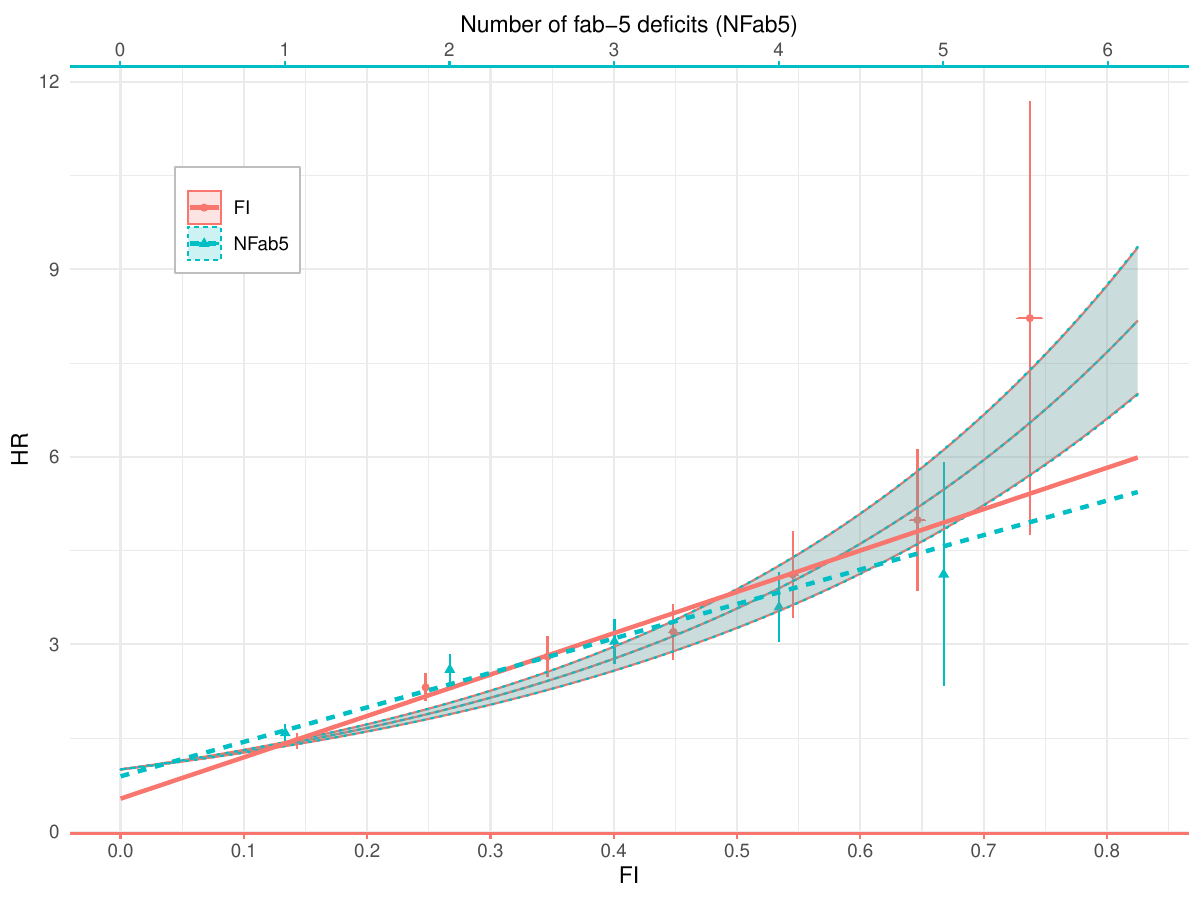} 
        \caption{Hazard rate per unit increase of the FP and FI (Cox coefficient, $\exp{(\beta)}$). Points: Cox model using ordinal scales. Both the FP and FI appear to be linearly proportional to the hazard (lines). Observe that the solid red FI line almost perfectly fits the FP data points (blue triangles). Meanwhile, the proportional hazard is also a good fit (shaded curves). Curves are Cox models treating FI and FP as continuous predictors. Note that the curves perfectly coincide. NHANES data.}
    \end{subfigure}
    \caption{\textbf{The FP and NFab5 detected a similar health phenomenon.} Both the FI and FP had linear relationships with the future number of fab-5 deficits (a) and mortality hazard (b). The NFab5 appears to be insensitive to very low and very high values (NFab5 $< 0$ and $> 5$), in contrast to the FI. This suggests that the FI is able to quantify a wider range of health states. In principal, the FI could extend to $1$ whereas the FP reaches its maximum at $5$. Empirically, the FI never reaches $1$, typically saturating near $0.7$ \cite{Mitnitski2015-ia}, nearly exactly the hazard of NFab5=5 (b).} \label{fig:calibration}
\end{figure*}

\begin{figure*}[!ht]
    \centering
    \includegraphics[width=0.5\textwidth]{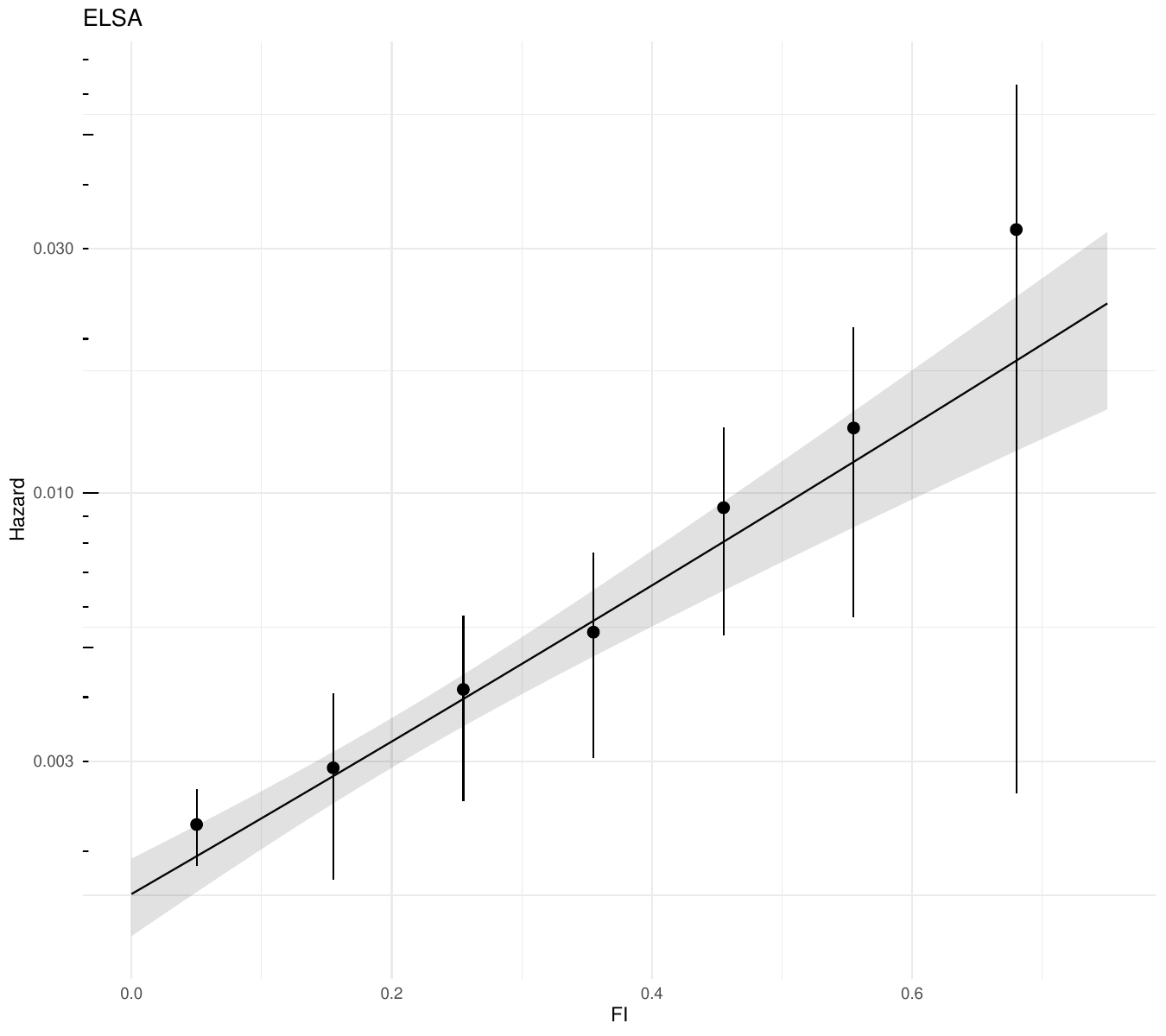} 
    \caption{Parametric hazard --- ELSA. $\beta = 3.472 \pm 0.405$, $\alpha = 0.105 \pm 0.000$, $\ln{h_0} =  -14.289 \pm 0.897$}
    \label{fig:hazard_elsa}
\end{figure*}

\subsection{Complete case analysis}

\begin{figure*}[!ht]
    \centering
    \begin{subfigure}[t]{0.5\textwidth}
        \centering
        \includegraphics[width=\textwidth]{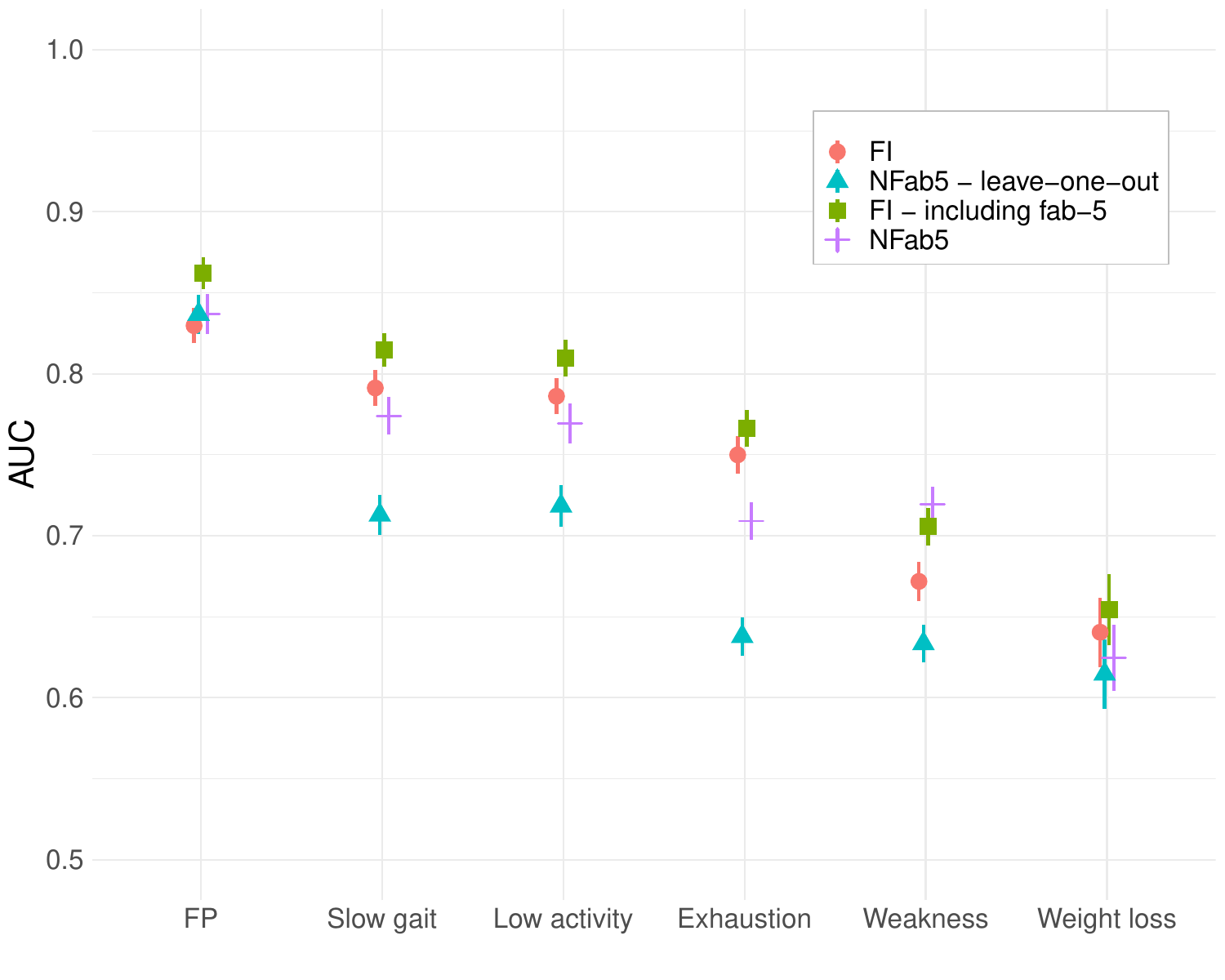} 
        \caption{Figure~\ref{fig:auc}(a) with complete-case data. We see the same relative performance as with the imputed data: the FI out-performs the leave-one-out NFab5 and the modified FI out-performs the NFab5. When the NFab5 is not greatly advantaged, the FI performs better at predicting the fab-5 deficits. Note the overall performance tends to be lower than in the imputed data, Figure~\ref{fig:auc}(a).}
    \end{subfigure}%
    \begin{subfigure}[t]{0.5\textwidth}
        \centering
        \includegraphics[width=\textwidth]{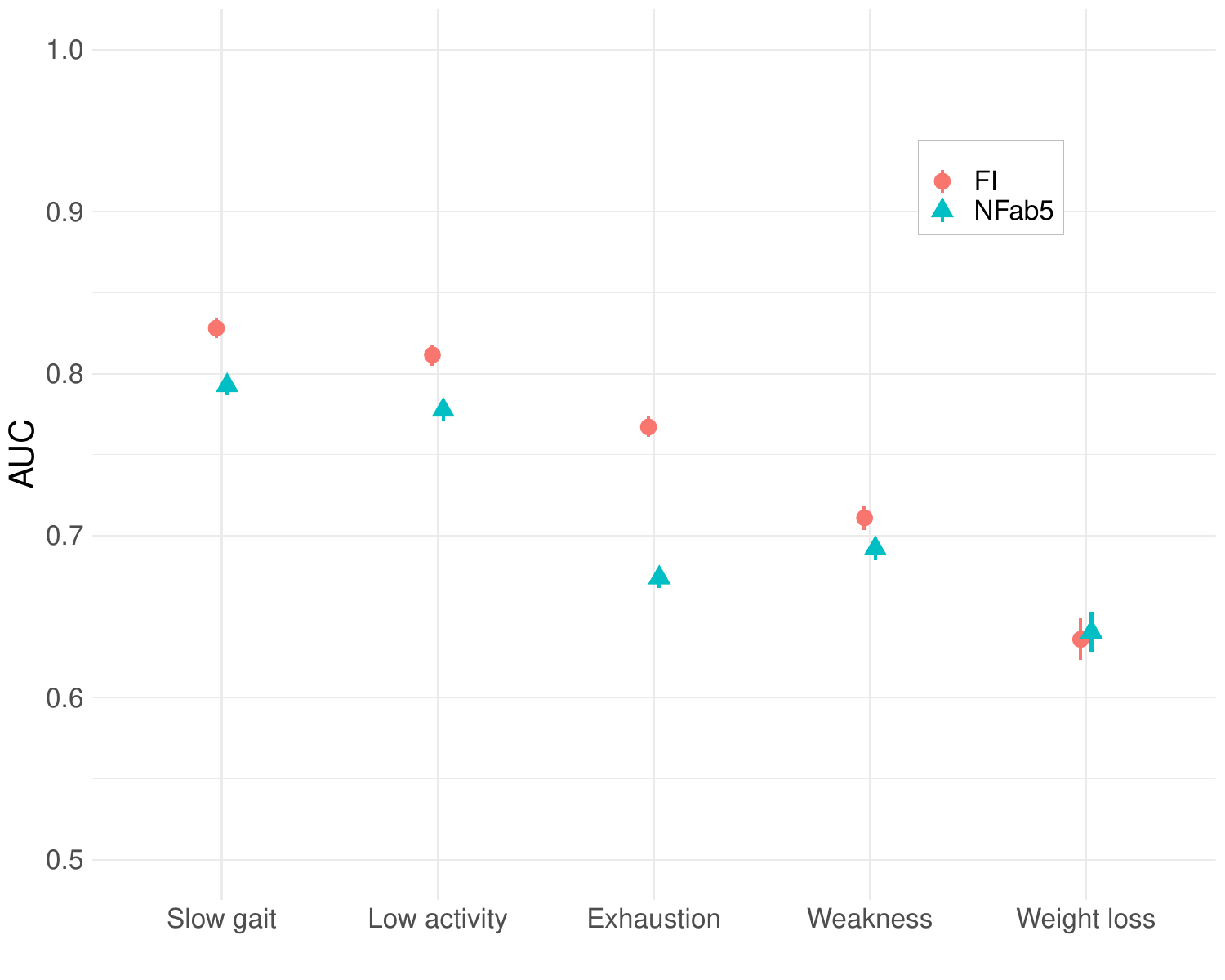} 
        \caption{Figure~\ref{fig:auc}(b) with complete-case data. AUC for current health deficits (cross-sectional). We see the same relative performance as with the imputed data, with the FI out-performing NFab5. The overall performance is notably lower, as was observed in (a).}
    \end{subfigure}%
    \caption{\textbf{The FI predicts key health deficits better than NFab5 (ELSA).} Complete-case data.} \label{fig:auc_cc}
\end{figure*}

\begin{figure*}[!ht]
    \centering
    \begin{subfigure}[t]{0.5\textwidth}
        \centering
        \includegraphics[width=\textwidth]{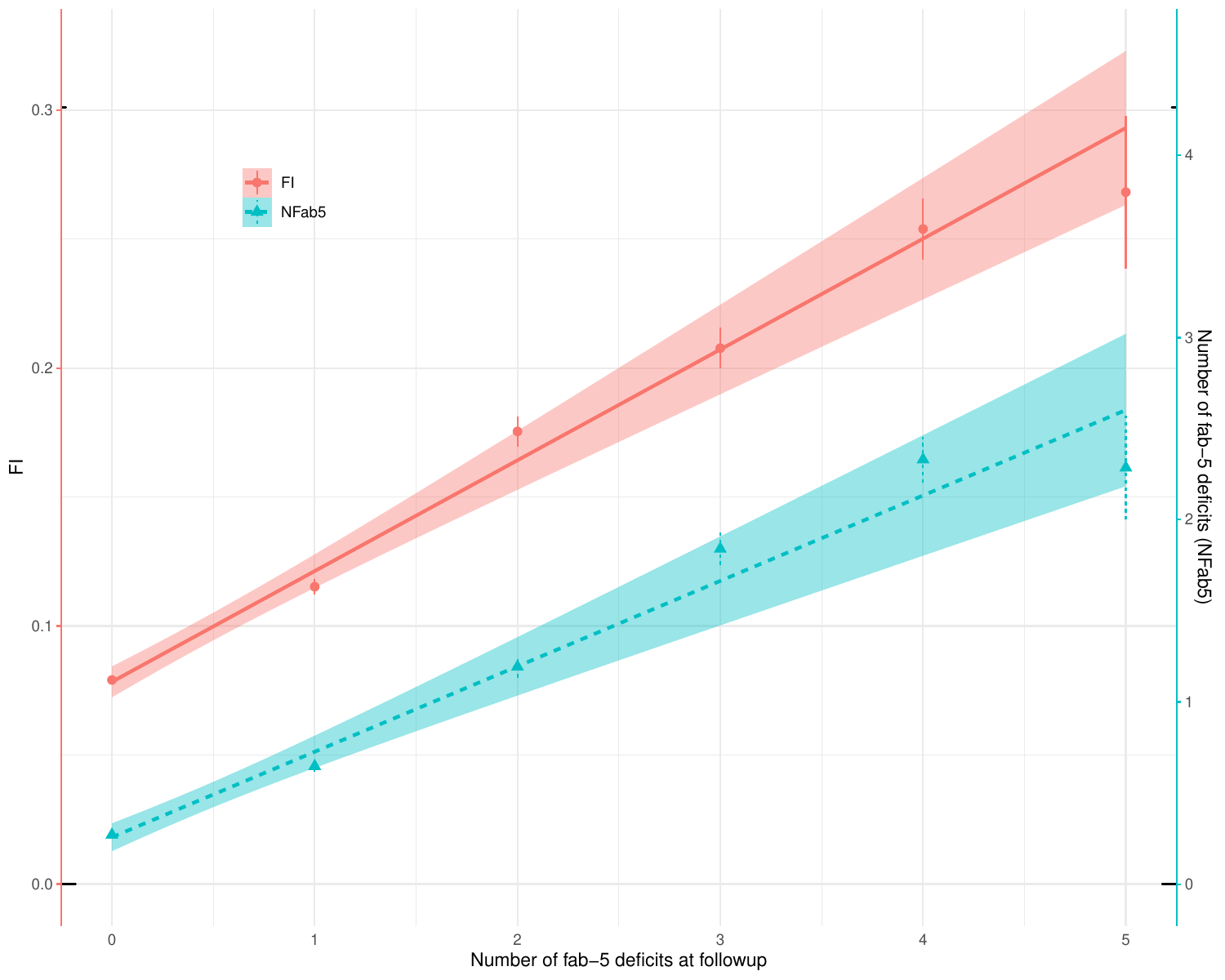} 
\caption{Calibration curves for FI and number of fab-5 deficits (NFab5), complete-case data. This is Figure~\ref{fig:calibration}(a) with complete-case data instead of imputed. We see nearly identical linear relationships as in the imputed data. Both the FI and NFab5 are linearly related to the the followup NFab5.}
    \end{subfigure}%
    ~ 
    \begin{subfigure}[t]{0.5\textwidth}
        \centering
        \includegraphics[width=\textwidth]{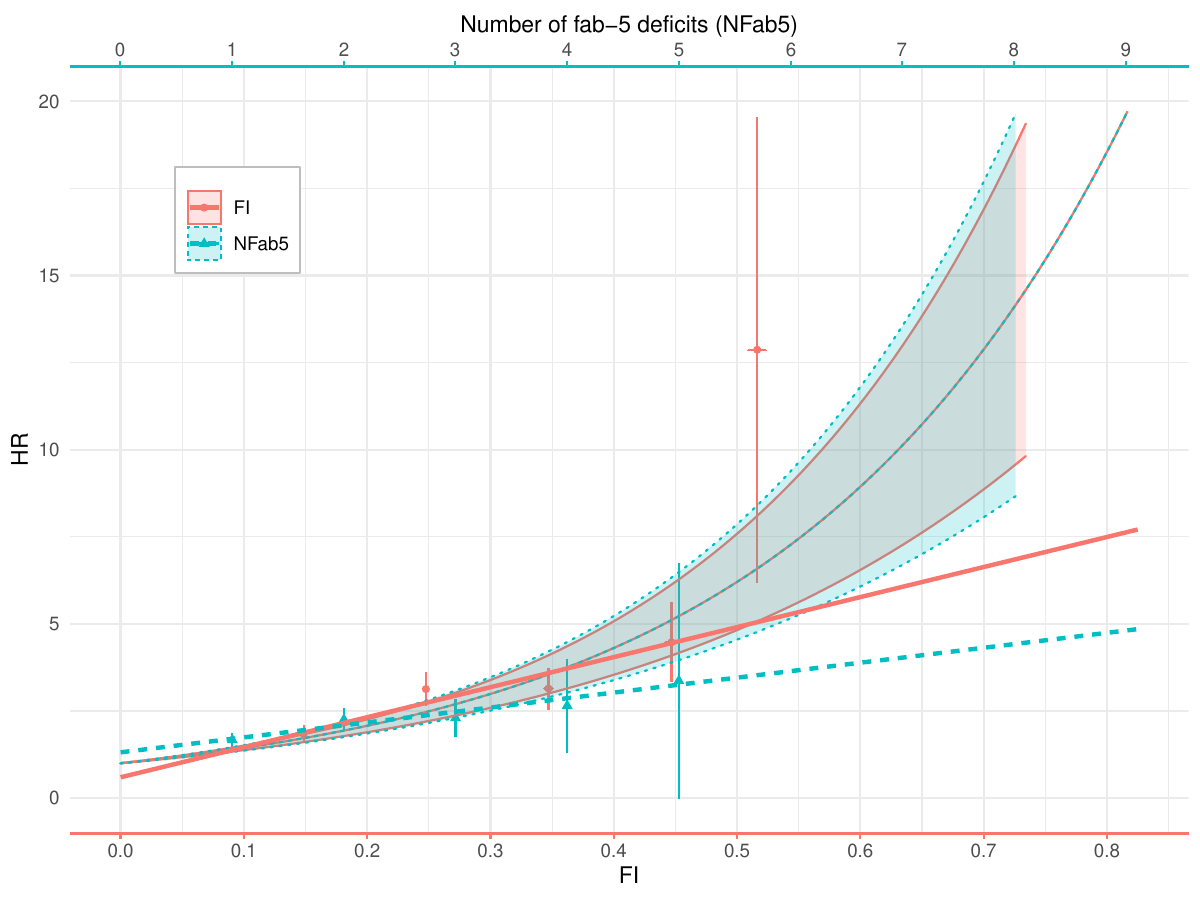} 
        \caption{Hazard rate per unit increase of the FP and FI (Cox coefficient, $\exp{(\beta)}$), complete-case data. This is Figure~\ref{fig:calibration}(b) with complete-case data instead of imputed. Points: Cox model using ordinal scales. Shaded curves: Cox proportional hazard model. We see that, relative to the imputed data, the errorbars are much larger, and the higher values of FI are very sparse. The NFab5 remains linear whereas the HR of the FI appears to grow faster than before, becoming superlinear at 0.5 whereas previously it did near 0.75. This suggests that the superlinear observation may simply be due to a lack of data (note the large errorbars).}
    \end{subfigure}
    \caption{\textbf{The FP and NFab5 detected a similar health phenomenon.} Complete-case data. } \label{fig:calibration_cc}
\end{figure*}

We confirm our results hold in the complete-case data. Our central result, that the FI out-performs the NFab5 for predicting the fab-5, is confirmed for the complete-case in Figure~\ref{fig:auc_cc}. Xue \textit{et al.}\cite{Xue2021-zx} have previously observed the presence of non-linear behaviour in the mortality hazard from NFab5 = 4 to NFab5 = 5, which we see neither in the imputed data nor in the complete case data, the latter is reported in Figure~\ref{fig:calibration_cc}. We do, however, observe that the errorbars are very large in Figure~\ref{fig:calibration_cc}(b) and, comparing with Figure~\ref{fig:calibration}(b), it is clear that a lack of data at large values can cause a sudden uptick in hazard which looks remarkably similar to the ``threshold effect'' reported by Xue \textit{et al.} (they did not impute).

\begin{figure*}[!ht]
    \centering
    \begin{subfigure}[t]{0.5\textwidth}
        \centering
        \includegraphics[width=\textwidth]{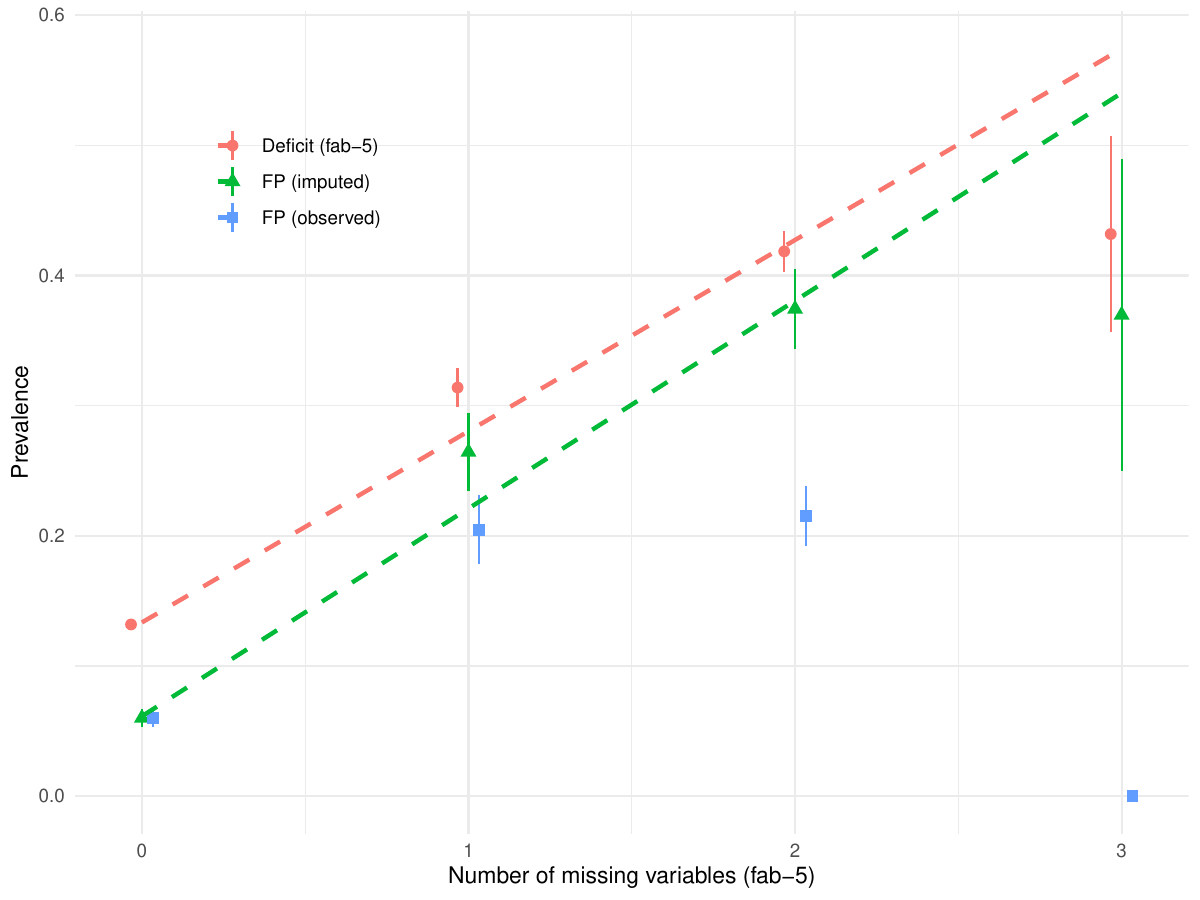} 
        \caption{Individuals with missing fab-5 variables were more likely to have deficits. Deficit frequency increased linearly with increasing missingness (red points). In parallel, the imputed FP prevalence also increases linearly (green triangles). This implies a reasonable imputation. Conversely, the observed FP prevalence appears to under-estimate the true prevalence, although it does increase from 0 to 1 missing variables.}
    \end{subfigure}%
    ~ 
    \begin{subfigure}[t]{0.5\textwidth}
        \centering
        \includegraphics[width=\textwidth]{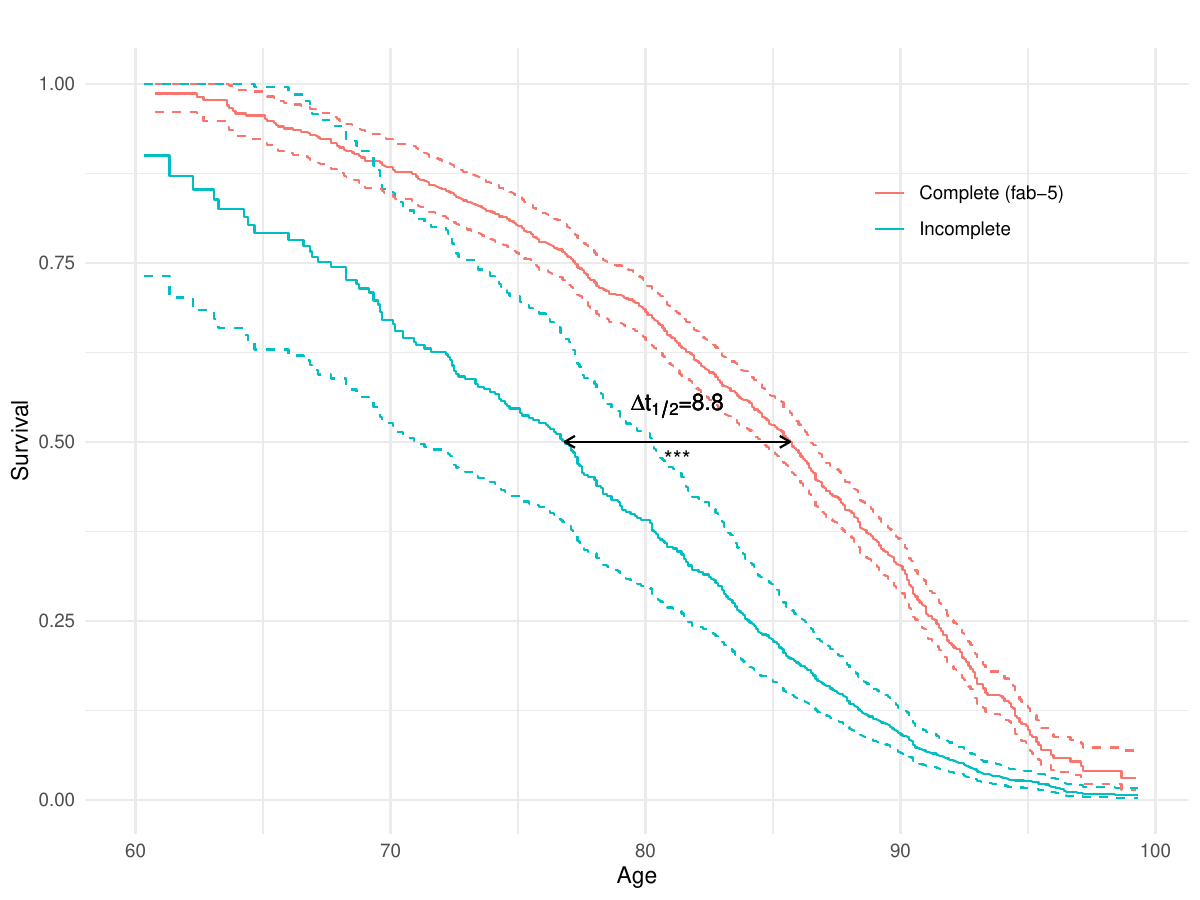} 
        \caption{Individuals with all of the fab-5 variables measured lived significantly longer, $p = 3\cdot10^{-21}$ (log-rank test). At the half-life, individuals with complete data would live 8.8 years long.}
    \end{subfigure}%
    \caption{\textbf{Failure to impute would lead to a biased population}. Individuals missing any fab-5 variables had significantly worse health, having more deficits and a higher risk of death. Similar results are seen for ELSA (supplemental).} \label{fig:missingness}
\end{figure*}

In Figure~\ref{fig:missingness} we illustrate how failure to impute can lead to a severe bias in the population. Individuals with missing fab-5 variables had worse health, as indicated by having a higher risk of death and deficits. Imputation appears to ameliorate this issue, at least partially. 

\FloatBarrier

\FloatBarrier

\section{Math --- I think this is depricated}

\subsection{Cox proportional hazard modelling}
We want to compare the Cox fit on ordinal scale --- encoding using dummy variables --- to that of a linear model for a single predictor, $f$. The linear Cox model is,
\begin{align}
    \frac{\lambda}{\lambda_0} = \exp{(\beta f)}
\end{align}
where $\lambda$ is the risk of death, $\lambda_0$ is the unoberved baseline risk at $f=0$ and $\beta$ is the Cox coefficient. Conversely, in the ordinal scale picture we instead have
\begin{align}
    \frac{\lambda_f}{\lambda_r} = \exp{(\beta_f)}    
\end{align}
where $\lambda_f$ is the hazard at the value $f$, $\lambda_r$ is the hazard at a reference value $f_r$ and $\beta_f$ is the Cox coefficient. In order to put $\lambda_f$ on the scale of baseline hazard, we need $\lambda_r$ which is unknown. Our solution is to assume that the reference hazard is close enough to linear that we can approximate it using the linear Cox model,
\begin{align}
    \frac{\lambda_r}{\lambda_0} \approx \exp{(\beta f_r)}
\end{align}
then we have
\begin{align}
    \frac{\lambda_f}{\lambda_0} \approx \exp{(\beta_f + \beta f_r)}.    
\end{align}
That is, we shift up each ordinal coefficient by $\beta f_r$. Since $f_r$ is typically small, this approximation should be valid so long as the linear Cox model is reasonably close to reality.
\bibliography{ref}